\documentclass[11pt]{article}
\usepackage[margin=1in]{geometry}
\usepackage{amsmath,amssymb,amsthm,mathtools}
\usepackage{enumitem}
\usepackage{booktabs}
\usepackage{float}
\usepackage{comment}
\usepackage{algorithm}
\usepackage{hyperref}
\usepackage{xcolor}
\usepackage{tikz}
\usepackage{microtype}
\usepackage{complexity}
\usepackage{natbib}
\usepackage{MnSymbol}
\usepackage[table]{xcolor}
\usepackage{tabularx}
\usepackage{aliascnt}
\DeclareFontShape{OT1}{cmr}{m}{scit}{<->ssub*cmr/m/sc}{}
\usepackage[framemethod=tikz]{mdframed}
\definecolor{niceMagenta}{HTML}{C2185B}

\NewCommandCopy{\originalcitep}{\citep}
\NewCommandCopy{\originalcitet}{\citet}
\NewCommandCopy{\originalcitealp}{\citealp}
\RenewDocumentCommand{\citep}{s}{\originalcitep*}
\RenewDocumentCommand{\citet}{s}{\originalcitet*}
\RenewDocumentCommand{\citealp}{s}{\originalcitealp*}

\hypersetup{colorlinks=true,linkcolor=niceMagenta,citecolor=green!40!black,urlcolor=niceMagenta,
 pdftitle={On the Role of Geometry in Solving Variational Inequalities},
 pdfauthor={Ioannis Anagnostides, Ioannis Panageas, Jingming Yan}}
\renewcommand{\R}{\mathbb R}
\newcommand{\Q}{\mathbb Q}
\newcommand{\inner}[2]{\langle #1,#2\rangle}
\newcommand{\norm}[1]{\left\lVert #1\right\rVert}
\newcommand{\calK}{\mathcal{K}}

\renewcommand{\Gap}{\operatorname{Gap}}
\renewcommand{\poly}{\operatorname{poly}}

\newcommand{\sgn}{\operatorname{sgn}}

\newcommand{\problem}[1]{\textnormal{\textsc{#1}}}
\newcommand{\PureCircuit}{\problem{PureCircuit}}
\newcommand{\LinVI}{\problem{LinVI}}

\newtheorem{theorem}{Theorem}[section]

\newaliascnt{lemma}{theorem}
\newtheorem{lemma}[lemma]{Lemma}
\aliascntresetthe{lemma}

\newaliascnt{proposition}{theorem}

\aliascntresetthe{proposition}

\newaliascnt{corollary}{theorem}
\newtheorem{corollary}[corollary]{Corollary}
\aliascntresetthe{corollary}

\theoremstyle{definition}
\newaliascnt{definition}{theorem}
\newtheorem{definition}[definition]{Definition}
\aliascntresetthe{definition}

\newaliascnt{remark}{theorem}
\newtheorem{remark}[remark]{Remark}
\aliascntresetthe{remark}

\newaliascnt{observation}{theorem}

\aliascntresetthe{observation}

\DeclareMathOperator*{\argmax}{arg\,max}

\renewcommand{\vec}[1]{\boldsymbol{#1}}
\newcommand{\mat}[1]{\mathbf{#1}}
\newcommand{\vx}{\vec x}
\newcommand{\vxp}{\vec x'}
\newcommand{\Fmax}{F_{\max}}
\newcommand{\deltapc}{\epsilon}
\newcommand{\norfun}{\phi_{\Nor}}
\newcommand{\purfun}{\phi_{\Purify}}
\newcommand{\defeq}{\coloneqq}
\newcommand{\Nor}{{\mathsf{NOR}} }
\newcommand{\Purify}{{\mathsf{PURIFY}}}

\usepackage{xcolor}

\usepackage[nameinlink,capitalise,noabbrev]{cleveref}
\newcommand{\ip}[2]{\langle #1,#2\rangle}
\newcommand{\abs}[1]{\lvert #1\rvert}

\newcommand{\one}{\mathbf{1}}
\newcommand{\eps}{\epsilon}
\newcommand{\Z}{\mathbb Z}

\setlist[itemize]{leftmargin=2.2em,itemsep=.2em,topsep=.35em}
\setlist[enumerate]{leftmargin=2.2em,itemsep=.2em,topsep=.35em}
\allowdisplaybreaks[1]
\title{On the Role of Geometry in Solving Variational Inequalities}
\author{
    Ioannis Anagnostides\\
    {\small Carnegie Mellon University}
    \and
    Ioannis Panageas\\
    {\small University of California, Irvine}
    \and
    Jingming Yan\\
    {\small University of California, Irvine}
}
\date{}
\begin{document}
\maketitle
\begin{abstract}
The geometry of constraints in variational inequalities (VIs) plays a central role in applications ranging from resource allocation and equilibrium computation to constrained optimization. Yet, its effect on computational complexity has received surprisingly little attention beyond special cases. We examine this question when the constraint set is an $\ell_p$ ball. This family of problems encompasses two of the most extensively studied classes of problems in algorithmic game theory: symmetric bimatrix games for $p = 1$ and multi-player polymatrix games for $p = \infty$.  Both of these problems are known to be \PPAD-complete even for affine operators.

% First, for the Euclidean $p = 2$ case, our results reveal a sharp separation. For affine operators, we give a polynomial-time algorithm, whereas for $p=1$ or $p=\infty$ the affine case is already \PPAD-hard. In contrast, for quadratic operators, we show that even the Euclidean case is \PPAD-hard, even at inverse polynomial precision. Beyond the Euclidean case, we show that for any $p \neq 2$, the problem becomes \PPAD-hard even for affine operators. This shows that the Euclidean case is special. 

We establish a comprehensive complexity landscape for VIs over \(\ell_p\) balls across the full range \(p\in[1,\infty]\). First, for any $p\neq 2$, we show that finding an inverse-polynomial approximate VI solution is $\PPAD$-complete even for affine operators.
In contrast, for \(p=2\), we give a polynomial-time algorithm for affine VIs, extending a classic result of Ye (1992) beyond symmetric operators. On the other hand, once the operator becomes quadratic, we establish \(\PPAD\)-completeness even at inverse polynomial precision.

From a technical standpoint, we develop a new toolkit for exploiting the geometry of the constraint set and the structure of the operator to characterize the complexity of equilibrium and optimization problems, which is of independent interest.
\end{abstract}

\clearpage
\tableofcontents
\clearpage
\section{Introduction}

\emph{Variational inequalities (VIs)} provide a unifying language for a host of applications in equilibrium computation, constrained optimization, and complementarity problems~\citep{Facchinei03:Finite}. In each of these applications, the geometry of the feasible sets plays a prominent role. For example, flow-conservation constraints arise in network routing, product constraints impose independent choices among multiple agents, while global norm constraints can enforce shared resource or capacity limits. Despite the
importance of these geometric constraints, their effect
on the complexity of VIs is not fully understood beyond a few special cases.

Formally, given a nonempty compact convex set $\calK\subset\R^d$ and a continuous operator $F:\calK\to\R^d$, the variational inequality problem asks for a point $\vx\in\calK$ such that
\begin{equation}\label{eq:VI}
 \inner{\vx-\vxp}{F(\vx)}\ge0\quad\text{for every }\vxp\in\calK.
 \tag{VI}
\end{equation}
A solution to this problem always exists~\citep{HartmanStampacchia1966}, as can be established through Brouwer's fixed-point theorem. This connection also makes VIs closely related to the complexity class $\PPAD$, which captures the computational difficulty of many fundamental fixed-point and equilibrium problems~\citep{Papadimitriou94:Complexity}. However, to operationalize this theory, a central algorithmic goal is to compute solutions efficiently under structural conditions, or otherwise delineate fundamental complexity barriers. Both of these directions have proven to be fruitful, engendering a rich literature~\citep{korpelevich1976extragradient,nemirovski2004prox,nesterov2007dual, Daskalakis09:The,Chen09:Settling, DSZ21, bernasconi2024role,Bernasconi26:Complexity}.

In this paper, we examine the effect of geometry on the complexity of solving variational inequalities. The question that drives our research is:

\begin{center}
\emph{How does the geometry of the constraint set in conjunction with the structure of the operator affect the complexity of variational inequalities?}
\end{center}

To answer this, we focus on the problem of solving VIs over an $\ell_p$ ball for $p \in [1, \infty]$, examining the entire spectrum of parameters. The two extreme cases are by now well understood, going back to celebrated results in the complexity of equilibria. In particular, at $p = 1$, the problem amounts to equilibrium computation in symmetric bimatrix games~\citep{Chen09:Settling}, a canonical $\PPAD$-complete problem. At $p=\infty$, the underlying feasible set is a hypercube, and the independent interval constraints naturally correspond to mixed strategies in two-action polymatrix games. In these games, computing a Nash equilibrium is $\PPAD$-hard even for a constant precision~\citep{Rubinstein15:Inapproximability,bernasconi2024role}. In fact, the hardness results at both endpoints already hold for affine VIs, where the operator takes the form $F(\vx)=\mat A\vx+\vec b$.

The fact that computational hardness arises even for affine VIs is particularly noteworthy. A central ingredient in many $\PPAD$-hardness reductions is the simulation of arithmetic and logical operations, often through piecewise polynomials or arithmetic circuits~\citep{Daskalakis09:The,Chen09:Settling,Rubinstein15:Inapproximability,Deligkas24:Pure}. In contrast, affine operators lack the algebraic expressiveness to implement such operations directly. Thus, hardness for affine VIs demonstrates how the nonlinear behavior needed for such reductions can instead emerge from the interaction between a simple operator and the geometry of the feasible set.

%For $\vx\in K$, the normal cone of $K$ at $x$ is
%$
% N_K(\vx)=\{\vec  v\in\R^n:\langle \vec v,\vec y-\vx\rangle\le 0
 %\text{ for every }y\in K\}.
%$
%Condition \eqref{eq:VI} is equivalent to
%\begin{equation*}\label{eq:NC}
% 0\in F(x^\star)+N_K(x^\star). \tag{NC}
%\end{equation*}
%Equation \eqref{eq:NC} separates the \textit{operator $F$} from the \textit{local geometry of $K$}.  Changing $F$ changes the first order or equilibrium conditions, while changing $K$ changes the feasible directions against which those conditions are tested. 

Previous hardness results for $p = 1$ or $p = \infty$, however, rely on the special structure of simplex and box constraints, and do not directly extend to the remaining $\ell_p$ balls. Among the intermediate values $p \in (1, \infty)$, the Euclidean case $p=2$ is particularly natural, motivated by the prevalence of Euclidean norm constraints in
optimization and machine learning. Indeed, bounding the Euclidean norm of a parameter vector is perhaps the most common constraint one encounters in many applications~\citep{hinton2012improving, goodfellow2016deep}, making it the natural domain used in classification and gradient-based methods. Recent work by \citet{Attias25:Fixed} further highlights the importance of this domain, motivated in part by learning problems such as generative adversarial networks and multi-agent reinforcement learning. However, their work studies general operators through smoothed analysis and query complexity over the Euclidean ball; it does not characterize the computational complexity of VIs with low-degree polynomial operators. More broadly, a comprehensive understanding of how the complexity of VIs depends on the geometry of $\ell_p$ balls and the degree of the operator is still lacking. In this paper, we characterize the complexity landscape for affine and quadratic VIs over $\ell_p$ balls across the entire spectrum $p\in[1,\infty]$.

\subsection{Our results}

For a parameter $p\ge1$, or $p=\infty$,
let $B_p^d=\{\vx\in\R^d:\norm{\vx}_p\le1\}$
denote the unit $\ell_p$ ball. We formulate the problem of finding an approximate VI solution over the $\ell_p$ ball as the following search problem.
%\textcolor{red}{We need to say something for $0<p<1$ because the reviewers might ask}

\medskip
\noindent
\problem{Ball$_p$-VI}
\begin{itemize}
    \item \emph{Input:}
    A 1-Lipschitz-continuous operator
    $F:B_p^d\to\R^d$, represented by a well-behaved
    arithmetic circuit with rational constants and operations $\{+, -, \times, \min, \max\}$, and an
    accuracy parameter $\epsilon\in(0,1)$.
    All inputs are encoded in binary.

    \item \emph{Output:}
    A point $\vx\in B_p^d$ satisfying
    \[
        \Gap_{B_p^d}(\vx;F)
        :=\max_{\vxp\in B_p^d}
        \inner{\vxp-\vx}{F(\vx)}
        \le\epsilon.
    \]
\end{itemize}

We call such a point an $\epsilon$-approximate VI solution.
The above definition adopts the standard notion of well-behaved arithmetic
circuits by~\citet{Fearnley23:Complexity}. In this paper, we
investigate the complexity of \problem{Ball$_p$-VI}
for affine and quadratic polynomial operators, specified
explicitly by their rational coefficients and monomials.
It is straightforward to establish that for every fixed exponent $p\ge1$, and also for $p=\infty$, \problem{Ball$_p$-VI} belongs to $\PPAD$ for arbitrary binary encoded $\epsilon$; for completeness we provide the proof of the membership in  Appendix~\ref{app:ball-vi-ppad}.

To begin with, we examine the case where the operator is affine. Our first main result extends the existing hardness results for the $\ell_1$ and $\ell_\infty$ balls to every $\ell_p$ ball with fixed $p \neq 2$.

\begin{theorem}[Abridged version of Theorem \ref{thm:main}]
\label{thm:classification}
For every fixed rational exponent $p\ge1$ with $p\ne2$, there exists a constant $c_p>0$ such that \problem{Ball$_p$-VI} is $\PPAD$-complete for $\epsilon=d^{-c_p}$. The hardness result holds even for affine operators.
\end{theorem}

For $p=\infty$, the above result holds even at constant
approximation accuracy \citep{Rubinstein15:Inapproximability, bernasconi2024role}. Notably, our hardness construction does not
extend to the Euclidean case $p=2$. This turns out to reflect something more fundamental than a limitation of the construction: we show that
every affine VI over the Euclidean ball can be solved
in polynomial time.

\begin{theorem}[Abridged version of Theorem \ref{thm:affine}]
\label{thm:classification-affine}
For affine operators, \problem{Ball$_2$-VI}
can be solved in time polynomial in the input
encoding length and $\log(1/\epsilon)$.
\end{theorem}

Importantly, this result does not require common structural assumptions that guarantee tractability. For example, when $F(\vx)=\mat A\vx+\vec b$ and
$\mat A$ is symmetric, polynomial-time solvability follows from classical algorithms for quadratic optimization over a Euclidean
ball~\citep{VavasisZippel1990,Ye1992Sphere}.
Indeed, a global maximizer of the quadratic function
$\frac12\vx^{\mathsf T}\mat A\vx+\vec b^{\mathsf T}\vx$
over the ball satisfies the corresponding VI.
\Cref{thm:classification-affine} extends this tractability to arbitrary, possibly nonsymmetric matrices $\mat A$.

% This is so despite that the fact that the underlying problem does not satisfy common structural assumptions that guarantee tractability, such as monotonicity or more broadly the Minty condition. When $\mat A$ is symmetric, polynomial-time solvability follows from classical algorithms for quadratic optimization over a Euclidean ball~\citep{VavasisZippel1990,Ye1992Sphere}, which is based on the observation that a global maximizer of $\frac12\vec x^{\mathsf T}\mat A\vec x+\vec b^{\mathsf T}\vec x$ over the ball satisfies the VI. \Cref{thm:classification-affine} extends this tractability to arbitrary, possibly nonsymmetric matrices $\mat A$.

One notable application of~\Cref{thm:classification-affine} concerns \emph{performative prediction}, where the deployment of a model changes the underlying data distribution~\citep{perdomo2021performativeprediction,Hardt25:Performative}, for example through strategic responses~\citep{Hardt16:Strategic}. Our algorithm yields a polynomial-time algorithm for finding (first-order) performatively stable points under quadratic loss functions and affine distribution shifts when the parameter vector lies in a Euclidean ball. This is in stark contrast to the case where the parameter vector is constrained to be on a box~\citep{Anagnostides26:Computational}. \Cref{thm:classification-affine} has further new implications in quadratic games as well as computing stationary points under linear \emph{temporal-difference updates}~\citep{Bhandari21:TemporalDifference}. We elaborate on these implications and their assumptions later in \Cref{sec:related}.

Having established the tractability of affine VIs over the Euclidean ball, a natural question arises: does this tractability extend beyond affine operators? Our next result shows that this tractability vanishes once the operator becomes quadratic, as computing an approximate VI solution becomes $\PPAD$-complete.

% \Cref{thm:classification-affine} raises a natural question: how far does the tractability of the Euclidean ball extend? Our next result shows that the moment we allow quadratic terms, computing solutions becomes $\PPAD$-complete.

\begin{theorem}[Abridged version of Theorem \ref{thm:degree2-unit}]
\label{thm:classification-quadratic}
There exists a constant $c>0$ such that
\problem{Ball$_2$-VI} is $\PPAD$-complete for
$\epsilon=d^{-c}$.
The hardness result holds even for quadratic operators.
\end{theorem}

Our main results are summarized in~\Cref{tab:classification}.

% Preamble: load xcolor with the table option and add tabularx:

\begin{table}[H]
\centering
\caption{Complexity of VIs on unit balls. In the $\ell_p$ row, $p\in(1,\infty)\backslash\{2\}$ is a fixed parameter. Our results are highlighted in the shaded cells. $\dagger$: \citet{Chen09:Settling}. $\ddagger:$ \citet{Rubinstein15:Inapproximability, bernasconi2024role}.}
\label{tab:classification}
\small
\renewcommand{\arraystretch}{1.15}
\begin{tabularx}{\linewidth}{
  >{\hsize=.95\hsize\linewidth=\hsize\centering\arraybackslash}X
  >{\hsize=.75\hsize\linewidth=\hsize\centering\arraybackslash}X
  >{\hsize=1.4\hsize\linewidth=\hsize\centering\arraybackslash}X
  >{\hsize=.9\hsize\linewidth=\hsize\centering\arraybackslash}X}
\toprule
\textbf{Domain} & \textbf{Operator} & \textbf{Complexity} & \textbf{Additive gap} \\
\midrule
\rowcolor{green!6}
Euclidean ball & Affine &
\colorbox{green!12}{\textcolor{green!45!black}{Polynomial time}} &
Any $\epsilon>0$ \\
\rowcolor{red!6}
Euclidean ball & Quadratic &
\colorbox{red!10}{\textcolor{red!70!black}{\PPAD-complete}} &
Inverse polynomial \\
\rowcolor{red!6}
$\ell_p$ ball ($p\neq 2$) & Affine &
\colorbox{red!10}{\textcolor{red!70!black}{\PPAD-complete}} &
Inverse polynomial \\
$\ell_1$ ball & Affine & \;\PPAD-complete\textsuperscript{$\dagger$} & Inverse polynomial \\
$\ell_\infty$ ball & Affine & \;\PPAD-complete\textsuperscript{$\ddagger$} & Constant \\
\bottomrule
\end{tabularx}
\end{table}

\begin{remark}[Case $0<p<1$] We restrict our attention to $p\ge1$, since for $0<p<1$, $\norm{.}_p$ does not satisfy the properties of a norm. In particular, $B_p^d$ is a nonconvex set and VI solutions in \eqref{eq:VI} need not exist, even for affine operators.
\end{remark}
% \begin{table}[H]
% \centering
% \caption{Complexity of VIs on unit balls. In the $\ell_p$ row, $p\in(1,\infty)\backslash\{2\}$ is fixed and rational. All hardness results hold with coefficients in $[-1,1]$. Our results are highlighted in the shaded cells.}
% \label{tab:classification}
% \small
% \renewcommand{\arraystretch}{1.15}
% \begin{tabularx}{\linewidth}{
%   >{\hsize=.95\hsize\linewidth=\hsize\centering\arraybackslash}X
%   >{\hsize=.75\hsize\linewidth=\hsize\centering\arraybackslash}X
%   >{\hsize=1.4\hsize\linewidth=\hsize\centering\arraybackslash}X
%   >{\hsize=.9\hsize\linewidth=\hsize\centering\arraybackslash}X}
% \toprule
% \textbf{Domain} & \textbf{Operator} & \textbf{Complexity} & \textbf{Additive gap} \\
% \midrule
% \rowcolor{green!6}
% Euclidean ball & Affine &
% \colorbox{green!12}{\textcolor{green!45!black}{Polynomial time}} &
% Any $\epsilon>0$ \\
% \rowcolor{red!6}
% Euclidean ball & Quadratic &
% \colorbox{red!10}{\textcolor{red!70!black}{\PPAD-complete}} &
% Inverse polynomial \\
% \addlinespace[4pt]
% \rowcolor{red!6}
% $\ell_p$ ball ($p\neq 2$) & Affine &
% \colorbox{red!10}{\textcolor{red!70!black}{\PPAD-complete}} &
% Inverse polynomial \\
% $\ell_1$ ball & Affine & \;\PPAD-complete~\originalcitep{Chen09:Settling} & Inverse polynomial \\
% $\ell_\infty$ ball & Affine & \;\PPAD-complete~\citep{Rubinstein15:Inapproximability} & Constant \\
% \bottomrule
% \end{tabularx}
% \end{table}

\subsection{Technical overview}

A common ingredient in typical $\PPAD$-hardness reductions is the simulation of logical and arithmetic gates through piecewise-linear functions~\citep{Daskalakis09:The, Chen09:Settling, Rubinstein15:Inapproximability,Deligkas24:Pure}. The main technical challenge in our setting is to implement such gates using affine or quadratic operators together with the geometry of $\ell_p$ balls. To this end, all our main hardness reductions start from \PureCircuit{}~\citep{Deligkas24:Pure} and approximate its threshold functions through weighted discretization,\footnote{
Related discretization techniques have appeared in
hardness reductions for team games and min--max
optimization~\citep{Anagnostides26:Team,Bernasconi26:Quadratic}. A key difference in our setting is that the weights must be implemented under a single $\ell_p$-ball constraint.} but implement the required weights in different ways. For $\ell_p$ balls with $p \neq 2$, we exploit the geometry of the constraints to simulate the threshold functions while keeping the operator affine. The Euclidean ball presents a contrasting picture, since it is possible to obtain a polynomial-time algorithm for affine VIs. Nevertheless, we show that quadratic operators provide sufficient expressiveness to simulate the same threshold functions and establish \(\PPAD\)-hardness. 

In what follows, we first explain why standard game reductions do not directly apply, and then develop these technical ideas in more detail.

\paragraph{Why standard game reductions do not directly apply.}

Many game-theoretic reductions enforce gate constraints through unilateral deviations~\citep{Daskalakis09:The, Chen09:Settling, Rubinstein15:Inapproximability}. Since the feasible set is a product of strategy spaces, each player can change their strategy while all other players' strategies remain fixed. Gate gadgets exploit this independence: an incorrect gate output gives some player a sufficiently profitable deviation, forcing every approximate equilibrium to simulate the gate correctly (up to the required tolerance). However, this product structure is not essential for all equilibrium reductions. In particular, a symmetric equilibrium of a symmetric bimatrix game can be characterized by a single mixed strategy, whose best-response conditions can be expressed as an affine VI over the simplex~\citep{Nisan07:Algorithmic}.

This argument does not directly extend to a single $\ell_p$ ball with $1<p<\infty$. On one hand, all coordinates share a common norm constraint, so changing one coordinate may in turn require further adjustments to others to preserve feasibility. As a result, the usual argument that hinges on independent deviations does not apply. Moreover, its curved boundary lacks the geometric properties needed for the symmetric encoding.

One possible alternative is to encode the desired gate or equilibrium conditions as polynomial equations and use their residuals as the VI operator. However, this approach introduces another challenge. If a VI solution lies in the interior, the operator must vanish, and hence all the intended equations are satisfied. At the boundary, however, the operator may be nonzero while still satisfying the VI optimality condition. A valid reduction must therefore ensure that every sufficiently accurate VI solution, including those on the boundary, can be decoded into an approximate solution of the original problem.

% An additional obstacle arises when one attempts to encode the desired gate or equilibrium conditions as polynomial equations and use their residuals as the VI operator. If a VI solution lies in the interior, the operator must vanish, and hence all the intended equations are satisfied. At the boundary, however, the operator may be nonzero while still satisfying the VI condition. A valid reduction must therefore ensure that every approximate VI solution, including those on the boundary, can be decoded to an approximate solution of the original problem.

\paragraph{Non-Euclidean geometry.}

We first consider affine VIs over an $\ell_p$ ball, where $1<p<\infty$ and $p\ne2$. Although affine operators cannot directly implement the threshold functions of \PureCircuit{}, the \emph{curvature} of these non-Euclidean balls induces nonlinear behavior through the VI optimality condition.  In particular, consider a coordinate $x_i$ whose operator component is $F_i(\vx)=h(\vx)-x_i$, where $h$ is an affine function. At a boundary solution $\vx^*$, the VI condition requires
$$
h(\vx^*)-x_i^*=\lambda |x_i^*|^{p-2}x_i^*
$$
for some $\lambda\ge0$. 
Let $\sigma_\lambda$ denote the inverse of the function $t\mapsto t+\lambda|t|^{p-2}t$, which is nonlinear whenever $p\ne2$ and $\lambda>0$. The boundary condition can therefore be written as
$$
x_i^*=\sigma_\lambda(h(\vx^*)).
$$
In other words, the boundary VI condition naturally implements a nonlinear inverse mapping $\sigma_\lambda$, even though the operator itself remains affine. At $p=2$, however, the inverse mapping reduces to $\sigma_\lambda(t)=t/(1+\lambda)$, which is linear. This distinction explains precisely why our reduction does not extend to the Euclidean case.

To simulate the threshold functions of \PureCircuit{}~\citep{Deligkas24:Pure, Bernasconi26:Complexity}, we seek to exploit the nonlinearity of the inverse mapping $\sigma_\lambda$. We first consider the finite difference
$$
\Delta_\lambda(t)
:=\sigma_\lambda(t+1)-2\sigma_\lambda(t)+\sigma_\lambda(t-1),
$$
which serves as a discrete analogue of the second derivative of $\sigma_\lambda$. In the Euclidean case, $\sigma_\lambda$ is linear, so $\Delta_\lambda(t)=0$ for every $t$. For $p\ne2$, however, $\Delta_\lambda$ captures the curvature of $\sigma_\lambda$ and decays to zero at a polynomial rate in the tails. This tail decay suggests a way to approximate the threshold functions: we sample their values on a finite grid and combine them through weighted averaging. The polynomial decay would assign small weights to distant samples, limiting their contributions to the approximation. However, $\Delta_\lambda(t)$ can take both positive and negative values, whereas our construction requires nonnegative weights.

To overcome this obstruction, we construct a weight function through another finite difference of $\sigma_\lambda$, namely
$$
k_\lambda(t)
\defeq \varsigma_p\left(
\sigma_\lambda\left(4+\frac{\Delta_\lambda(t)}{4}\right)
+\sigma_\lambda\left(4-\frac{\Delta_\lambda(t)}{4}\right)
-2\sigma_\lambda(4)
\right),
$$
where $\varsigma_p=\operatorname{sgn}(2-p)$. Since $\sigma_\lambda$ is convex when $1<p<2$ and concave when $p>2$ on the domain of interest, the sign factor $\varsigma_p$ ensures that $k_\lambda$ is nonnegative for every $t$. Moreover, when $\lambda$ is sufficiently close to 1, this construction yields a polynomial tail bound $k_\lambda(t)=O(|t|^{-\beta_p})$ for some $\beta_p>2$. 

These properties allow us to control the error of the weighted approximation. 
By Lipschitz continuity, samples within distance $1/A$
of the input have function values differing from the
target by at most $O(1/A)$. Meanwhile, the polynomial
tail bound ensures that the total normalized contribution
of other distant samples is at most $O(A^{1-\beta_p})$.
Since $\beta_p>2$, both errors become small as $A$
increases. Choosing a sufficiently fine grid then yields the desired approximation accuracy. Thus, by replacing the
threshold functions with compactly supported versions,
we can approximate the desired gates through weighted
averaging using $k_\lambda$.

Two issues remain in implementing this construction. First, each coordinate can implement the nonlinear inverse mapping $\sigma_\lambda$ only through the boundary VI condition. Second, the threshold function approximation requires the common multiplier $\lambda$ to be sufficiently close to one. We address both issues by introducing an anchor coordinate $z_0$ with a constant positive operator component $F_0=c_R$. This positive component prevents the operator from vanishing, forcing every exact VI solution to lie on the boundary. Moreover, by carefully choosing the radius $R$ and the constant $c_R$, we ensure that $\lambda$ remains sufficiently close to one, where the weighted approximations achieve the
desired accuracy. For approximate VI solutions, a sufficiently small VI gap ensures that the simulated \PureCircuit{} gates satisfy their constraints within the required tolerance.

\begin{figure}[tbp]
    \centering
    \input{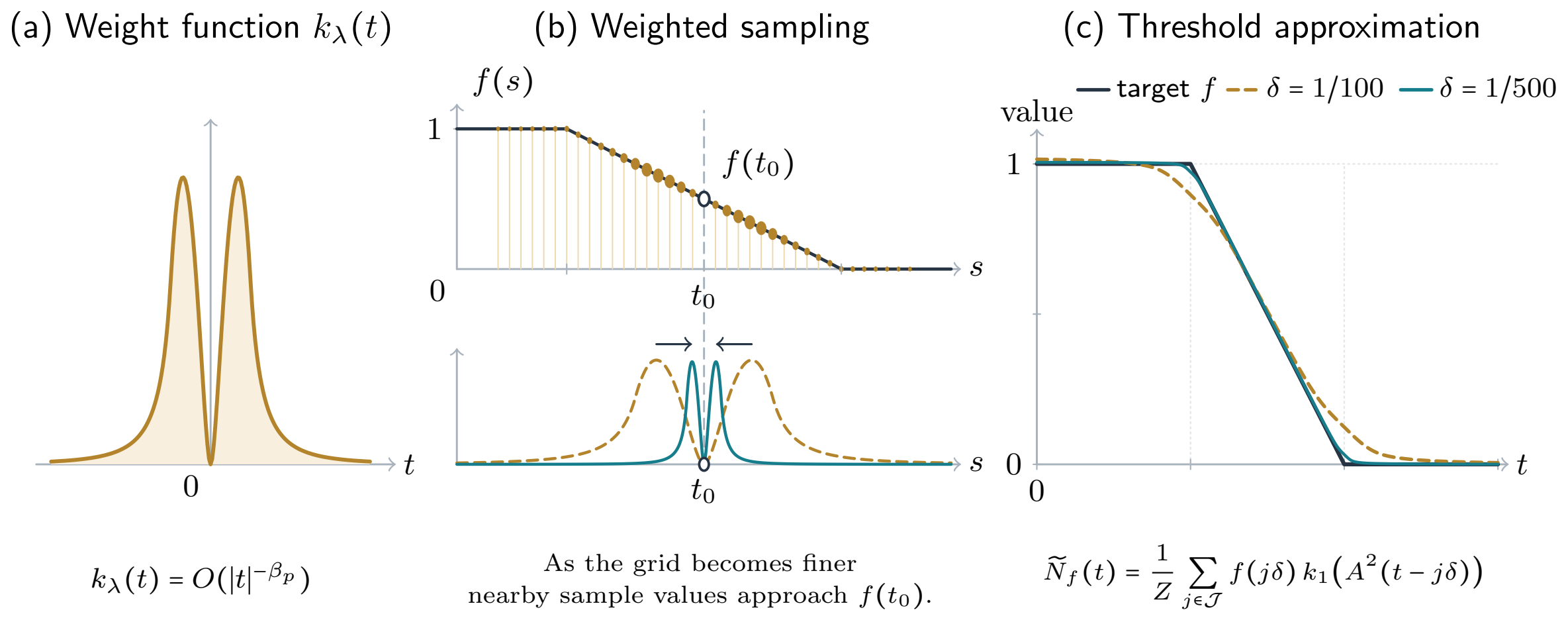}
    \caption{Illustration of the weighted approximation in the non-Euclidean affine VI reduction (schematic). Left: the weight function $k_\lambda$.
    Middle: weighted sampling at two different grid resolutions.
    Right: the weighted approximation function $\widetilde{N}_{f}(t)$.}
    \label{fig:non-euclidean-geometry}
\end{figure}

Finally, we rescale the construction to the unit $\ell_p$ ball and normalize the operator coefficients, preserving $\PPAD$-hardness at inverse-polynomial accuracy. The endpoint cases $p=1$ and $p=\infty$ follow from hardness results for bimatrix Nash equilibrium~\citep{Chen09:Settling} and affine VIs over the hypercube~\citep{Rubinstein15:Inapproximability,bernasconi2024role}, respectively.
\Cref{fig:non-euclidean-geometry} illustrates how weighted averaging yields an approximation $\widetilde{N}_{f}$ of a threshold function.

% More broadly, this reduction illustrates how the geometry of a feasible set can serve as a source of computational expressiveness. By translating geometric properties into nonlinear mappings and subsequently into approximations of logical gates, we develop a toolkit for designing hardness gadgets that may find applications beyond the specific problems studied here. These techniques provide a promising approach to characterizing the complexity of other equilibrium and optimization problems.

\paragraph{Bisection search for the Euclidean case.}

As discussed earlier, the Euclidean ball has a particular geometric structure. Specifically, for an affine operator $F(\vx)=\mat A\vx+\vec b$, an exact VI solution either satisfies $\mat A\vx+\vec b=\vec 0$ or lies on the boundary and satisfies
\begin{equation}\label{eq:overview-affine-boundary}
    \mat A\vx+\vec b=\lambda\vx,\quad \lambda>0,\quad \norm{\vx}_2=1.
\end{equation}
% For a fixed $\lambda$, we can efficiently determine whether this system has a solution through convex programming, by computing a minimum-norm solution and then evaluating feasibility.
When $\mat A$ is symmetric, condition~\eqref{eq:overview-affine-boundary} arises naturally in maximizing the function $\frac12\vx^{\mathsf T}\mat A\vx+\vec b^{\mathsf T}\vx$ over the Euclidean ball. Although this quadratic optimization problem need not be convex, the classic result of \citet{Ye1992Sphere} allows us to compute a global optimum efficiently, which is in turn a VI solution. For nonsymmetric $\mat A$, however, the gradient of this objective no longer matches the given operator. 
Our algorithm instead exploits the boundary condition~\eqref{eq:overview-affine-boundary} directly, without reverting to a quadratic potential.

The key observation is that, for a fixed $\lambda$, we can efficiently determine whether the system $(\lambda\mat I-\mat A)\vx=\vec b$ has a solution through convex programming, by computing a minimum-norm solution and then evaluating feasibility. In particular, we first check whether $\mat A\vx+\vec b=\vec 0$ has a solution in the ball, in which case we can immediately return it. Otherwise, feasibility fails at $\lambda=0$ and succeeds for sufficiently large $\lambda$. We can then apply bisection search to obtain two nearby multipliers $\lambda_\ell<\lambda_u$ for which the feasibility check fails and succeeds, respectively. Notably, this feasibility need not be monotone in $\lambda$, in that success at one multiplier does not guarantee success at every larger multiplier. Nevertheless, the bisection only needs to maintain an infeasible lower endpoint and a feasible upper endpoint. 

% To do so, at the midpoint $\lambda_m=(\lambda_\ell+\lambda_u)/2$, we set $\lambda_u=\lambda_m$ if the feasibility check succeeds and $\lambda_\ell=\lambda_m$ if it fails. This maintains the invariance the endpoint outcomes are opposite while halving the interval in each step, yielding arbitrarily close infeasible and feasible multipliers.

Once $\lambda_u-\lambda_\ell$ is sufficiently small, the remaining challenge is to obtain a point near the boundary. The solution $\vx_u$ at $\lambda_u$ lies in the ball and satisfies $(\lambda_u\mat I-\mat A)\vx_u=\vec b$ exactly. As a result, its residual in the linear system at $\lambda_\ell$ satisfies $\norm{(\lambda_\ell\mat I-\mat A)\vx_u-\vec b}_2\le \lambda_u-\lambda_\ell$. However, this alone does not guarantee a small VI gap, because $\vx_u$ may still lie well inside the ball.

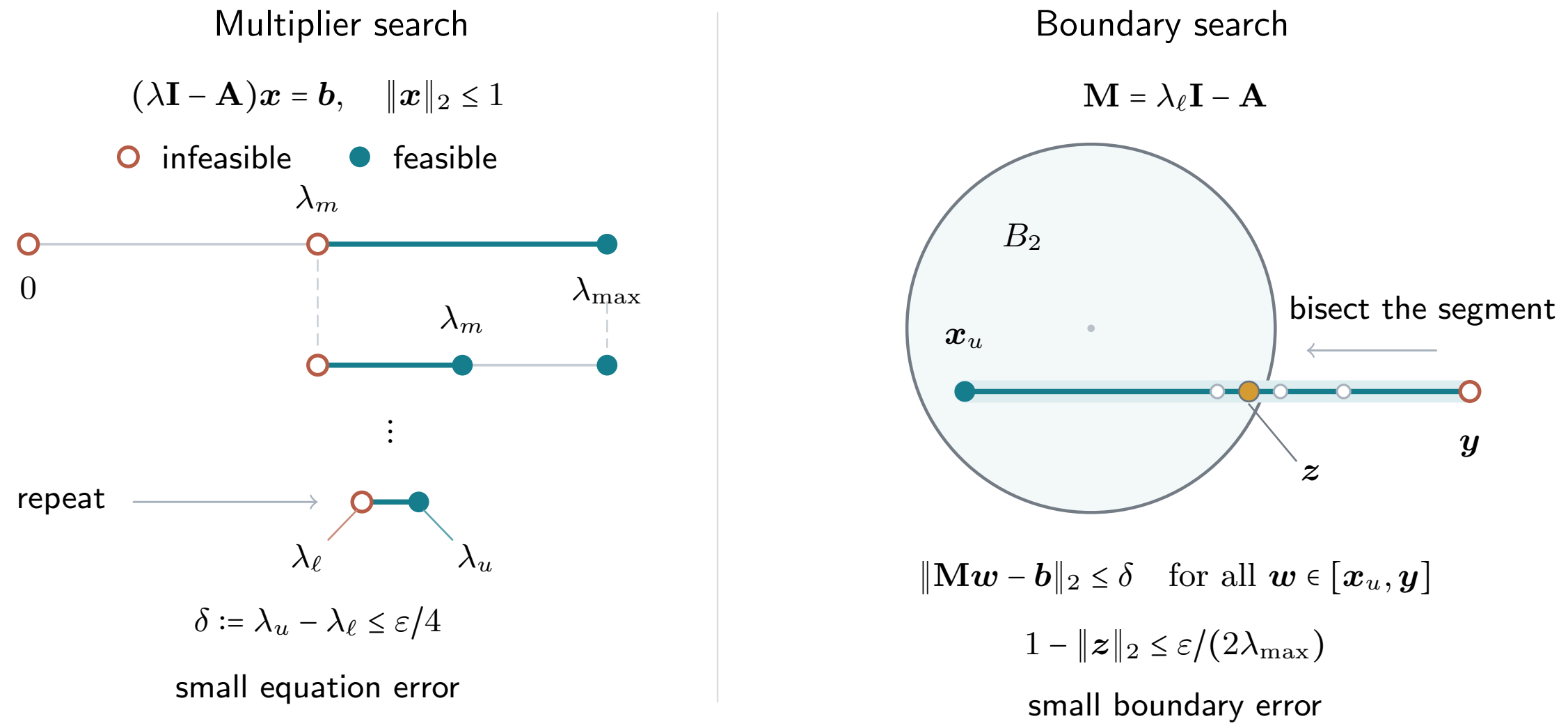
\begin{figure}[tbp]
    \centering
    % Reusable figure: requires amsmath, amssymb, and tikz.
% Uses the manuscript's \mat and \vec macros for consistent notation.
% The standalone wrapper supplies the same definitions.
\begingroup
\definecolor{bisectInk}{HTML}{273444}
\definecolor{bisectTeal}{HTML}{167D8D}
\definecolor{bisectRust}{HTML}{B65B45}
\definecolor{bisectGold}{HTML}{D39B32}
\definecolor{bisectGray}{HTML}{A8B2BD}
\begin{tikzpicture}[
    x=1cm,y=1cm,
    font=\fontsize{9}{11}\selectfont,
    text=black,
    line cap=round,line join=round,
    title/.style={font=\fontsize{10}{12}\selectfont\sffamily,anchor=center},
    note/.style={font=\fontsize{9}{11}\selectfont\sffamily,text=black},
    guide/.style={draw=bisectGray!65,densely dashed,line width=.45pt},
    rail/.style={draw=bisectGray!65,line width=.7pt},
    kept/.style={draw=bisectTeal,line width=1.4pt},
    fail/.style={circle,draw=bisectRust,fill=white,line width=1pt,
                 minimum size=5.2pt,inner sep=0pt},
    pass/.style={circle,draw=bisectTeal,fill=bisectTeal,
                 minimum size=5.2pt,inner sep=0pt},
    trial/.style={circle,draw=bisectGray,fill=white,line width=.6pt,
                  minimum size=3.6pt,inner sep=0pt}
]
\path[use as bounding box] (0,0) rectangle (15.6,7.05);

% Panel titles and the two linear systems.
\node[title] at (3.575,6.72) {Multiplier search};
\node[title] at (11.375,6.72) {Boundary search};
\draw[bisectGray!45,line width=.45pt] (7.15,.3) -- (7.15,6.85);
\node at (3.35,6.05)
    {$(\lambda\mat{I}-\mat{A})\vec{x}=\vec{b},\quad \|\vec{x}\|_2\le1$};
\node at (11.5,6.05) {$\mat{M}=\lambda_\ell\mat{I}-\mat{A}$};

% Outcome legend. Filled/open markers remain distinct in grayscale.
\node[fail] at (1.55,5.48) {};
\node[note,anchor=west] at (1.73,5.48) {infeasible};
\node[pass] at (3.75,5.48) {};
\node[note,anchor=west] at (3.93,5.48) {feasible};

% Two illustrative iterations. Only tested points have a feasibility label.
% A thick interval indicates the retained bracket, not a feasible region.
\draw[rail] (.6,4.65) -- (6.1,4.65);
\draw[kept] (3.35,4.65) -- (6.1,4.65);
\node[fail] at (.6,4.65) {};
\node[fail] at (3.35,4.65) {};
\node[pass] at (6.1,4.65) {};
\node[below=5pt] at (.6,4.65) {$0$};
\node[above=5pt] at (3.35,4.65) {$\lambda_m$};
\node[below=5pt] at (6.1,4.65) {$\lambda_{\max}$};

\draw[guide] (3.35,4.5) -- (3.35,3.67);
\draw[guide] (6.1,4.1) -- (6.1,3.67);
\draw[rail] (3.35,3.5) -- (6.1,3.5);
\draw[kept] (3.35,3.5) -- (4.725,3.5);
\node[fail] at (3.35,3.5) {};
\node[pass] at (4.725,3.5) {};
\node[pass] at (6.1,3.5) {};
\node[above=5pt] at (4.725,3.5) {$\lambda_m$};
\node at (4.04,2.87) {$\vdots$};

% Final bracket after further iterations; short leaders separate the labels.
\draw[kept] (3.77,2.2) -- (4.31,2.2);
\node[fail] at (3.77,2.2) {};
\node[pass] at (4.31,2.2) {};
\draw[bisectRust!70,line width=.5pt] (3.72,2.12) -- (3.45,1.84);
\draw[bisectTeal!70,line width=.5pt] (4.36,2.12) -- (4.63,1.84);
\node[anchor=north east,inner sep=1pt] at (3.45,1.84) {$\lambda_\ell$};
\node[anchor=north west,inner sep=1pt] at (4.63,1.84) {$\lambda_u$};
\node[note,anchor=west] at (.35,2.2) {repeat};
\draw[->,bisectGray,line width=.55pt] (1.6,2.2) -- (3.35,2.2);

% The ball and a nonradial segment: the start may be well inside the ball.
\fill[bisectTeal!5] (10.7,3.85) circle[radius=1.75];
\draw[bisectInk!65,line width=.85pt]
    (10.7,3.85) circle[radius=1.75];
\node at (10.05,4.72) {$B_2$};
\fill[bisectGray!80] (10.7,3.85) circle[radius=1pt];

% The pale band denotes a small equation residual on the entire segment.
\draw[bisectTeal!14,line width=6pt] (9.5,3.25) -- (14.3,3.25);
\draw[kept] (9.5,3.25) -- (14.3,3.25);
\node[pass] at (9.5,3.25) {};
\node[fail] at (14.3,3.25) {};
\node[above=8pt] at (9.5,3.25) {$\vec{x}_u$};
\node[below=8pt] at (14.3,3.25) {$\vec{y}$};

% Illustrative segment midpoints, followed by the returned inside point.
\node[trial] at (11.9,3.25) {};
\node[trial] at (13.1,3.25) {};
\node[trial] at (12.5,3.25) {};
\filldraw[fill=bisectGold,draw=bisectInk!70,line width=.5pt]
    (12.2,3.25) circle[radius=2.7pt];
\draw[bisectInk!65,line width=.5pt]
    (12.2,3.13) -- (12.65,2.59);
\node[anchor=north west,inner sep=1pt] at (12.65,2.59) {$\vec{z}$};
\node[note] at (13.85,4.02) {bisect the segment};
\draw[->,bisectGray,line width=.55pt] (13.98,3.64) -- (12.75,3.64);

% Keep both guarantees on aligned baselines, separate from the geometry.
\node at (3.35,1.05)
    {$\delta:=\lambda_u-\lambda_\ell\le\varepsilon/4$};
\node[note] at (3.35,.42) {small equation error};
\node at (11.5,1.48)
    {$\|\mat{M}\vec{w}-\vec{b}\|_2\le\delta\quad\text{for all }\vec{w}\in[\vec{x}_u,\vec{y}]$};
\node at (11.5,.83)
    {$1-\|\vec{z}\|_2\le\varepsilon/(2\lambda_{\max})$};
\node[note] at (11.5,.25) {small boundary error};
\end{tikzpicture}
\endgroup
    \caption{Two bisections for affine VIs on the Euclidean ball (schematic).
    Left: each midpoint test retains endpoints with opposite feasibility outcomes;
    highlighted intervals are retained brackets, not regions of feasibility.
    Right: an exterior point $\vec y$ is chosen so that the equation residual
    stays small along the entire segment from $\vec x_u$ to $\vec y$.
    It decreases to zero when $\vec y$ solves the system at $\lambda_\ell$,
    and stays unchanged when moving along a kernel direction.
    The second bisection returns a point $\vec z$ inside the ball and near its
    boundary. Together, the two error bounds imply $\Gap(\vec z; F)\le\epsilon$.}
    \label{fig:euclidean-bisection}
\end{figure}

To address this, we move from $\vx_u$ toward an outside point while keeping the equation error at $\lambda_\ell$ small along that segment. If $\lambda_\ell\mat I-\mat A$ is invertible, we take its exact solution as the outside point: it lies outside the ball because the feasibility test fails at $\lambda_\ell$, and the error decreases linearly to zero along the segment. If the matrix is singular, we instead move along a kernel direction, leaving the error unchanged. In either case, the segment crosses the sphere, allowing us to perform a second bisection and find a feasible point arbitrarily close to its boundary. In other words, the first bisection search controls the equation error, while the second controls the distance to the boundary. Together, they guarantee a small VI gap, with logarithmic dependence on the inverse accuracy. The two searches are illustrated in \Cref{fig:euclidean-bisection}. 

\paragraph{Quadratic operators suffice for Euclidean hardness.} We next turn to quadratic operators on the Euclidean ball, where we show that the tractability of the affine case no longer applies. A natural first attempt is to exploit the connection between equilibrium computation and linear complementarity problems (LCPs)~\citep{Facchinei03:Finite}, whose complementarity conditions take the form $z_iw_i=0$. These product terms naturally suggest a quadratic formulation. However, as discussed earlier, a boundary VI solution need not solve the intended complementarity equations, since the VI condition only requires $F(\vx)=\lambda\vx$ for some $\lambda\ge0$. Thus, directly encoding the complementarity conditions may introduce unintended boundary solutions.

Our reduction instead starts from \PureCircuit{}. As in the non-Euclidean reduction, we approximate these threshold functions through weighted averaging over a discretized domain. The difference lies in how the weights are implemented: rather than exploiting the nonlinear boundary optimality conditions to obtain a weight function with sufficient decay, we introduce \emph{selector variables} and encode the desired weights directly at interior VI solutions.

To make this more precise, let $t$ denote the displacement between the input value and a grid point, normalized by the grid spacing $\delta$. We assign each grid point a weight $1/(1+t^2)$, whose quadratic decay limits the contributions of distant samples. Specifically, samples $k$ grid steps away receive weight $O(k^{-2})$, while their threshold values differ from the target by at most $O(k\delta)$. Combining these estimates, we show that the normalized weighted average approximates the threshold function with error $O(\delta\log (1/\delta))$. Thus, a sufficiently fine grid yields the approximation accuracy required for our reduction.

To implement the weights with quadratic decay, one might try to encode $\alpha(1+t^2)=1$ directly into the operator. However, this would introduce a degree-3 term. Our key observation is that introducing an auxiliary variable $\beta$ allows the desired weight to be obtained \emph{implicitly} through two coupled quadratic equations:
$$
\alpha-t\beta=1,\quad \beta+t\alpha=0
\quad\Rightarrow\quad
\alpha=\frac{1}{1+t^2}.
$$
These equations can be written as
$\mat{A}\vec v=\vec e_1$,
where $\vec v=(\alpha,\beta)^\top$,
$\vec e_1=(1,0)^\top$, and
$$
\mat{A}=\begin{pmatrix}
1 & -t\\
t & 1
\end{pmatrix}.
$$
The off-diagonal part of $\mat{A}$ is skew-symmetric, a property we will exploit later to control approximation errors.

The quadratic equations above enforce the desired weights whenever the VI operator vanishes, which is guaranteed only for interior solutions. We therefore design our construction to ensure that every exact VI solution lies inside the Euclidean ball. In particular, at any exact VI solution $\vec z^*$, the selector variables satisfy
$${\alpha^*}^2+{\beta^*}^2\le 1.$$ Moreover, the selector weights are positive, ensuring all other coordinates lie in \([0,1]\). Together, these properties provide an upper bound on the Euclidean norm $\norm{\vec z^*}_2.$ By choosing a sufficiently large radius $R$, we ensure that $\| \vec z^* \|_2<R$, so every exact VI solution lies in the interior and satisfies the intended equations.

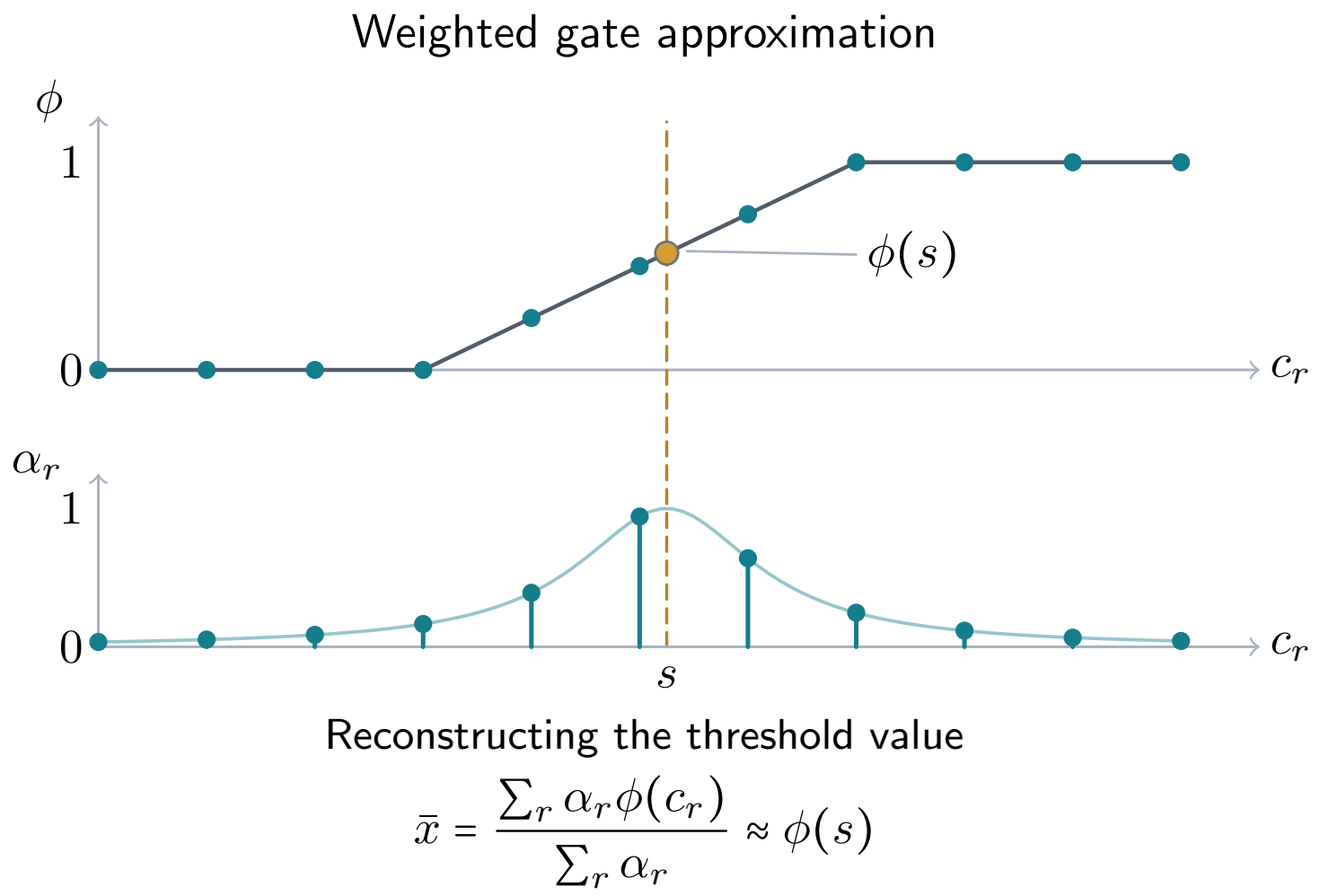
\begin{figure}[!t]
    \centering
    % Requires amsmath, amssymb, and tikz.
% The standalone wrapper supplies the manuscript math fonts.
\begingroup
\definecolor{quadInk}{HTML}{273444}
\definecolor{quadTeal}{HTML}{167D8D}
\definecolor{quadGold}{HTML}{D39B32}
\definecolor{quadGray}{HTML}{A8B2BD}
\begin{tikzpicture}[
    x=1.45cm,y=.85cm,
    font=\fontsize{10}{12}\selectfont,
    text=black,
    line cap=round,line join=round,
    title/.style={font=\fontsize{10}{12}\selectfont\sffamily},
    note/.style={font=\fontsize{9}{11}\selectfont\sffamily},
    axis/.style={->,draw=quadGray,line width=.55pt},
    sample/.style={draw=quadTeal,fill=quadTeal}
]
\path[use as bounding box] (0,0) rectangle (7.2,7.65);
\node[title] at (3.575,7.25) {Weighted gate approximation};

% A representative continuous threshold, sampled on a uniform grid.
% The input s=5.25 is deliberately between grid points.
\draw[axis] (.8,4.35) -- (6.7,4.35);
\draw[axis] (.8,4.35) -- (.8,6.55);
\node[anchor=south east,inner sep=1pt] at (.65,6.45) {$\phi$};
\node[anchor=west,inner sep=2pt] at (6.7,4.35) {$c_r$};
\node[anchor=east,inner sep=3pt] at (.8,4.35) {$0$};
\node[anchor=east,inner sep=3pt] at (.8,6.15) {$1$};
\draw[quadInk!80,line width=.85pt]
    (.8,4.35) -- (2.45,4.35) -- (4.65,6.15) -- (6.3,6.15);

% A shared input guide aligns the samples and their weights.
\draw[quadGold!85!black,densely dashed,line width=.65pt]
    (3.6875,1.95) -- (3.6875,6.5);
\foreach \r in {0,...,10} {
    \pgfmathsetmacro{\quadX}{.8+.55*\r}
    \pgfmathsetmacro{\quadValue}{max(0,min(1,(\r-3)/4))}
    \filldraw[sample,line width=.4pt]
        (\quadX,{4.35+1.8*\quadValue}) circle[radius=1.7pt];
}
\filldraw[fill=quadGold,draw=quadInk!70,line width=.5pt]
    (3.6875,5.3625) circle[radius=2.4pt];
\draw[quadGray,line width=.5pt] (3.79,5.38) -- (4.65,5.35);
\node[anchor=west,inner sep=2pt] at (4.65,5.35) {$\phi(s)$};

% Below: ideal selector weights, on exactly the same grid.
\draw[axis] (.8,1.95) -- (6.7,1.95);
\draw[axis] (.8,1.95) -- (.8,3.45);
\node[anchor=south east,inner sep=1pt] at (.65,3.35) {$\alpha_r$};
\node[anchor=west,inner sep=2pt] at (6.7,1.95) {$c_r$};
\node[anchor=east,inner sep=3pt] at (.8,1.95) {$0$};
\node[anchor=east,inner sep=3pt] at (.8,3.15) {$1$};
\draw[quadTeal!45,line width=.7pt,domain=0:10,samples=101]
    plot ({.8+.55*\x},{1.95+1.2/(1+(5.25-\x)^2)});
\foreach \r in {0,...,10} {
    \pgfmathsetmacro{\quadX}{.8+.55*\r}
    \pgfmathsetmacro{\quadWeight}{1/(1+(5.25-\r)^2)}
    \draw[quadTeal,line width=.95pt]
        (\quadX,1.95) -- (\quadX,{1.95+1.2*\quadWeight});
    \filldraw[sample,line width=.4pt]
        (\quadX,{1.95+1.2*\quadWeight}) circle[radius=1.7pt];
}
\node[anchor=north,inner sep=4pt] at (3.6875,1.95) {$s$};
\node[note] at (3.575,1.16)
    {Reconstructing the threshold value};
\node at (3.575,.35)
    {$\displaystyle \bar{x}=
      \frac{\sum_r\alpha_r\phi(c_r)}{\sum_r\alpha_r}
      \approx\phi(s)$};

\end{tikzpicture}
\endgroup
    \caption{Approximating a threshold function using selectors (schematic).
    Top: samples of the threshold function on a uniform grid.
    Bottom: the ideal weights $\alpha =1/(1+t^2)$
    concentrate around the input value $s$, marked by the dashed line.
    Nearby samples dominate the normalized weighted average, which
    approximates $\phi(s)$ as the grid is refined.}
    \label{fig:quadratic-hardness}
\end{figure}

To extend the argument to approximate VI solutions, we construct an interior comparison point using the ideal selector values. The skew-symmetric part of $\mat{A}$ ensures that the cross terms cancel in the VI comparison. \citet{Anagnostides25:Complexity} used a similar technique exploiting the skew-symmetric structure of the VI in their analysis of symmetric equilibria. We extend this idea to obtain quantitative bounds on selector errors. By appropriately scaling the selector equations, we show that a small VI gap forces the approximate solution to be close to this comparison point. Combining this proximity bound with the grid approximation yields an approximate circuit solution that can be decoded into a valid \PureCircuit{} assignment. A final normalization, as in the non-Euclidean case, establishes $\PPAD$-hardness on the unit Euclidean ball at inverse-polynomial accuracy. \Cref{fig:quadratic-hardness} illustrates how the selector weights approximate a threshold gate from grid samples.

\paragraph{Broader applicability of techniques.}

Our hardness reductions share a discretization-based approach to approximating threshold functions, but realize the sample weights in two complementary ways. The non-Euclidean construction exploits the geometry of the feasible set to obtain nonlinear inverse mappings, which are then transformed into weight functions suitable for function approximation. This approach suggests a possible route to establishing hardness over other general constraint sets, even with affine operators. In contrast, the Euclidean construction introduces selector variables to implement the desired weights directly through quadratic operators. The use of discretization in both our hardness reductions, together with similar techniques in prior works (see \Cref{sec:related}), highlights its versatility and suggests potential approaches to characterizing the complexity of other equilibrium and optimization problems.

\subsection{Further related work}
\label{sec:related}

\paragraph{The role of geometry and operator structure.} In variational inequalities, there are certain structural assumptions on the operator that guarantee tractability. For example, \citet{Anagnostides25:Polynomial} established polynomial-time algorithms for VIs under a \emph{Minty condition}, which generalizes monotonicity. Our positive result does not hinge on such a condition; in fact, \Cref{appendix:minty} gives an explicit example for which the Minty condition fails. In terms of the role of the constraint set, it is worth mentioning recent work on min-max optimization~\citep{bernasconi2024role}, where the product structure played a key role in developing reductions. Another result highlighting the role of geometry in
computational complexity is due to
\citet{goldberg2021hairy}, who established $\PPAD$-completeness of the Hairy Ball problem on even-dimensional spheres.

\paragraph{Fixed-point computation under $\ell_p$ norms.} 
Another related line of work concerns
fixed-point computation under different $\ell_p$ norms. \citet{chen2025computing} established polynomial query complexity for computing approximate fixed points of
$\ell_\infty$-contractions, which was subsequently extended to arbitrary $\ell_p$ norms by \citet{haslebacher2026query}. More recently, \citet{Daskalakis26:Polynomial} gave a polynomial-time algorithm for approximate fixed points of $\ell_p$-nonexpansive mappings when $p$ is an even integer. From a complexity-theoretic perspective,
\citet{fearnley2020unique} showed that finding fixed points
of piecewise-linear $\ell_p$-contractions lies in
$\mathsf{UEOPL}$, while \citet{batziou2025monotone}
established $\mathsf{UEOPL}$ membership for monotone
$\ell_\infty$-contractions. In these works, the norm determines the contraction or nonexpansiveness condition, whereas in ours, it defines the feasible region without
imposing such a restriction on the operator.

\paragraph{Quadratic optimization.}
\citet{Fearnley25:KKT} showed that computing a Karush-Kuhn-Tucker (KKT) point of a quadratic objective over a box is $\mathsf{CLS}$-complete. \citet{ghosh2024complexitysymmetricbimatrixgames} extended the corresponding hardness for symmetric equilibria in symmetric common-payoff bimatrix games. These results concern operators derived from quadratic potentials and establish hardness at exponentially small accuracy, whereas our $\PPAD$-hardness results hold at inverse-polynomial precision for operators
that need not admit a potential.

\paragraph{Polynomial VIs.} More broadly, there has been much work examining VIs in which the operator is a polynomial. In such settings, sum-of-squares (SoS) methods provide semidefinite relaxations for systems of polynomial equalities and inequalities~\citep{Parrilo03:Semidefinite}. Building on such techniques, \citet{Nie25:PolynomialVI} developed algorithms for polynomial VIs using Lagrange multipliers and SoS relaxations. In earlier work, \citet{Parrilo06:PolynomialGames} introduced a semidefinite programming formulation for computing the value and optimal mixed strategies of \emph{two-player zero-sum} polynomial games. Subsequently, \citet{Stein11:Correlated} developed algorithms for approximating correlated equilibria in continuous games with polynomial utilities.

\paragraph{Quadratic games.} A direct application of~\Cref{thm:affine} concerns quadratic games with shared constraints~\citep{Hou12:Normalized}. In particular, let us consider a game in which each player $i$ chooses a scalar $x_i$ with payoff
\[
 u_i(\vx)=b_i x_i-\frac{q_i}{2}x_i^2+x_i\sum_{j\ne i}w_{ij}x_j,
 \quad q_i\ge0,
\]
subject to the global constraint $\sum_i x_i^2\le R^2$. By applying~\Cref{thm:affine}, we can compute a point with VI gap at most $\epsilon$ in time polynomial in the input length and $\log(1/\epsilon)$, without requiring symmetry or monotonicity assumptions on the interactions. If in addition each player's utility is concave in that player's strategy, an approximate VI solution is also an $\epsilon$-generalized Nash equilibrium.

%On a related note, \citet{Benenati26:DynamicGames} examined the connection between linear-quadratic dynamic games and nonsymmetric VIs. \cref{thm:affine} also implies tractability for finite-horizon open-loop games whose sole constraint on the stacked control inputs is $\sum_{i,t}\norm{\vec u_{i,t}}_2^2\le R^2$. With a fixed initial state, eliminating the states through the linear dynamics makes each player's quadratic cost a quadratic function of the control inputs. If each reduced cost is convex in that player's own control sequence, the negative own-action gradients form an affine operator on the joint ball. The preceding argument therefore yields an approximate generalized Nash equilibrium, in polynomial time.

\paragraph{Temporal-difference updates.} Finally,
\citet{Bhandari21:TemporalDifference} analyzed temporal-difference learning with linear function approximation, where the expected update takes the form $g(\vec\theta)=\vec b-\mat A\vec\theta$. \Cref{thm:affine} computes a point with VI gap at most $\epsilon$ for $g$ on $B_2^d(R)$. Exact VI solutions are precisely stationary points of the projected update
\[
 \vec\theta=\Pi_{B_2^d(R)}\bigl(\vec\theta+\eta g(\vec\theta)\bigr),
\]
where $\eta > 0$ and $\Pi_{B_2^d(R)}$ denotes Euclidean projection.
\section{Preliminaries}\label{sec:preliminaries}
\paragraph{Notation.}
We use $d$ for the dimension of a VI and $n$ for the number of nodes in a source \PureCircuit{} instance. For a positive integer $d$, let $[d]=\{1,\ldots,d\}$. We write $\norm{\cdot}_p$ for the $\ell_p$ norm, with the usual convention $\norm{\vx}_\infty=\max_i|x_i|$. An unsubscripted norm denotes the Euclidean norm. 
For $R>0$, define the ball over $\R^d$ to be
\[
 B_p^d(R)=\{\vx\in\R^d:\norm{\vx}_p\le R\}.
\]
For the unit ball, we write $B_p^d=B_p^d(1)$. For any exponent $p$, the dual exponent is $q=p/(p-1)$ when $1<p<\infty$, and $q=\infty$ for $p=1$, $q=1$ for $p=\infty$.

\subsection{Variational inequalities and computational complexity}
Let $\calK\subset\R^d$ be nonempty, compact, and convex, and let $F:\calK\to\R^d$ be continuous.
\begin{definition}[VI solution]\label{def:vi-gap}
For $\vx\in\calK$, define
\[
 \Gap_{\calK}(\vx;F)=\max_{\vxp\in\calK}\inner{\vxp-\vx}{F(\vx)}.
\]
A point $\vx$ is an exact VI solution if $\Gap_{\calK}(\vx;F)=0$. It is an $\epsilon$-approximate VI solution if $\Gap_{\calK}(\vx;F)\le\epsilon$.
\end{definition}
We use the maximization convention in \eqref{eq:VI}. Accordingly, if
\[
 N_{\calK}(\vx)=\{\vec v\in\R^d:\inner{\vec v}{\vec y-\vx}\le0\text{ for every }\vec y\in\calK\}
\]
is the outward normal cone, the exact condition is $F(\vx)\in N_{\calK}(\vx)$, or equivalently $0\in-F(\vx)+N_{\calK}(\vx)$. This fixes the sign convention throughout the paper.

% \paragraph{Input, accuracy, and fixed exponents.}

% All affine and quadratic hardness results on unit balls hold with operator coefficients in $[-1,1]$. For an affine operator $F(\vx)=\mat A\vx+\vec b$, this applies to every entry of $\mat A$ and $\vec b$. For a quadratic operator, it applies to the coefficients of the constant, linear, and quadratic monomials in each coordinate, after expanding and collecting identical monomials. For the case where the exponent is a fixed constant, every finite exponent $p\ge1$ is a rational number chosen independently of the dimension. 

% \Cref{sec:uniform} considers the case where the finite rational exponent is given in a binary encoding.

\paragraph{The class \PPAD.}
The canonical problem \problem{End-of-Line} is given by Boolean predecessor and successor circuits describing a directed graph with indegree and outdegree at most one, together with a distinguished source. The task is to find another source or a sink. The class \PPAD{} consists of search problems that are polynomial-time reducible to this problem~\citep{Papadimitriou94:Complexity}. 

\subsection{PureCircuit and threshold functions}
\label{sec:purecircuit}
Our hardness constructions reduce from \PureCircuit{}, which was introduced by~\citet{Deligkas24:Pure}. A \PureCircuit{} instance has a node set $V$ with $|V|=n$ and gates of type \textsf{NOR} and \textsf{PURIFY}. The three nodes appearing in any gate are distinct, and each node in the circuit is the output of exactly one gate. A solution of \PureCircuit{} is an assignment $g$ that assigns each node a value in $\{0,1,\bot\}$. A \textsf{NOR} gate takes as input two nodes, say $u$ and $v$, and produces an output node $w$. We denote this by $w = \Nor(u, v).$ The rule of the \textsf{NOR} gate is that
\[
    g(u)= g(v) =0 \Rightarrow g(w) = 1
    \quad \text{and} \quad
    g(u) = 1 \text{ or } g(v) = 1 \Rightarrow g(w) = 0.
\]
A \textsf{PURIFY} gate takes as input a node $u$, and produces two outputs $v$ and $w$, denoted by $(v, w) = \Purify(u) $. The rule of the \textsf{PURIFY} gate is that
\[
    g(u) \in\{0,1\}\Rightarrow g(v) = g(w) = g(u),
\]
and if $g(u)=\bot$, at least one of $g(v) , g(w)$ lies in $\{0,1\}$.
\begin{theorem}[\citealp{Deligkas24:Pure}]
    \label{thm:purecircuit}
    Finding a solution to \PureCircuit{} is \PPAD-complete.
\end{theorem}
We define the following two piecewise linear threshold functions:
\[
    \norfun(t):=
    \begin{cases}
        1 & \text{if } t \le 1/4,\\
        2-4t & \text{if } 1/4\le t \le 1/2,\\
        0 & \text{if } t \ge 1/2,
    \end{cases} \quad \text{and} \quad
    \purfun(s):=
    \begin{cases}
        0 & \text{if } s \le 5/12,\\
        6 s - 5/2 & \text{if } 5/12 \le s \le 7/12,\\
        1 & \text{if } s \ge 7/12.
    \end{cases}
\]
Both functions map $\R$ to $[0,1]$ and are $6$-Lipschitz continuous.
\begin{figure}[t]
    \centering
    \begin{tikzpicture}[
    x=4.6cm,
    y=2.8cm,
    axis/.style={->, thick},
    graph/.style={very thick, niceMagenta},
    guide/.style={densely dashed, gray},
    ticklabel/.style={font=\scriptsize},
    title/.style={font=\small}
]

\newcommand{\thresholdaxes}[1]{
    \draw[axis] (-0.08,0) -- (0.88,0) node[right] {$t$};
    \draw[axis] (0,-0.08) -- (0,1.18) node[above] {$#1(t)$};
    \foreach \y/\label in {0/0,1/1} {
        \draw (-0.012,\y) -- (0.012,\y)
            node[left=7pt,ticklabel] {$\label$};
    }
}

\begin{scope}
    \thresholdaxes{\norfun}
    \draw[graph] (-0.05,1) -- (0.25,1) -- (0.5,0) -- (0.85,0);
    \foreach \x/\label in {0.25/\frac14,0.5/\frac12} {
        \draw[guide] (\x,0) -- (\x,1);
        \draw (\x,-0.012) -- (\x,0.012)
            node[below=3pt,ticklabel] {$\label$};
    }
\end{scope}

\begin{scope}[xshift=6.2cm]
    \thresholdaxes{\purfun}
    \draw[graph] (-0.05,0) -- (0.4166667,0) -- (0.5833333,1) -- (0.85,1);
    \foreach \x/\label in {0.4166667/\frac{5}{12},0.5833333/\frac{7}{12}} {
        \draw[guide] (\x,0) -- (\x,1);
        \draw (\x,-0.012) -- (\x,0.012)
            node[below=3pt,ticklabel] {$\label$};
    }
\end{scope}

\end{tikzpicture}
    \caption{The two piecewise linear threshold functions used to encode the \textsf{NOR} and \textsf{PURIFY} gates.}
    \label{fig:threshold-functions}
\end{figure}
We associate each node $v\in V$ with a scalar value
$x_v$ and let $\vx \in\R^n$
be the collection of all node values. From a \PureCircuit{} instance, we define a piecewise linear operator $T:\R^n \to [0,1]^n$ as follows. If node $w$ is the output of a $\Nor$ gate with input nodes $u$ and $v$, then the $w$-th coordinate of $T(\vx)$ is defined to be
\begin{align} \label{eq:piecewise_nor_operator}
    T_w(\vec x)=\norfun(x_u+x_v),
\end{align}
where $x_u$ and $x_v$ are the $u$-th and $v$-th coordinates of $\vx$, respectively. On the other hand, if $v$ and $w$ are the output nodes of a $\Purify$ gate with input node $u$, we define
\begin{align} \label{eq:piecewise_purify_operator}
    T_v(\vec x)=\purfun(x_u+1/4) \quad 
    \text{ and } \quad
    T_w(\vec x)=\purfun(x_u-1/4).
\end{align}
In the sequel, we will adopt the following notation. If $w = \textsf{NOR}(u, v)$, then we write $s_w(\vec x) = x_{u} + x_v$. If $(v, w) = \textsf{PURIFY}(u)$, then $s_v(\vec{x}) = x_u + 1/4$ and $s_w(\vec{x}) = x_u - 1/4$.
Writing $\phi_v=\norfun$ for a  $\Nor$ output and $\phi_v=\purfun$ for a $\Purify$ output, every coordinate then has the form
\begin{equation}\label{eq:gate-input}
 T_v(\vx)=\phi_v(s_v(\vx)).
\end{equation}
We proceed to show that solving~\PureCircuit{} reduces to finding an approximate solution of the induced fixed-point problem.
\begin{lemma}
\label{lem:rounding}
Let $0\le\deltapc \leq \frac{1}{32}$. If a vector $\vec x\in\R^n$ satisfies $\norm{\vec x-T(\vec x)}_\infty\le \deltapc,$ then the assignment
\[
    g(v)=
    \begin{cases}
        0 & \text{if } x_v\le 1/8,\\
        1 & \text{if } x_v\ge 7/8,\\
        \bot & \text{if }  1/8 < x_v < 7/8
    \end{cases}
\]
is a solution to~\PureCircuit{}.
\end{lemma}
\begin{proof}
Since $T(\vx) \in [0,1]^n$, $\norm{\vec x-T(\vec x)}_\infty\le \deltapc$ implies $x_v\in[-1/32,1+1/32]$ for every node $v \in V$.
For a \textsf{NOR} gate with output node $w$ and input nodes $u$ and $v$, we consider the following two cases. 
\begin{itemize}
    \item If $g(u)=g(v)=0$, from the definition of $g(\cdot)$, we have $x_u+x_v\le1/4$. Hence, $T_w(\vec x)=1$ and $x_w\ge T_w(\vx)-1/32\ge7/8$, so $g(w)=1$.
    \item If $g(u)=1$ or $g(v)=1$, without loss of generality, assume $g(u)=1$. Then $x_u\ge7/8$, so $x_u+x_v \geq 7/8 - 1/32 \geq 1/2 $. Hence, $T_w(\vec x)=0$ and $x_w\le T_w(\vx) + 1/32\le 1/8$, so $g(w)=0$.
\end{itemize}
For a \textsf{PURIFY} gate with output nodes $v$ and $w$ and input node $u$, consider the following cases.
\begin{itemize}
    \item If $g(u)=0$, then $x_u\le1/8$, we have $x_u+1/4<5/12$ and $x_u-1/4<5/12$. It holds that $T_v(\vec x)=T_w(\vec x)=0$. This implies $x_v \leq 1/32$ and $x_w \leq 1/32$, and both $g(v)$ and $g(w)$ have value 0.
    \item If $g(u)=1$, then $x_u \ge 7/8$, this gives $x_u+1/4 > 7/12$ and $x_u- 1/4 > 7/12$. Hence, $T_v(\vec x)=T_w(\vec x)=1$, which in turn implies that $x_v \geq 7/8$ and $x_w \geq 7/8$. Therefore we have $g(v) = g(w) = 1.$ 
    \item If $g(u)=\bot$, then $x_u\in(1/8,7/8)$. First consider the case where $x_u\le2/3$, then it holds that $x_u-1/4\le5/12$ and $T_w(\vec x)=0.$ Therefore, $x_w \leq 1/8$ and $g(w) = 0$. On the other hand, if $x_u > 2/3$, then $x_u+1/4\ge7/12$ and $T_v(\vec x)=1$. Hence, $x_v \geq 7/8$ and $g(v) = 1$. We show that in this case, at least one of $g(w)$ and $g(v)$ takes value in $\{0, 1\}.$
\end{itemize} We conclude that $g$ yields a valid assignment for \PureCircuit{} in all cases. The proof is complete.
\end{proof}
% \paragraph{Fixed padding.}
% The reductions may assume that the number of circuit nodes exceeds any fixed constant. To see this, adjoin disjoint copies of the three gates
% \[
%  a=\Nor(b,c),\quad b=\Nor(a,c),\quad c=\Nor(a,b).
% \]
% This component has the Boolean solution $(g(a),g(b),g(c))=(1,0,0)$, and restricting any solution of the enlarged instance to its original nodes solves the original instance. The padding does not change any gate in the original component.
\section{Geometry of an \texorpdfstring{$\ell_p$}{lp} ball}\label{sec:ball-geometry}
In this section, we establish the geometric properties of $\ell_p$ balls needed to analyze VI solutions. Let $1<p<\infty$ be a constant, and let $q= {p}/{(p-1)}$ be the dual exponent.  Define the duality map
\[
        J_p(\vec z)_i=\abs{z_i}^{p-2}z_i.
\]
We next characterize the support point of an $\ell_p$ ball in terms of the duality map.
\begin{lemma}\label{lem:support_point}
    For any vector $\vec z \in \R^d$, $\vec z \neq \vec{0},$ the support point, which is the unique maximizer of $\ip{\vec y}{\vec z}$ over $B_p^d(R)$ is
    \begin{equation}\label{eq:support-map}
        S_R(\vec z) := \argmax_{\vec y \in B_p^d(R)} \left\langle \vec y , \vec z \right\rangle
        =R\frac{J_q(\vec z)}{\norm{\vec z}_q^{q-1}}.
    \end{equation}
\end{lemma}
\begin{proof}
    Since $(q - 1)p = q,$ we have
    \begin{align*}
        \norm{J_q(\vec z)}_p^p = \sum_{i = 1}^d |z_i|^{(q - 1)p} = \sum_{i = 1}^d |z_i|^q = \norm{\vec z}_q^q.
    \end{align*}
    Thus $\norm{J_q(\vec z)}_p = \norm{\vec z}^{q - 1}_q$. Let $\vec s = R J_q(\vec z)/\norm{\vec z}^{q-1}_q,$ it then yields $\norm{\vec s}_p = R$ and therefore $\vec s \in B_p^d(R).$ Moreover, it holds that
    \begin{align*}
        \left\langle \vec s, \vec z \right\rangle = R \frac{\norm{\vec z}_q^q}{\norm{\vec z}_q^{q -1}} = R \norm{\vec z}_q.
    \end{align*}
    On the other hand, for any vector $\vec y \in B_p^d(R),$ Hölder's inequality gives $\left\langle \vec y, \vec z \right\rangle \leq \norm{\vec y}_p \norm{\vec z}_q \leq R \norm{\vec z}_q.$ Thus, we conclude that \begin{align*}
        \left\langle \vec s, \vec z \right\rangle = \max_{\vec y \in B_p^d(R)} \left\langle \vec y, \vec z \right\rangle.
    \end{align*}
    The uniqueness of the maximizer then follows from the fact that for $1 < p < \infty$, the $\ell_p$ ball is strictly convex.
\end{proof}
For any $1 < p < \infty$ and any $\vec z\in B_p^d(R)$, \Cref{lem:support_point} gives
\begin{equation}\label{eq:p-gap}
        \Gap_{B_p^d(R)}(\vec z;F)
        =R\norm{F(\vec z)}_q-\ip{\vec z}{F(\vec z)}.
\end{equation}
Moreover, the outward normal cone at $\vec z\in B_p^d(R)$ is
\[
 N_{B_p^d(R)}(\vec z)=
 \begin{cases}
 \{\vec 0\},&\norm{\vec z}_p<R,\\
 \{\lambda J_p(\vec z):\lambda\ge0\},&\norm{\vec z}_p=R.
 \end{cases}
\]
Indeed, any nonzero normal vector $\vec v \in N_{B_p^d(R)}(\vec z)$ satisfies $\vec z=S_R(\vec v)$, and the support formula gives
$\vec v=\norm{\vec v}_q J_p(\vec z)/R^{p-1}$. Conversely, H\"older's inequality shows that every nonnegative multiple of $J_p(\vec z)$ is a normal at a boundary point. We have the following characterization.
\begin{lemma}\label{lem:exact-normal}
Let $1<p<\infty$ be the exponent. A point $\vec z^*\in B_p^d(R)$ is an exact VI solution if and only if either $F(\vec z^*)=0$, or
\[
        \norm{\vec z^*}_p=R
        \quad\text{and}\quad
        F(\vec z^*)=\lambda J_p(\vec z^*)
\]
for some $\lambda>0$.
\end{lemma}
\begin{proof}
The case where $F(\vec z^*) = 0$ follows directly from \eqref{eq:p-gap}. If $F(\vec z^*) \neq 0$, \eqref{eq:p-gap} implies that $\vec z^*$ attains the support function of $B_p^d(R)$ in direction $F(\vec z^*)$.  Strict convexity gives $\vec z^*=S_R(F(\vec z^*))$.  From \eqref{eq:support-map}, we then have 
\[
        J_p(\vec z^*)=R^{p-1}\frac{F(\vec z^*)}{\norm{F(\vec z^*)}_q},
\]
so $F(\vec z^*)=\lambda J_p(\vec z^*)$ with $\lambda={\norm{F(\vec z^*)}_q}/{R^{p - 1}}$. The converse follows by substitution into \eqref{eq:p-gap}.
\end{proof}
For the Euclidean ball, the duality map is the identity, so
\[
 N_{B_2^d(R)}(\vec z)=
 \begin{cases}
 \{\vec 0\},&\norm{\vec z}_2<R,\\
 \{\lambda\vec z:\lambda\ge0\},&\norm{\vec z}_2=R.
 \end{cases}
\]
Thus the normal condition is linear in $\vec z$. Moreover, since $\max_{\norm{\vec z'}_2\le R}\inner{\vec z'}{F(\vec z)}=R\norm{F(\vec z)}_2$, \eqref{eq:p-gap} becomes
\begin{equation}
    \Gap_{B_2^d(R)}(\vec z;F)=R\norm{F(\vec z)}_2-\inner{\vec z}{F(\vec z)}.
    \label{eq:ball-gap}
\end{equation} We have the following corollary for the Euclidean case.
\begin{corollary}
\label{lem:normal-form}
A point $\vec z^*\in B_2^d(R)$ is an exact VI solution if and only if either $F(\vec z^*)=\vec 0$, or $\norm{\vec z^*}_2=R$ and there is $\lambda > 0$ such that
$
    F(\vec z^*) =\lambda \vec z^*.
$
\end{corollary}
% \begin{proof}
% By~\eqref{eq:ball-gap}, $\vec z \in B_2^d(R)$ is an exact VI solution if and only if
% \[
%     R\norm{F(\vec z)}_2=\inner{\vec z}{F(\vec z)}.
% \]
% If $F(\vec z)=\vec 0$, we are done. Otherwise, the Cauchy–Schwarz inequality yields
% \[
%     \inner{\vec z}{F(\vec z)}\le \norm{\vec z}_2\norm{F(\vec z)}_2\le R\norm{F(\vec z)}_2 = \langle \vec z, F(\vec z) \rangle.
% \]
% The above inequality implies $\norm{\vec z}_2=R$ and $\vec z/R=F(\vec z)/\norm{F(\vec z)}_2$. We conclude that $F(\vec z)=\lambda\vec z$ for $\lambda=\norm{F(\vec z)}_2/R > 0$. The converse follows immediately from~\eqref{eq:ball-gap}.
% \end{proof}
Next, we extend \Cref{lem:exact-normal} to approximate VI solutions.  Set
\begin{equation}\label{eq:mu-r}
 r_p=\begin{cases}2,&1<p\le2,\\ p,&p > 2,\end{cases}
 \quad
 \mu_p=\begin{cases}
        (p-1)/2,&1<p\le2,\\
        2^{2-p}/p,&p > 2,
 \end{cases}
 \quad
 C_p = \begin{cases}
     2^{2 - p}, & 1<p\leq 2,\\
     p - 1, & p > 2.
 \end{cases}
\end{equation}
The following lemma establishes that an approximate VI solution is also close to the support point $S_R(F(\vec z)).$ 
\begin{lemma}\label{lem:approx-normal}
For any exponent $1<p<\infty$ and for any vector $\vec z\in B_p^d(R)$, if $\Gap_{B_p^d(R)}(\vec z;F)\le\eps$ and $F(\vec z)\neq0$, then
\begin{equation}\label{eq:normality-gap}
 \norm{\vec z-S_R(F(\vec z))}_p
 \le
 R\left(
 \frac{\eps}{\mu_p R \norm{F(\vec z)}_q}
 \right)^{1/r_p}.
\end{equation}
Moreover, letting $\theta_p = \min \{1, p- 1\},$ for some $\lambda > 0,$
we have
\begin{align}
\label{eq:operator_align_with_gradient}
    \norm{F(\vec z) - \lambda J_p(\vec z)}_q \leq C_p \norm{F(\vec z)}_q \left(\frac{\eps}{\mu_p R \norm{F(\vec z)}_q}\right)^{\theta_p / r_p}.
\end{align}
\end{lemma}
\begin{proof}
    We denote the normalized vectors
    \begin{align*}
        \bar{\vec z} = \frac{\vec z}{R}, \quad \text{and} \quad \bar{\vec s} = \frac{S_R(F(\vec z))}{R} = \frac{J_q(F(\vec z))}{\norm{F(\vec z)}_q^{q - 1}}.
    \end{align*}
    Recall that $S_R(F(\vec z))$ lies on the boundary of $B_p^d(R),$ we then have $\norm{\bar{\vec z}}_p \leq 1$ and $\norm{\bar{\vec s}}_p = 1.$ Moreover, applying the duality map to $\bar{\vec s}$ gives
    \begin{align}\label{eq:gradient_normal_support}
        J_p(\bar{\vec s}) = \frac{F(\vec z)}{\norm{F(\vec z)}_q}.
    \end{align}
    We consider the following two cases:
    \begin{itemize}
        \item For $1 < p \leq 2,$ the function $\phi(\vec z) = \frac{1}{2} \norm{\vec z}_p^2$ is $( p - 1)$-strongly convex with respect to $\norm{\cdot}_p$. Thus, it holds that
        \begin{align*}
            \frac{1}{2} \norm{\bar{\vec z}}_p^2 \geq \frac{1}{2} \norm{\bar{\vec s}}_p^2 + \left\langle J_p(\bar{\vec s}), \bar{\vec z} - \bar{\vec s}\right\rangle + \frac{p - 1}{2} \norm{\bar{\vec z} - \bar{\vec s}}_p^2
        \end{align*}
        Using the fact that $\left\langle J_p(\bar{\vec s}), \bar{\vec s}\right\rangle = \norm{\bar{\vec s}}_p = 1$ and $\norm{\bar{\vec z}}_p \leq 1,$ we have
        \begin{align*}
            1 - \left\langle J_p(\bar{\vec s}), \bar{\vec z}\right\rangle \geq \frac{p - 1}{2} \norm{\bar{\vec z} - \bar{\vec s}}_p^2.
        \end{align*}
        Therefore,
        \begin{align*} 
            \frac{R \norm{F(\vec z)}_q-\ip{\vec z}{F(\vec z)}}{R\norm{F(\vec z)}_q}
            =1-\ip{\bar{\vec z}}{J_p(\bar{\vec s})}
            \ge\mu_p\norm{\bar{\vec z}-\bar {\vec s}}_p^{r_p}.
        \end{align*}
        \item For $p > 2,$ the standard strong monotonicity inequality
        \[
         \ip{J_p(\vec u)-J_p(\vec v)}{\vec u- \vec v}
         \ge2^{2-p}\norm{\vec u-\vec v}_p^p
        \]
        implies
        \[
         \frac1p\norm{\bar{\vec z}}_p^p
         \ge
         \frac1p+\ip{J_p(\bar {\vec s})}{\bar{\vec z}-\bar {\vec s}}
         +\frac{2^{2-p}}p\norm{\bar{\vec z}-\bar {\vec s}}_p^p.
        \]
        Indeed, this inequality follows by integrating the strong monotonicity of $J_p$ along the line segment
        $\bar{ \vec s}+ t(\bar{\vec z}-\bar{ \vec s}).$
        We defer the detailed proof to \cref{appendix:convexity}. Using the fact that $\norm{\bar{\vec z}}_p \leq 1,$ it holds that
        \begin{align*} 
            \frac{R \norm{F(\vec z)}_q-\ip{\vec z}{F(\vec z)}}{R\norm{F(\vec z)}_q}
            =1-\ip{\bar{\vec z}}{J_p(\bar{\vec s})}
            \ge\mu_p\norm{\bar{\vec z}-\bar {\vec s}}_p^{r_p}.
        \end{align*}
    \end{itemize}
    Applying \eqref{eq:p-gap} to both cases gives \eqref{eq:normality-gap}.
    It remains to establish \eqref{eq:operator_align_with_gradient}. Let
    \[
        \lambda = \frac{\norm{F(\vec z)}_q}{R^{p-1}}.
    \]
    By the definition of $\bar{\vec s}$ and \eqref{eq:gradient_normal_support}, we have
    \[
        F(\vec z)
        = \norm{F(\vec z)}_q J_p(\bar{\vec s})
        = \lambda J_p(R\bar{\vec s}).
    \]
    Since $R\bar{\vec s}=S_R(F(\vec z))$, it follows that
    \[
        \norm{F(\vec z)-\lambda J_p(\vec z)}_q
        =
        \lambda
        \norm{J_p(S_R(F(\vec z)))-J_p(\vec z)}_q.
    \]
    Again we consider the following two cases:

    \begin{itemize}
        \item For $1<p\leq 2$, the duality map $J_p$ is $(p-1)$-H\"older continuous with
    \[
        \norm{J_p(\vec u)-J_p(\vec v)}_q
        \leq
        2^{2-p}\norm{\vec u-\vec v}_p^{p-1}.
    \]
    Therefore,
    \begin{align*}
        \norm{F(\vec z)-\lambda J_p(\vec z)}_q
        &\leq
        \lambda
        2^{2-p}
        \norm{S_R(F(\vec z))-\vec z}_p^{p-1} \\
        &=
         \frac{\norm{F(\vec z)}_q}{R^{p-1}} 2^{2-p}
        \left(
            \norm{S_R(F(\vec z))-\vec z}_p
        \right)^{p-1} \\
        &\leq
        2^{2-p}\norm{F(\vec z)}_q
        \left(
            \frac{\eps}
            {\mu_p R\norm{F(\vec z)}_q}
        \right)^{(p-1)/2},
    \end{align*}
    where the last step follows from \eqref{eq:normality-gap}.
    \item Similarly, for $p>2$, $J_p$ is Lipschitz on $B_p^d(R)$ with
    \[
        \norm{J_p(\vec u)-J_p(\vec v)}_q
        \leq
        (p-1)R^{p-2}\norm{\vec u-\vec v}_p.
    \]
    Hence
    \begin{align*}
        \norm{F(\vec z)-\lambda J_p(\vec z)}_q
        &\leq
        \lambda
        (p-1)R^{p-2}
        \norm{S_R(F(\vec z))-\vec z}_p \\
        &=
        (p-1)\norm{F(\vec z)}_q
        \frac{\norm{S_R(F(\vec z))-\vec z}_p}{R} \\
        &\leq
        (p-1)\norm{F(\vec z)}_q
        \left(
            \frac{\eps}
            {\mu_p R\norm{F(\vec z)}_q}
        \right)^{1/p}.
    \end{align*}
    \end{itemize}
    Substituting $C_p$ and $\theta_p$ completes the proof.
\end{proof}

For the Euclidean ball, the following corollary holds.
\begin{corollary}\label{lem:approx-normality}
Consider a point $\vec z\in B_2^d(R)$ such that $\Gap_{B_2^d(R)}(\vec z;F)\le\epsilon$, and suppose $\norm{F(\vec z')}_2\le \Fmax$ for all $\vec z'\in B_2^d(R)$. Then there are $\lambda\ge0$ and $\vec e\in\R^d$ such that
\[
    F(\vec z)=\lambda\vec z+\vec e,
    \quad \text{and} \quad
    \norm{\vec e}_2\le \sqrt{\frac{2\epsilon \Fmax}{R}}.
\]
\end{corollary}
\begin{proof}
If $F(\vec z)=\vec 0$, we take $\lambda=0$ and $\vec e=\vec 0$. Otherwise, let $\vec u=F(\vec z)/\norm{F(\vec z)}_2$ be a unit vector and $\lambda=\norm{F(\vec z)}_2/R$. Then
\[
    \norm{F(\vec z)-\lambda\vec z}_2^2
    = \norm{ \norm{F(\vec z)}_2 \vec{u} - \frac{ \norm{F(\vec z)}_2 }{R} \vec{z}}_2^2 = \frac{\norm{F(\vec z)}_2^2}{R^2}\norm{R\vec u-\vec z}_2^2.
\]
Since $\norm{\vec z}\le R$, it holds that
\[
    \norm{R\vec u-\vec z}_2^2
    =R^2+\norm{\vec z}_2^2-2R\inner{\vec z}{\vec u}
    \le 2R(R-\inner{\vec z}{\vec u}).
\]
Moreover, \eqref{eq:ball-gap} gives
\[
    R-\inner{\vec z}{\vec u}
    =\frac{R\norm{F(\vec z)}_2-\inner{\vec z}{F(\vec z)}}{\norm{F(\vec z)}_2}
    =\frac{\Gap_{B_2^d(R)}(\vec z;F)}{\norm{F(\vec z)}_2}
    \le \frac{\epsilon}{\norm{F(\vec z)}_2}.
\]
Combining gives 
\begin{align*}
    \norm{F(\vec z)-\lambda\vec z}^2_2\le \frac{2\epsilon\norm{F(\vec z)}_2}{R}\le \frac{2\epsilon \Fmax}{R}.
\end{align*} Taking the square root completes the proof.
\end{proof}
% \subsection{The Euclidean ball as a special case}\label{sec:euclidean-normal}
% For the Euclidean ball $B_2^d(R)$. The duality map is the identity, so
% \[
%  N_{B_2^d(R)}(\vec z)=
%  \begin{cases}
%  \{\vec 0\},&\norm{\vec z}_2<R,\\
%  \{\lambda\vec z:\lambda\ge0\},&\norm{\vec z}_2=R.
%  \end{cases}
% \]
% Thus the normal condition is linear in $\vec z$ once the multiplier is fixed. We record both the exact condition and the Euclidean estimate used below.

% \begin{lemma}
% \label{lem:normal-form}
% A point $\vec z\in B_2^d(R)$ is an exact VI solution of $\textsc{VI}(B_2^d(R), F)$ if and only if either $F(\vec z)=\vec 0$, or $\norm{\vec z}_2=R$ and there is $\lambda > 0$ such that
% \[
%     F(\vec z) =\lambda \vec z.
% \]
% \end{lemma}
% \begin{proof}
% By~\eqref{eq:ball-gap}, $\vec z \in B_2^d(R)$ is an exact VI solution if and only if
% \[
%     R\norm{F(\vec z)}_2=\inner{\vec z}{F(\vec z)}.
% \]
% If $F(\vec z)=\vec 0$, we are done. Otherwise, the Cauchy–Schwarz inequality yields
% \[
%     \inner{\vec z}{F(\vec z)}\le \norm{\vec z}_2\norm{F(\vec z)}_2\le R\norm{F(\vec z)}_2 = \langle \vec z, F(\vec z) \rangle.
% \]
% The above inequality implies $\norm{\vec z}_2=R$ and $\vec z/R=F(\vec z)/\norm{F(\vec z)}_2$. We conclude that $F(\vec z)=\lambda\vec z$ for $\lambda=\norm{F(\vec z)}_2/R > 0$. The converse follows immediately from~\eqref{eq:ball-gap}.
% \end{proof}
\section{Affine VIs over a non-Euclidean ball}\label{sec:non-euclidean}
In this section, we show that finding approximate VI solutions for affine operators is \PPAD-hard on every non-Euclidean $\ell_p$ ball. For any constant exponent $1<p<\infty$, $p\ne2$, the reduction exploits the nonlinear normal map induced by the boundary VI condition. The endpoint cases $p = 1$ and $p = \infty$ follow from known hardness results~\citep{Chen09:Settling, Rubinstein15:Inapproximability, bernasconi2024role}. 

The starting point of our reduction is \PureCircuit{}. To encode the piecewise-linear threshold functions from \PureCircuit{}, we discretize the
domain and approximate their values through weighted
averages of finitely many samples. Related discretization techniques have been used by~\citet{Anagnostides26:Team} and~\citet{Bernasconi26:Quadratic}. In our reduction, we construct a nonnegative weight function whose polynomial tail decay controls the contributions of distant samples.

We start by stating the main theorem of this section.
% \begin{theorem}\label{thm:main}
% For every fixed rational exponent $p$ such that $1 \leq p < \infty$ or $p=\infty$, there exists a rational affine operator $F(\vx)=\mat A\vx+\vec b$, where $\mat A\in\Q^{d\times d}$ and $\vec b\in\Q^d$, with every entry of $\mat A$ and $\vec b$ in $[-1,1]$ such that
% \begin{enumerate}[label=(\roman*)]
% \item If $p\in [1,\infty)\backslash\{2\}$, there is a constant $c_p>0$ such that it is $\PPAD$-hard to find $\vx\in B_p^d$ with
% \[
%         \Gap_{B_p^d}(\vx;F)\le d^{-c_p}.
% \]
% % \item There is a constant $c_1>0$ such that, for $p=1$, it is $\PPAD$-hard to find $\vx\in B_1^d$ with
% % \[
% %         \Gap_{B_1^d}(\vx;F)\le d^{-c_1}.
% % \]
% \item If $p=\infty$, there is an absolute constant $\eps_\infty>0$ such that it is $\PPAD$-hard to find $\vx\in B_\infty^d$ with
% \[
%         \Gap_{B_\infty^d}(\vx;F)\le \eps_\infty.
% \]
% \end{enumerate}
% All constructed operators have rational coefficients of polynomial bit complexity.
% \end{theorem}
% \jy{Define the computational problem}

\begin{theorem}\label{thm:main}
The following results hold for
\problem{Ball$_p$-VI}, even when the input operator
is affine, $F(\vec z)=\mat A\vec z+\vec b$, with
$\mat A\in\Q^{d\times d}$ and $\vec b\in\Q^d$,
and all coefficients lie in $[-1,1]$.
\begin{enumerate}[label=(\roman*)]
    \item For every fixed rational exponent
    $p\in[1,\infty)\backslash\{2\}$, there exists
    a constant $c_p>0$ such that
    \problem{Ball$_p$-VI} is $\PPAD$-complete for $\epsilon = d^{-c_p}$.

    \item For $p=\infty$, there exists a
    constant $c > 0$ such that
    \problem{Ball$_\infty$-VI} is $\PPAD$-complete
    for $\epsilon = c$.
\end{enumerate}
\end{theorem}

\subsection{The inverse mapping and weight functions}
We first consider the case where $1<p<\infty$ and $p\ne2$. The main ingredient of our reduction is a nonlinear inverse mapping induced by the boundary VI optimality condition.
\subsubsection{The inverse mapping \texorpdfstring{$\sigma_\lambda$}{sigma-lambda}}
One crucial observation is that although the operator $F$ is affine, the duality map $J_p$ is nonlinear whenever $p\neq 2$. Our reduction
exploits this to encode nonlinearity
through a careful design of an affine operator. Specifically, we consider an affine
operator of the form $F(\vec z)=h(\vec z)-\vec z$, where
$h:\R^d\to\R^d$ is affine. Suppose $\vec z^*\in B_p^d(R)$
is an exact VI solution with $F(\vec z^*)\neq\vec 0$.
By \Cref{lem:exact-normal}, there exists $\lambda>0$ such that
\[
    h(\vec z^*)=\vec z^*+\lambda J_p(\vec z^*).
\]
As we will show, this equation allows us to express each
coordinate of $\vec z^*$ through the inverse of a nonlinear mapping, which serves as the main building block
of our reduction.

% We note that if an exact VI solution $\vec z^*$ lies in the interior of $B_p^d(R)$, it implies that $F(\vec z^*)=\vec 0$. In this case, the optimality condition reduces to a system of affine equations, and the nonlinearity vanishes. To avoid this degeneracy, our construction introduces a positive anchor coordinate whose corresponding VI operator is strictly positive. This ensures that $F(\vec z)\neq\vec 0$ for every $\vec z$. Consequently, every exact VI solution must lie on the boundary of $B_p^d(R)$.

Since the duality map $J_p$ acts on each coordinate independently, we can analyze the nonlinearity by considering the following scalar mapping. For any $\lambda>0$, we define $h_\lambda:\R\to\R$ by
\begin{equation}\label{eq:h-sigma}
h_\lambda(t)=t+\lambda|t|^{p-2}t.
\end{equation}
We begin by establishing several basic properties of $h_\lambda$, including its invertibility.
\begin{lemma} \label{lem:h_scalar_property}
    For every $\lambda>0$, $h_\lambda: \R \to \R$ is odd, strictly increasing, and bijective. Moreover, its inverse is 1-Lipschitz.
\end{lemma}
% \begin{proof}
%     The term $\left|x\right|^{p - 2} x$ is strictly increasing for $1 < p <\infty$.  Hence, for $u,v\in\R$,
%     \[
%      (h_\lambda(u)-h_\lambda(v))(u-v)
%      =(u-v)^2+\lambda(J_p(u)-J_p(v))(u-v)
%      \ge (u-v)^2.
%     \]
%     The inequality is strict when $u\ne v$. Since $h_\lambda$ is continuous and tends to $\pm\infty$ as its argument tends to $\pm\infty$, it is onto. Its inverse is $1$-Lipschitz by the same inequality.  Oddness is immediate.
% \end{proof}
\begin{proof}
    Since both $t$ and $|t|^{p-2}t$ are strictly
    increasing for $p>1$, $h_\lambda$ is strictly
    increasing. Moreover, for any $u>v$, we have
    \[
        h_\lambda(u)-h_\lambda(v)\ge u-v.
    \]
    Together with the continuity of $h_\lambda$ and the fact that
    $\lim_{t\to\pm\infty}h_\lambda(t)=\pm\infty$,
    we show that $h_\lambda$ is bijective
    and its inverse is $1$-Lipschitz.
    Oddness follows directly from the definition.
\end{proof}
Since $h_\lambda$ is a bijection, we define the inverse mapping by
\begin{align*}
    \sigma_\lambda=h_\lambda^{-1}.
\end{align*}
We proceed to show some important properties of $\sigma_\lambda.$
\begin{lemma} \label{lem:sigma_property}
The inverse mapping $\sigma_\lambda$ is odd, increasing, and $1$-Lipschitz.  For any $t \in \R$, we have $\abs{\sigma_\lambda(t)}\le\abs{t}.$
Moreover, $\sigma_\lambda$ is strictly convex on $(0,\infty)$ when $1<p<2$, and strictly concave on $(0,\infty)$ for $p>2$.
\end{lemma}
\begin{proof}
    The oddness, monotonicity, and Lipschitz continuity of $\sigma_\lambda$ follow directly from \cref{lem:h_scalar_property}. The pointwise bound $\abs{\sigma_\lambda(t)}\le\abs{t}$ follows from the fact that $h_\lambda(u)$ has the same sign as $u$ and $\abs{h_\lambda(u)}\ge\abs{u}$.  For $t>0,$ implicit differentiation gives
    \begin{align}
         \sigma_\lambda'(t)
         &=\frac{1}{1+\lambda(p-1)\sigma_{\lambda}(t)^{p-2}},\label{eq:sigma-prime}
    \end{align}
    and
    \begin{align}
         \sigma_\lambda''(t)
         &=-\frac{\lambda(p-1)(p-2)\sigma_{\lambda}(t)^{p-3}}
         {\bigl(1+\lambda(p-1)\sigma_\lambda(t)^{p-2}\bigr)^3}.
         \label{eq:sigma-second}
    \end{align}
    It remains straightforward to check that when $1 < p < 2,$ $\sigma''_\lambda(t) > 0$ and $\sigma''_\lambda(t) < 0$ when $p > 2.$
\end{proof}
\Cref{lem:sigma_property} reveals a key distinction between the Euclidean and non-Euclidean cases. When $p=2$, the inverse mapping $\sigma_\lambda(t)=t/(1+\lambda)$ is linear, whereas for $p\neq2$, it exhibits nonzero curvature on $(0,\infty)$. As we will show, this nonlinearity in the curvature plays a central role in our hardness reduction.
\subsubsection{A weight function from finite differences}
We now exploit the curvature of the inverse mapping $\sigma_\lambda$ to construct a nonnegative weight function through successive finite differences. In fact, we will show that this weight function is positive on a fixed interval and satisfies a polynomial tail bound of order \(O(|t|^{-\beta_p})\) for some \(\beta_p>2\). This tail bound controls the contributions of distant samples, allowing us to approximate the threshold functions of \PureCircuit{} through weighted averaging. Specifically, we set the sign indicator
\[
        \varsigma_p=\sgn(2 - p)\in\{-1,1\},
\]
and define the finite difference
\begin{align}
 \Delta_\lambda(t)
 &:=\sigma_\lambda(t+1)-2\sigma_\lambda(t)+\sigma_\lambda(t-1).
 \label{eq:Delta}
\end{align}
Intuitively, $\Delta_\lambda(t)$ is a discrete analogue of the second derivative of $\sigma_\lambda$ and captures its local curvature.
We further construct a weight function through another finite difference of $\sigma_\lambda$, with a step size determined by $\Delta_\lambda(t)$
\begin{align}
        k_\lambda(t)
     &:=\varsigma_p\left(
     \sigma_\lambda\left(4+\frac14\Delta_\lambda(t)\right)
     +\sigma_\lambda\left(4-\frac14\Delta_\lambda(t)\right)
     -2\sigma_\lambda(4)
     \right).
     \label{eq:kernel}
\end{align}
The key advantage of this construction is that, unlike $\Delta_\lambda$, the resulting weight function $k_\lambda$ is always nonnegative while retaining sufficiently strong polynomial decay in the tails.

Our construction considers the multiplier $\lambda$ to be close to $1$. To this end, we establish the properties of $k_\lambda$ over the neighborhood $\lambda\in[1/2,2]$ and show that $k_\lambda$ remains close to $k_1$ when $\lambda$ is sufficiently close to $1$. Later in the reduction, we design an anchor coordinate to ensure this condition holds.

We move on to show some important properties of the weight function $k_\lambda$.
\begin{lemma}\label{lem:kernel}
The weight function $k_\lambda$ has the following properties.
\begin{enumerate}[label=(\roman*)]
\item For every $\lambda>0$, $k_\lambda$ is nonnegative and $2$-Lipschitz, with $|k_\lambda(t)|\le17$ for all $t\in\R$.
% \item For every $\lambda>0$, $k_\lambda(t)\ge0$ for all $t$, and $k_\lambda$ is $2$-Lipschitz.
\item There exist positive constants
$c_0,r_0\in\Q$, $c_k>0$, and $\beta_p>2$,
depending only on $p$, such that for every
$\lambda\in[1/2,2]$, we have 
\begin{align}
    k_\lambda(t)\ge c_0
    \quad \forall |t-2|\le r_0, \quad \text{and} \quad
    k_\lambda(t)\le c_k|t|^{-\beta_p}
    \quad \forall |t|\ge4.
    \label{eq:k-tail}
\end{align}
\item For every $\eps_k>0$, there exists
$\kappa_k = \poly(\eps_k) > 0$ such that
\begin{equation}\label{eq:k-lambda-stability}
    |\lambda-1|\le\kappa_k
    \quad\Longrightarrow\quad
    \|k_\lambda-k_1\|_\infty\le\eps_k.
\end{equation}
\end{enumerate}
\end{lemma}
\begin{proof}
Because $\sigma_\lambda$ is $1$-Lipschitz, we have
\[
 \abs{\Delta_\lambda(t)}
 \le
 \abs{\sigma_\lambda(t+1)-\sigma_\lambda(t)}
 +\abs{\sigma_\lambda(t)-\sigma_\lambda(t-1)}
 \le2.
\]
Therefore in \eqref{eq:kernel}, $4 + \Delta_\lambda(t)/{4} \leq {9}/{2}$ and $4 - \Delta_\lambda(t)/{4} \leq {9}/{2}$.  Since
$\abs{\sigma_\lambda(s)}\le\abs{s}$,
\[
        \abs{k_\lambda(t)}
        \le\frac92+\frac92+2\cdot4=17.
\]
By \Cref{lem:sigma_property}, $\sigma_\lambda$ is strictly convex on the interval $(0, \infty)$ when $1 < p<2$ and strictly concave when $p>2$. The quantity
\begin{align*}
    \sigma_\lambda\left(4+\frac14\Delta_\lambda(t)\right)
     +\sigma_\lambda\left(4-\frac14\Delta_\lambda(t)\right)
     -2\sigma_\lambda(4)
\end{align*}
is therefore nonnegative for $1 < p < 2$ and nonpositive when $p > 2$.  The sign $\varsigma_p$ then yields $k_\lambda\ge0$. Moreover, since $\sigma_{\lambda}$ is $1$-Lipschitz continuous, we have
\[
        \abs{\Delta_\lambda(t)-\Delta_\lambda(s)}\le4\abs{t-s}.
\]
As a result, the weight function $k_\lambda$ is $2$-Lipschitz.

For part~(ii), we first show that for all $\lambda \in [\frac{1}{2}, 2],$ it holds that $k_\lambda(2) > 0.$ We choose $t=2$ so that all three points $t+1, t,$ and $t - 1$ in $\Delta_\lambda(2)$ lie in $(0,\infty)$, where $\sigma_\lambda$ is strictly convex or strictly concave.
Indeed, by \Cref{lem:sigma_property}, $\sigma_\lambda$ is strictly convex
on $(0,\infty)$ when $1<p<2$ and strictly concave there when $p>2$.
Hence
\[
    \Delta_\lambda(2)
    =
    \sigma_\lambda(3)-2\sigma_\lambda(2)+\sigma_\lambda(1)
    \neq 0.
\]
Moreover, since $\abs{\Delta_\lambda(t)}\le 2$ for any $t$, we have
$
    4\pm \Delta_\lambda(2)/4 \in \left[{7}/{2}, {9}/{2}\right].
$
Applying again the strict convexity or
concavity of $\sigma_\lambda$ over $(0, \infty)$ implies that the quantity
\begin{align*}
    \sigma_\lambda\left(4+\frac14\Delta_\lambda(2)\right)
     +\sigma_\lambda\left(4-\frac14\Delta_\lambda(2)\right)
     -2\sigma_\lambda(4) 
\end{align*}
is strictly positive for $1 < p <2$ and strictly negative for $p > 2$. By the choice of the sign
$\varsigma_p$, this gives $k_\lambda(2)>0.$ Moreover, from the continuity of $k_\lambda$ with respect to $\lambda$ and the fact that $[{1}/{2}, 2]$ is compact, we have 
\[
    q_p
    =
    \min_{\lambda\in[1/2,2]} k_\lambda(2)
    >0.
\]
Choose numbers $c_0,r_0\in(0,1)$ such that
\[
    c_0\le \frac{q_p}{2}
    \quad \text{and} \quad 
    r_0\le \frac{q_p}{4}.
\]
Since $q_p>0$ depends only on the fixed exponent $p$,
we can choose $c_0$ and $r_0$ as fixed positive rationals independent of the circuit size $n$.
This property will be crucial in our subsequent PPAD-hardness
reduction, as it allows the gate approximations to be constructed using a constant number of fixed rational coefficients, ensuring that the overall reduction runs in polynomial time and produces operators with coefficients of polynomial bit length.

Since $k_\lambda$ is $2$-Lipschitz for every $\lambda>0$, whenever $\lambda\in[1/2,2]$ and $\abs{t-2}\le r_0$,
\[
\begin{aligned}
    k_\lambda(t)
    &\ge
    k_\lambda(2)-2\abs{t-2} \\
    &\ge
    q_p-2r_0 \\
    &\ge
    \frac{q_p}{2}
    \ge c_0.
\end{aligned}
\]
This proves the first bound of part~(ii).
To prove the tail bound, we fix
$\lambda \in [1/2,2]$ and $t \geq 3$. Since $h_\lambda(1)=1+\lambda\leq 3$ and $h_\lambda$ is increasing,
we have $\sigma_\lambda(t) \geq 1$. We consider the following two cases:
\begin{itemize}
    \item Suppose first that $p>2$. From the definition of $h_\lambda,$
    \[
        t
        =
    \sigma_\lambda(t)+\lambda \sigma_\lambda(t)^{p-1}
        \leq
        3\sigma_\lambda(t)^{p-1},
    \]
    and hence
    \[
        \sigma_\lambda(t)
        \geq
        \left(\frac{t}{3}\right)^{1/(p-1)}.
    \]
    Using~\eqref{eq:sigma-second} and plugging in $\lambda \in [1/2, 2],$ we obtain
    \[
        \left|\sigma_\lambda''(t)\right|
        \leq
        \frac{16(p-2)}{(p-1)^2}
        \left(\frac{t}{3}\right)^{\frac{3 - 2p}{p - 1}}.
    \]
    \item If $1<p<2$, then $\sigma_\lambda(t)^{p-1}\leq \sigma_\lambda(t)$, so
    \[
        t
        =
        \sigma_\lambda(t)+\lambda \sigma_\lambda(t)^{p-1}
        \leq
        3\sigma_\lambda(t),
    \]
    and therefore $\sigma_\lambda(t)\geq t/3$. The denominator in
    \eqref{eq:sigma-second} is at least one, and thus
    \[
        \left|\sigma_\lambda''(t)\right|
        \leq
        2(p-1)(2-p)\left(\frac{t}{3}\right)^{p - 3}.
    \]
\end{itemize}
Since the function $\sigma_\lambda(\cdot)$ is odd, the same absolute-value estimates hold for $t\leq -3$.
Set
\begin{align*}
    \gamma_p
    =
    \begin{cases}
        3-p,
        & 1<p<2, \\[4pt]
        \displaystyle
        \frac{2p - 3}{p -1},
        & p>2,
    \end{cases} \quad \text{and} \quad C_{\sigma,p}
    =
    \begin{cases}
        2(p-1)(2-p)3^{\gamma_p},
        & 1<p<2, \\[4pt]
        \displaystyle
        \frac{16(p-2)}{(p-1)^2}
        3^{\gamma_p},
        & p>2.
    \end{cases}
\end{align*}
Then for $\lambda\in[1/2,2]$ and $|t|\geq 3$, we have 
\begin{align} \label{eq:bound_on_second_order}
    \left|\sigma_\lambda''(t)\right|
    \leq
    C_{\sigma,p}|t|^{-\gamma_p},
\end{align}
with some constant $\gamma_p \in (1, 2).$
For $|t|\geq 4$, the mean-value theorem implies that there exists a point
$\xi_t\in(t-1,t+1)$ such that
\[
    \Delta_\lambda(t)
    =
    \sigma_\lambda''(\xi_t).
\]
Since $|\xi_t|
    \geq
    |t|-1
    \geq
    \frac{|t|}{2},$
\eqref{eq:bound_on_second_order} yields
\[
    |\Delta_\lambda(t)|
    \leq
    C_{\Delta,p}|t|^{-\gamma_p},
\]
with $
    C_{\Delta,p}
    =
    2^{\gamma_p}C_{\sigma,p}.$
Moreover, since $\Delta_\lambda(t)\in[-2,2]$, applying the mean-value theorem again implies that there exists some point $r\in\left[{7}/{2},{9}/{2}\right]$
such that
\begin{align}\label{eq:second_order_diff_bound}
    \sigma_\lambda\left(4+\frac{\Delta_\lambda(t)}{4}\right)
    +
    \sigma_\lambda\left(4-\frac{\Delta_\lambda(t)}{4}\right)
    -
    2\sigma_\lambda(4)
    =
    \frac{\Delta_\lambda(t)^2}{16}\sigma_\lambda''(r).
\end{align}
For
$r\in[7/2,9/2]$, we have 
    $1< \sigma_\lambda(r)\leq 9/2.$
Hence \eqref{eq:sigma-second} gives the explicit bound
\[
    \left|\sigma_\lambda''(r)\right|
    \leq
    M_p
    =
    2(p-1)|p-2|
    \max\left\{
        1,
        \left(\frac{9}{2}\right)^{p-3}
    \right\}.
\]
Combining the preceding bound with \eqref{eq:second_order_diff_bound}, we get
\[
    0
    \leq
    k_\lambda(t)
    \leq
    \frac{M_p}{16}
    C_{\Delta,p}^2
    |t|^{-2\gamma_p}.
\]
The desired tail bound then holds with
\[
    \beta_p
    =
    2\gamma_p
    >2,
    \quad
    c_k
    =
    \frac{M_p}{16}C_{\Delta,p}^2.
\]

Finally, for any $s \in \R$, let $
    u = \sigma_\lambda(s),$ and $ v = \sigma_1(s).
$
From the definition of $\sigma_\lambda$ and $h_\lambda,$ we have
\begin{align*}
    s = h_1(v) = v + J_p(v).
\end{align*}
On the other hand, we have
\begin{align*}
    h_\lambda(v) = v + \lambda J_p(v) = s + (\lambda - 1) J_p(v) = h_\lambda(u) + (\lambda - 1) J_p(v).
\end{align*}
Thus,
\begin{align}
 \abs{\sigma_\lambda(s)-\sigma_1(s)}
 & = \left|\sigma_\lambda(h_\lambda(u)) - \sigma_\lambda(h_\lambda(v))\right| \notag\\
 & \leq \left|h_\lambda(u) - h_\lambda(v)\right| \notag\\
 & \leq
 \abs{\lambda-1}\abs{J_p(v)} \notag\\
 & \le
 \abs{\lambda-1}\abs{s}^{p-1} \label{eq:sigma-lambda-finite},
\end{align}
where in the last step we use the fact that $|J_p(v)| = |v|^{p - 1}$ and $|v| = |\sigma_1(s)| \leq |s|.$
Now fix any accuracy $\eps_k>0$, and choose the constant
\[
 M=\max\left\{4,\left\lceil\left(\frac{4c_k}{\eps_k}\right)^{1/\beta_p}\right\rceil\right\}.
\]
For $\lambda\in[1/2,2]$ and $\abs{t}\ge M$, \eqref{eq:k-tail} gives
\[
 \abs{k_\lambda(t)-k_1(t)}\le2c_kM^{-\beta_p}\le\frac{\eps_k}{2}.
\]
On the other hand, if $\abs{t}\le M$, the definition of \(\Delta_\lambda\), together with \eqref{eq:sigma-lambda-finite}, yields
\[
 \abs{\Delta_\lambda(t)-\Delta_1(t)}
 \le4\abs{\lambda-1}(M+1)^{p-1}.
\]
Moreover, for $|t|\leq M$, recall that
\[
    k_\lambda(t)
    =
    \varsigma_p
    \left(
        \sigma_\lambda\left(4+\frac{\Delta_\lambda(t)}{4}\right)
        +
        \sigma_\lambda\left(4-\frac{\Delta_\lambda(t)}{4}\right)
        -
        2\sigma_\lambda(4)
    \right).
\]
For the first term $\sigma_\lambda\left(4+\frac{\Delta_\lambda(t)}{4}\right)$, we have
\begin{align*}
&\left|
    \sigma_\lambda\left(4+\frac{\Delta_\lambda(t)}{4}\right)
    -
    \sigma_1\left(4+\frac{\Delta_1(t)}{4}\right)
\right| \\
&\leq
\left|
    \sigma_\lambda\left(4+\frac{\Delta_\lambda(t)}{4}\right)
    -
    \sigma_\lambda\left(4+\frac{\Delta_1(t)}{4}\right)
\right| 
+
\left|
    \sigma_\lambda\left(4+\frac{\Delta_1(t)}{4}\right)
    -
    \sigma_1\left(4+\frac{\Delta_1(t)}{4}\right)
\right|.
\end{align*}
Since $\sigma_\lambda$ is $1$-Lipschitz, the first term is at most
\[
    \frac{1}{4}
    \left|
        \Delta_\lambda(t)-\Delta_1(t)
    \right|
    \leq
    |\lambda-1|(M+1)^{p-1}.
\]
For the second term, using \eqref{eq:sigma-lambda-finite}
together with the fact that $|\Delta_1(t)|\leq 2$, we have
\[
    \left|
        \sigma_\lambda\left(4+\frac{\Delta_1(t)}{4}\right)
        -
        \sigma_1\left(4+\frac{\Delta_1(t)}{4}\right)
    \right|
    \leq
    |\lambda-1|
    \left(\frac{9}{2}\right)^{p-1}.
\]
Therefore,
\[
\left|
    \sigma_\lambda\left(4+\frac{\Delta_\lambda(t)}{4}\right)
    -
    \sigma_1\left(4+\frac{\Delta_1(t)}{4}\right)
\right|
\leq
|\lambda-1|
\left(
    (M+1)^{p-1}
    +
    \left(\frac{9}{2}\right)^{p-1}
\right).
\]
Similarly we also get
\[
\left|
    \sigma_\lambda\left(4-\frac{\Delta_\lambda(t)}{4}\right)
    -
    \sigma_1\left(4-\frac{\Delta_1(t)}{4}\right)
\right|
\leq
|\lambda-1|
\left(
    (M+1)^{p-1}
    +
    \left(\frac{9}{2}\right)^{p-1}
\right).
\]
Finally,
\[
    2|\sigma_\lambda(4)-\sigma_1(4)|
    \leq
    2|\lambda-1|4^{p-1}.
\]
Combining the three estimates yields
\begin{align}\label{eq:k-lambda-local-finite}
    |k_\lambda(t)-k_1(t)|
    \leq
    L_{p,M}|\lambda-1|,
\end{align}
where
\[
    L_{p,M}
    =
    2\left((M+1)^{p-1}
    +
    \left(\frac{9}{2}\right)^{p-1}
    +
    4^{p-1}\right).
\]
Therefore setting
\[
        \abs{\lambda-1} \leq \kappa_k =
        \min\left\{\frac12,\frac{\eps_k}{L_{p,M}}\right\}
\]
ensures that the right-hand side of \eqref{eq:k-lambda-local-finite} is at most
$\eps_k$. The choice of $\kappa_k$ also ensures that $\lambda$ lies in $[1/2,2]$. We have now proved that in both cases where $|t| \geq M$ and $|t| < M,$
\eqref{eq:k-lambda-stability} holds and the proof is complete.
\end{proof}

% The construction has been arranged so that the weight function disappears exactly at the Euclidean exponent. 
% \begin{remark}
%     If $p=2$, then $k_\lambda\equiv0$ for every $\lambda>0$.
% \end{remark}

\subsubsection{Weighted approximation functions}
We introduce the notion of compactly supported functions.

\begin{definition}
    A function $f:\R\to\R$ is called \emph{compactly supported} if there exists a constant $S>0$ such that $f(t)=0$
for all $|t|\ge S$.
\end{definition}

We next show that any compactly supported piecewise-linear function can be approximated through weighted averaging of grid samples, with weights determined by $k_\lambda$.

\begin{lemma}\label{lem:compiler}
Let $f:\R\to\R$ be an $L_f$-Lipschitz, compactly supported, rational piecewise-linear function. For every $\eta>0$, there exist a positive integer $m \in \Z_+$, coefficients
\[
        \alpha_j,\ \beta_j\in\Q,
        \quad a_j\in\Q_{>0},
        \quad j=1,\ldots,m,
\]
and constants $\kappa, B > 0$, such that the weighted approximation function
\begin{equation}\label{eq:N-lambda}
        N_{f,\lambda}(t)
        :=\sum_{j=1}^m\alpha_j k_\lambda(a_jt+\beta_j)
\end{equation}
satisfies (i) $\norm{N_{f,\lambda}-f}_\infty\le\eta$ whenever $\abs{\lambda-1}\le\kappa$, and (ii) $\abs{N_{f,\lambda}(t)}\le B$ for all $t\in\R$ and $\lambda>0$.
% \begin{align}
%  \norm{N_{f,\lambda}-f}_\infty&\le\eta
%         &&\text{whenever }\abs{\lambda-1}\le\kappa,\label{eq:compiler-approx}\\
%  \abs{N_{f,\lambda}(t)}&\le B
%         &&\text{for every }t\in\R,\ \lambda>0.\label{eq:compiler-3}
% \end{align}
% All quantities depend only on $p,f,\eta$.
\end{lemma}
\begin{proof}
If $f$ is identically zero, take $N_{f, \lambda} \equiv 0$ and we are done.
Otherwise, since $f$ is compactly supported, we can choose an integer $S\ge1$ such that $f(t)=0$ whenever
$|t|\ge S$, and put $M_f=\norm{f}_\infty$.

% We first construct the approximation at $\lambda=1$, using the
% the properties established in \Cref{lem:kernel}.
% We then use its $2$-Lipschitz property to control the error between grid
% points, and finally account for perturbations of $\lambda$.
% \jy{Revisit this paragraph, consider defining a general definition of a nice class of functions}

We first approximate $f$ through discretization using the weight function $k_1$. Specifically, we sample $f$ on a finite grid and assign each sample a weight determined by $k_1$, evaluated at the scaled distance from the input $t$.  The nearby
sample points have values close to $f(t)$, whereas the contribution of distant samples is controlled by the polynomial tail bound established in \Cref{lem:kernel}.

Let $c_0,r_0,c_k,\beta_p$ be the constants from
\Cref{lem:kernel}~(ii) and
take integers $A\ge4$ and $U\ge1/r_0$. The exact values of $A$ and $U$ will be specified after
we derive the error bounds. The parameter $A$ controls the local concentration of the weight functions $k_1$, while $A$ and $U$ together determine the spacing of the finite grid. In particular, we set the spacing to be $\delta=1/{(A^2U)}$ and define the finite grid
\[
 \mathcal J= \left\{j\in\Z:|j\delta|\le S+
\frac{1}{A}\right\}.
\]
The extra margin $1/A$ ensures that enough neighboring sample points are included, even near the endpoints. 

For each grid point $t_i=i\delta$, we consider a window of nearby samples $j\delta$ satisfying $|t_i-j\delta|\le 1/A$. This window contains exactly $2AU+1$ grid points.
We then define the normalizing sum
\begin{equation*}
 Z =\sum_{r=-AU}^{AU}k_1(A^2r\delta) =\sum_{r=-AU}^{AU}k_1\left(\frac{r}{U}\right).
\end{equation*}
The scaling factor $A^2$ ensures the weight function $k_1$ has sufficient local concentration such that the contribution of a sample point outside the window is small. Recall from \Cref{lem:kernel}~(ii) that $k_1$ is bounded below by $c_0>0$ on the interval $[2-r_0,2+r_0]$. Since $U\ge1/r_0$, the positive interval $[2-r_0,2+r_0]$ contains at
least $r_0U$ points of the set $\left\{r/U: r \in \mathbb{Z}\right\}$ and each of them contributes at least $c_0$ to $Z$. Thus we conclude that
$Z\ge c_0 r_0 U>0$.

For any $i \in \Z,$ we write the corresponding function value $f_i=f(i\delta).$ For any $t \in \R$, consider the normalized weighted approximation function
\begin{equation*}
 \widetilde N_f(t)
 =\frac1Z\sum_{j\in\mathcal J}f_jk_1\bigl(A^2(t-j\delta)\bigr).
\end{equation*}
The normalization factor $Z$ ensures that, at each grid point $t_i=i\delta\in[-S,S]$, the weights of samples in the window satisfying $|t_i-j\delta|\le 1/A$ sum to one. Notably,
$Z$ does not depend on $t$, so the coefficients of the weighted approximation function remain fixed.

For any $i\in \Z$, we call a sample point $j$ nearby when $|i\delta-j\delta|\le1/A$, or equivalently $|i-j|\le AU$.
Since $A^2(i\delta-j\delta)=(i-j)/U$, setting $r=i-j$ gives
\[
 f_i = \frac{1}{Z}\sum_{j \in \mathcal{J}:|i-j|\le AU} f_ik_1\left(\frac{i - j}{U}\right).
\]
Consequently, adding and subtracting $f_i$ yields
\begin{align*}
 |\widetilde N_f(i\delta)-f_i|
 &\le\frac1Z\sum_{j \in \mathcal{J} : |i-j|\le AU}
 |f_j-f_i|k_1\left(\frac{i - j}{U}\right)+\frac1Z
 \sum_{\substack{j\in\mathcal J :|i-j|>AU}}
 |f_j|k_1\left(\frac{i - j}{U}\right).
\end{align*}
For nearby samples, Lipschitz continuity gives
$|f_j-f_i|\le L_f/A$, thus the total contribution of the first term is at most $L_f/A$.

For distant samples, the tail bound from \Cref{lem:kernel} applies because
$|i-j|/U>A\ge4$. Setting $r=i-j$, we obtain
\begin{align}
 \frac1Z\sum_{\substack{j\in\mathcal{J} :|i-j|>AU}}
 |f_j|k_1\left(\frac{i - j}{U}\right)
 &\le\frac{M_fc_kU^{\beta_p}}{Z}
       \sum_{j \in \mathcal{J} : |i - j| > AU} |i - j|^{-\beta_p}\notag\\
       & \leq \frac{2M_fc_kU^{\beta_p}}{c_0r_0U} \frac{(AU)^{1-\beta_p}}{\beta_p - 1} \label{eq:weight_M_cancle_out}\\
 &=\frac{2M_fc_k}{c_0r_0(\beta_p-1)}A^{1-\beta_p},
 \label{eq:weighted-grid-tail}
\end{align}
where \eqref{eq:weight_M_cancle_out} follows from \Cref{lem:polynomial-tail} and the fact that $Z \geq c_0r_0U.$
% Now we move to the case where $|i\delta|>S$ and $f_i=0$. Each nearby sample in $\mathcal J$ satisfies
% $|f_j|\le L_f/A$, and the total contribution is again at most $L_f/A$.
% Moreover, \eqref{eq:weighted-grid-tail} still bounds the distant contribution.
Thus we conclude that for every $i\in\Z$,
\begin{equation}\label{eq:weighted-grid-lattice}
 |\widetilde N_f(i\delta)-f_i|
 \le\frac{L_f}{A}+\frac{2M_fc_k}{c_0r_0(\beta_p-1)}A^{1-\beta_p}.
\end{equation}
% For the weight function, this part of the argument uses only
% nonnegativity, the lower bound on $Z$, and a polynomial tail exponent
% greater than one. We will reuse it for the normalized weight functions in
% \Cref{unif:sec:compiler}.

To extend the estimate to any real input $t \in \R$, we control how much the
approximation changes between an input and its nearest grid point.
Only inputs in $[-S,S]$ can have nonzero function value, so
\begin{equation}\label{eq:weighted-grid-coeffsum}
 \frac1Z\sum_{j\in\mathcal J}|f_j|
 \le\frac{M_f(2SA^2U+1)}{c_0r_0U}
 \le\frac{M_f(2SA^2+1)}{c_0r_0}.
\end{equation}
For any $t\in\R$, we can find the grid point $i\in\Z$ with $|t-i\delta|\le \delta/2$.
Thus we have
$A^2|t-i\delta|\le1/(2U)$. Since $k_1$ is $2$-Lipschitz, it holds that
\begin{align}\label{eq:sample_value_diff}
&\left|k_1\bigl(A^2(t-j\delta)\bigr)
-k_1\bigl(A^2(i\delta-j\delta)\bigr)\right|
\le 2A^2|t-i\delta|
\le\frac1U.
\end{align}
Using \eqref{eq:weighted-grid-coeffsum}  and \eqref{eq:sample_value_diff}, we have
\[
\begin{aligned}
|\widetilde N_f(t)-\widetilde N_f(i\delta)| \le\frac1Z\sum_{j\in\mathcal J}|f_j|
\left|k_1\bigl(A^2(t-j\delta)\bigr)
-k_1\bigl(A^2(i\delta-j\delta)\bigr)\right| \leq \frac{M_f(2SA^2+1)}{c_0r_0U}.
\end{aligned}
\]
Moreover, from the Lipschitz continuity of $f$, we obtain
\begin{equation*}
 \begin{aligned}
 |f(t)-f_i| \le \frac{L_f}{2A^2U}.
 \end{aligned}
\end{equation*}
Therefore, combining with \eqref{eq:weighted-grid-lattice}, it holds that
\begin{align}
 \abs{\widetilde N_f(t)-f(t)} & \leq |\widetilde N_f(t)-\widetilde N_f(i\delta)|
+|\widetilde N_f(i\delta)-f_i|
+|f_i-f(t)| \notag\\
 &\le \frac{M_f(2SA^2+1)}{c_0r_0U}+\frac{L_f}{A}+\frac{2M_fc_k}{c_0r_0(\beta_p-1)}A^{1-\beta_p}+\frac{L_f}{2A^2U} \label{eq:weighted-grid-estimate}.
\end{align}
We set the integer $A\ge4$ large enough that
\begin{equation*}
 \frac{L_f}{A}\le\frac{\eta}{12}\quad \text{and} \quad
 \frac{2M_fc_k}{c_0r_0(\beta_p-1)}A^{1-\beta_p}
 \le\frac{\eta}{12}.
\end{equation*}
Moreover, we choose an integer $U\ge1/r_0$ large enough that
\begin{equation*}
 \frac{M_f(2SA^2+1)}{c_0r_0U}+\frac{L_f}{2A^2U}
 \le\frac{\eta}{6}.
\end{equation*}
The resulting approximation satisfies
\begin{equation}\label{eq:finite-grid-uniform-error}
 \norm{\widetilde N_f-f}_\infty\le\frac{\eta}{3}.
\end{equation}

While the sample values and all other coefficients in $\widetilde N_f$ are rational, the normalization $Z^{-1}$ may
be irrational as it depends on the value of $k_1$. We replace it by a sufficiently accurate rational number,
using the global bound on $k_1$ to control the resulting error. Note that if $Q = \sum_{j\in\mathcal J}|f_j|=0$, then \eqref{eq:finite-grid-uniform-error} implies
$\norm{f}_\infty\le\eta/3$, so $N_{f,\lambda} \equiv 0$
satisfies both conclusions. Otherwise  we can choose a positive rational
number $q$ such that
\begin{equation*}
 |q-Z^{-1}|\le\frac{\eta}{102Q}.
\end{equation*}
For each $j\in\mathcal J$, set
\[
    \alpha_j=qf_{j},\quad
    a_j=A^2,\quad
    \beta_j=-\frac{j}{U}.
\]
These numbers are rational, since $\delta$ is rational and $f$ is
piecewise linear. We also have $a_jt+\beta_j=A^2(t-j\delta)$. Since $\mathcal J$ is finite, we may relabel
its elements by $[m]$, where $m=|\mathcal J|$. By
\Cref{lem:kernel}~(i), replacing $Z^{-1}$ by $q$ yields
\[
\begin{aligned}
|N_{f,1}(t)-\widetilde N_f(t)|
&=\left|
\left(q-\frac1Z\right)
\sum_{j\in\mathcal J}
f_jk_1\bigl(A^2(t-j\delta)\bigr)
\right|\\
&\le 17 \left|q - \frac{1}{Z}\right| \sum_{j \in \mathcal{J}} \left|f_j\right| \leq 17Q\left|q-\frac{1}{Z}\right|\le\eta/6
\end{aligned}
\]
Together with
\eqref{eq:finite-grid-uniform-error}, this gives
\[
 \norm{N_{f,1}-f}_\infty\le\frac{\eta}{2}.
\]

The coefficients are now fixed. Let $S_\alpha=\sum_j|\alpha_j|>0$ and
apply \Cref{lem:kernel}~(iii) with
$\eps_k=\eta/(2S_\alpha)$ and choose 
$\kappa = \min\{\kappa_k, 1/2\}$. Whenever $|\lambda-1|\le\kappa$, we have
\[
 \norm{N_{f,\lambda}-N_{f,1}}_\infty
 \le S_\alpha\norm{k_\lambda-k_1}_\infty\le\frac{\eta}{2}.
\]
Combining the approximation error and the multiplier perturbation gives
\[
 \norm{N_{f,\lambda}-f}_\infty
 \le\norm{N_{f,\lambda}-N_{f,1}}_\infty
     +\norm{N_{f,1}-f}_\infty
 \le\frac{\eta}{2}+\frac{\eta}{2}=\eta.
\]
This proves (i). Finally, the global bound from \Cref{lem:kernel}~(i) holds for every
positive multiplier, hence $|N_{f,\lambda}(t)|\le17S_\alpha$ for every $t \in \R$ and $\lambda > 0.$ Setting $B=17S_\alpha$ completes the proof for (ii).
\end{proof}

Since $p$, $\eta$, and $f$ are fixed independently of the
input circuit size $n$, the parameters in \Cref{lem:compiler} are also constants. Consequently, given the exponent $p$, we may fix an integer $m$ and parameters $(\alpha_j,a_j,\beta_j)_{j=1}^m$ satisfying the lemma and hardwire them into the reduction.
% The proof of \Cref{lem:compiler} is existential only in constants that are independent of the input circuit.  Once $p$, $f$, and $\eta$ are fixed, one may choose a single finite list of rational parameters $m,(\alpha_j,a_j,\beta_j)_{j=1}^m$ satisfying the lemma and hardwire that list into the reduction.  The same applies to the derived constants $\kappa,B,S_*,C_*,\rho_0,K,U$, and $c_\eps$.  No real number depending on the input instance is given as advice.

We next show how to simulate the weight function $k_\lambda$
through the solutions of VIs with affine operators.
Recall that the VI optimality condition
$F(\vec z^*)=\lambda J_p(\vec z^*)$ allows us to realize
$\sigma_\lambda$ through an affine operator.
Since $k_\lambda$ can be expressed through successive evaluations of $\sigma_\lambda$,
the weight assigned to each sample point can be implemented using a constant number of auxiliary variables while keeping the VI operator affine.

In particular, consider the weighted approximation function
\[
    N_{f,\lambda}(t)
    =\sum_{j=1}^m
    \alpha_j k_\lambda(a_jt+\beta_j)
\]
from \Cref{lem:compiler}.
For each $j\in[m]$, we introduce six auxiliary variables
$r_j^-,r_j^0,r_j^+,w_j^-,w_j^0,w_j^+$.
Given an input $t \in \R$, let $s_j=a_jt+\beta_j$ be the input and write
$d_j=r_j^+-2r_j^0+r_j^-$.
The auxiliary variables are intended to satisfy
\begin{align}
    r_j^- &= \sigma_\lambda(s_j-1),
    &
    r_j^0 &= \sigma_\lambda(s_j),
    &
    r_j^+ &= \sigma_\lambda(s_j+1),
    \label{eq:network-first-layer}\\
    w_j^+ &= \sigma_\lambda\left(4+\frac{d_j}{4}\right),
    &
    w_j^0 &= \sigma_\lambda(4),
    &
    w_j^- &= \sigma_\lambda\left(4-\frac{d_j}{4}\right).
    \label{eq:network-second-layer}
\end{align}
Define the
realization by
\begin{equation}\label{eq:network-readout}
    \widehat N_{f,\lambda}(t)
    =\varsigma_p
    \sum_{j=1}^m
    \alpha_j(w_j^++w_j^--2w_j^0).
\end{equation}
When \eqref{eq:network-first-layer} and
\eqref{eq:network-second-layer} hold exactly,
the definition of $k_\lambda$ yields
$\widehat N_{f,\lambda}(t)=N_{f,\lambda}(t)$. We proceed to show that this realization remains accurate even when the conditions \eqref{eq:network-first-layer} and \eqref{eq:network-second-layer} are satisfied approximately.

\begin{lemma}
\label{lem:network-stability}
Suppose that for every $j\in[m]$, all six equations in
\eqref{eq:network-first-layer} and
\eqref{eq:network-second-layer} hold with
additive error at most $\epsilon\ge0$.
Then the realization $\widehat N_{f,\lambda}(t)$ in \eqref{eq:network-readout} satisfies
\begin{equation}\label{eq:network-error}
    |\widehat N_{f,\lambda}(t)-N_{f,\lambda}(t)|
    \le 6\epsilon\sum_{j=1}^m|\alpha_j|.
\end{equation}
Moreover, for any $\lambda > 0$ and $|t|\le S$,
every auxiliary variable is bounded in absolute value by
\[
    5+2\epsilon+\max_{j\in[m]}\left(a_jS+|\beta_j|\right),
\]
\end{lemma}

\begin{proof}
Fix $j\in[m]$. The additive error $\epsilon$ in
\eqref{eq:network-first-layer} gives
\[
    |r_j^- -\sigma_\lambda(s_j-1)|\le\epsilon,\quad
    |r_j^0-\sigma_\lambda(s_j)|\le\epsilon,\quad
    |r_j^+-\sigma_\lambda(s_j+1)|\le\epsilon.
\]
By the definition of $\Delta_\lambda$, we have
\begin{equation}\label{eq:network-d-error}
\begin{aligned}
    |d_j-\Delta_\lambda(s_j)|
    &\le
    |r_j^+-\sigma_\lambda(s_j+1)|
    +2|r_j^0-\sigma_\lambda(s_j)|\\
    &\quad+|r_j^--\sigma_\lambda(s_j-1)|
    \le4\epsilon.
\end{aligned}
\end{equation}

Using the Lipschitz continuity of $\sigma_\lambda$
and \eqref{eq:network-d-error}, we obtain
\begin{align*}
    \left|w_j^--\sigma_\lambda\left(4-\frac{\Delta_\lambda(s_j)}{4}\right)\right|\le
    \left|w_j^--
    \sigma_\lambda\left(4-\frac{d_j}{4}\right)\right|
    +
    \frac14|d_j-\Delta_\lambda(s_j)| \le2\epsilon
\end{align*}
and similarly $|w_j^+-\sigma_\lambda(4+\Delta_\lambda(s_j)/4)| \leq 2\epsilon.$
Together we have
\begin{align*}
    &\left|
    \varsigma_p(w_j^++w_j^--2w_j^0)
    -k_\lambda(s_j)
    \right|\le
    \left|w_j^+-\sigma_\lambda\left(4+\frac{\Delta_\lambda(s_j)}{4}\right)\right|
    +\left|w_j^--\sigma_\lambda\left(4-\frac{\Delta_\lambda(s_j)}{4}\right)\right|+2|w_j^0-\sigma_\lambda(4)|\le6\epsilon.
\end{align*}
Summing over $j\in[m]$ with coefficients $\alpha_j$
gives
\[
    |\widehat N_{f,\lambda}(t)-N_{f,\lambda}(t)|
    \le
    \sum_{j=1}^m|\alpha_j|
    \left|
    \varsigma_p(w_j^++w_j^--2w_j^0)
    -k_\lambda(s_j)
    \right|
    \le6\epsilon\sum_{j=1}^m|\alpha_j|.
\]

It remains to show that the auxiliary variables are bounded.
Since $|\sigma_\lambda(s)|\le|s|$ for every
$\lambda>0$, \eqref{eq:network-first-layer} implies
\[
    |r_j^-|,\ |r_j^0|,\ |r_j^+|
    \le a_jS+|\beta_j|+1+\epsilon.
\]
Thus
\[
    |d_j|
    \le |r_j^+|+2|r_j^0|+|r_j^-|
    \le4 \left(a_jS+|\beta_j|+1+\epsilon\right).
\]
Conditions in \eqref{eq:network-second-layer} then give
\[
    |w_j^+|,\ |w_j^-|
    \le4+\frac{|d_j|}{4}+\epsilon
    \le a_jS+|\beta_j|+5+2\epsilon,
\]
and
\[
    |w_j^0|\le|\sigma_\lambda(4)|+\epsilon\le4+\epsilon.
\]
Thus every auxiliary variable in \eqref{eq:network-first-layer} and \eqref{eq:network-second-layer} is bounded by
\[
    5+2\epsilon+\max_{j\in[m]}
    \left(a_jS+|\beta_j|\right).
\]
We note that the above bound depends only on the fixed coefficients $a_j, \beta_j$,
$S$, and $\epsilon$, and is independent of $\lambda$. The proof is complete.
\end{proof}

\subsection{Affine VI construction over an $\ell_p$ ball}\label{sec:smooth-affine}
We now use the weighted approximation functions developed in the previous subsection to construct an affine VI that encodes a \PureCircuit{} instance. Throughout this subsection, we fix a constant exponent $p\in(1,\infty)\backslash\{2\}$. All fixed constants in this section may depend on $p$ but are independent of the input circuit size $n$.
\subsubsection{Compactly supported \PureCircuit{} threshold functions}
Since \Cref{lem:compiler} holds with compactly supported functions, we introduce compactly supported
versions of $\phi_{\Nor}$ and $\phi_{\Purify}$. Specifically,
we define the following piecewise-linear functions:
\begin{equation}\label{eq:gbar}
 \hat\phi_{\Nor}(t)=
 \begin{cases}
 0,&t\le-2,\\
 t+2,&-2\le t\le-1,\\
 \norfun(t),&-1\le t\le3,\\
 0,&t\ge3,
 \end{cases}
\quad \text{and} \quad
 \hat\phi_{\Purify}(t)=
 \begin{cases}
 0,&t\le-1,\\
 \purfun(t),&-1\le t\le2,\\
 3-t,&2\le t\le3,\\
 0,&t\ge3.
 \end{cases}
\end{equation}

\begin{figure}[t]
    \centering
    \begin{tikzpicture}[
    x=1.15cm,
    y=2.8cm,
    axis/.style={->, thick},
    graph/.style={very thick, niceMagenta},
    guide/.style={densely dashed, gray},
    ticklabel/.style={font=\scriptsize}
]

\newcommand{\compactthresholdaxes}[1]{
    \draw[axis] (-2.35,0) -- (3.45,0) node[right] {$t$};
    \draw[axis] (0,-0.08) -- (0,1.18) node[above] {$#1(t)$};
    \draw (-0.04,0) -- (0.04,0)
        node[below left=3pt,ticklabel] {$0$};
    \draw (-0.04,1) -- (0.04,1);
    \node[above left=3pt,ticklabel] at (0,1) {$1$};
}

\begin{scope}
    \compactthresholdaxes{\hat\phi_{\Nor}}
    \foreach \x in {-2,-1,0.25,0.5} {
        \draw[guide] (\x,0) -- (\x,1);
    }
    \draw[graph] (-2.25,0) -- (-2,0) -- (-1,1)
        -- (0.25,1) -- (0.5,0) -- (3.3,0);
    \foreach \x/\label in {-2/-2,-1/-1,3/3} {
        \draw (\x,-0.018) -- (\x,0.018)
            node[below=3pt,ticklabel] {$\label$};
    }
    \draw (0.25,-0.018) -- (0.25,0.018);
    \node[below=3pt,ticklabel] at (0.25,0) {$\frac14$};
    \draw (0.5,-0.018) -- (0.5,0.018);
    \node[below=3pt,ticklabel] at (0.5,0) {$\frac12$};
\end{scope}

\begin{scope}[xshift=7.5cm]
    \compactthresholdaxes{\hat\phi_{\Purify}}
    \foreach \x in {5/12,7/12,2,3} {
        \draw[guide] ({\x},0) -- ({\x},1);
    }
    \draw[graph] (-2.25,0) -- ({5/12},0) -- ({7/12},1)
        -- (2,1) -- (3,0) -- (3.3,0);
    \foreach \x/\label in {-1/-1,2/2,3/3} {
        \draw (\x,-0.018) -- (\x,0.018)
            node[below=3pt,ticklabel] {$\label$};
    }
    \draw ({5/12},-0.018) -- ({5/12},0.018);
    \node[below=3pt,ticklabel] at ({5/12},0) {$\frac{5}{12}$};
    \draw ({7/12},-0.018) -- ({7/12},0.018);
    % \draw[thin] ({7/12},-0.018) -- ({7/12},-0.16);
    % \node[below=3pt,ticklabel] at ({7/12},-0.16) {$\frac{7}{12}$};
\end{scope}

\end{tikzpicture}
    \caption{The compactly supported threshold functions
    $\hat\phi_{\Nor}$ and $\hat\phi_{\Purify}$
    defined in \eqref{eq:gbar}.}
    \label{fig:compact-threshold-functions}
\end{figure}
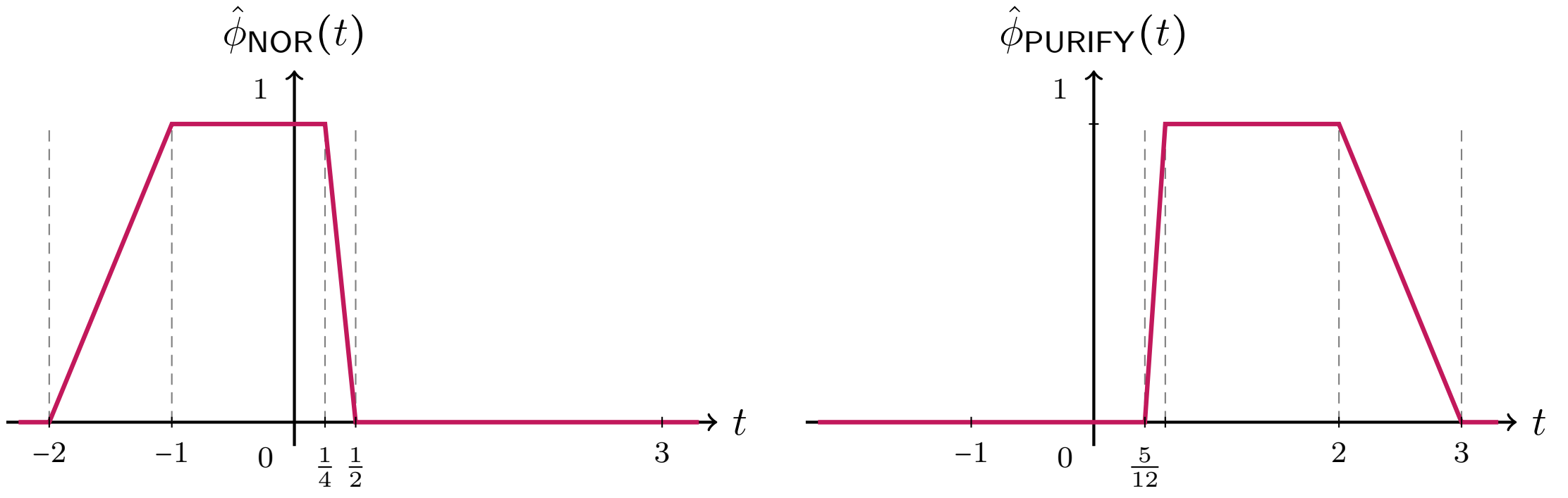

Recall that \Cref{lem:rounding} holds with error
$\epsilon_{\rm pc}=1/32$, we set
\[
        \eta=\frac{\eps_{\rm pc}}4=\frac1{128}.
\]
We apply \Cref{lem:compiler} to $\hat\phi_{\Nor}$ and
$\hat\phi_{\Purify}$ with accuracy $\eta$, and denote the resulting
weighted approximation functions by $N^{\Nor}_\lambda$ and
$N^{\Purify}_\lambda$.
By the same lemma, there exists a constant
$\kappa > 0$ such that whenever $\abs{\lambda - 1} \leq \kappa$, both weighted approximation
functions satisfy
\[
    \norm{N^\Nor_\lambda-\hat\phi_\Nor}_\infty
    \le\eta
    \quad\text{and} \quad \norm{N^\Purify_\lambda-\hat\phi_\Purify}_\infty
    \le\eta.
\]
% We now place copies of these two fixed realizations at the circuit
% nodes. The remaining task will be to construct an affine VI that
% approximately enforces their defining equations and keeps the common
% multiplier in this range.

\subsubsection{The auxiliary variables and affine inputs}
Let a \PureCircuit{} instance have $n\ge2$ nodes. For each node $v$, let $N^v_\lambda$ denote
$N^{\Nor}_\lambda$ if $v$ is the output of a $\Nor$
gate, and $N^{\Purify}_\lambda$ if $v$ is an output
of a $\Purify$ gate. Recall that $x_v$ denotes the scalar value associated with node $v$, while $s_v(\vx)$ is its scalar input, as defined in \eqref{eq:gate-input}. The realization of $N^v_\lambda$ therefore takes $s_v(\vx)$ as the input.

For each node $v\in[n]$ and each sample point
$j\in[m]$, introduce six auxiliary variables
\[
    r_{v,j}^-,\quad r_{v,j}^0,\quad r_{v,j}^+,
    \quad w_{v,j}^-,\quad w_{v,j}^0,\quad w_{v,j}^+,
\]
corresponding to the variables in
\eqref{eq:network-first-layer} and
\eqref{eq:network-second-layer}.
The total number of auxiliary variables is $d_{\mathrm{aux}}=6mn$.
Since $m$ is fixed independently of the circuit size
$n$, we have $d_{\mathrm{aux}}=O(n)$. We collect all the auxiliary variables into a vector
$\vec \zeta\in\R^{d_{\mathrm{aux}}}$.

Given the values of the auxiliary variables, one can compute
the scalar value $x_v$ as in
\eqref{eq:network-readout}.
For a $\Nor$ output, this is
\begin{align}\label{eq:node_v_realization}
    x_v
    :=\varsigma_p\sum_{j=1}^m\alpha_{\Nor,j}
    (w_{v,j}^++w_{v,j}^--2w_{v,j}^0).
\end{align}
For a $\Purify$ output, use the same expression
with $\alpha_{\Purify,j}$ in place of
$\alpha_{\Nor,j}$.

We next specify the inverse mapping equations that
our affine VI will be designed to enforce.
If $v$ is an output of a $\Nor$ gate, the three auxiliary variables from \eqref{eq:network-first-layer} are intended to satisfy
\begin{align*}
    r_{v,j}^- &=
\sigma_\lambda(a_{\Nor,j}s_v(\vx)+\beta_{\Nor,j}-1),\\
    r_{v,j}^0 &=
\sigma_\lambda(a_{\Nor,j}s_v(\vx)+\beta_{\Nor,j}),\\
    r_{v,j}^+ &=
\sigma_\lambda(a_{\Nor,j}s_v(\vx)+\beta_{\Nor,j}+1).
\end{align*}
Similarly, for the variables from \eqref{eq:network-second-layer},
let $d_{v,j}=r_{v,j}^+-2r_{v,j}^0+r_{v,j}^-$.
The variables $w_{v,j}^-$, $w_{v,j}^0$, and
$w_{v,j}^+$ are intended to satisfy
\[
\begin{aligned}
    w_{v,j}^-
    &=\sigma_\lambda\left(4-\frac{d_{v,j}}4\right),\quad
    w_{v,j}^0
    =\sigma_\lambda(4),\quad
    w_{v,j}^+
    =\sigma_\lambda\left(4+\frac{d_{v,j}}4\right).
\end{aligned}
\]

Since each node value $x_v$ is completely determined by the auxiliary variables in \eqref{eq:node_v_realization}, the preceding conditions can be expressed as a system of fixed-point equations in $\vec\zeta$. Specifically, for each auxiliary variable $\zeta_\ell,$ $\ell \in [d_{\mathrm{aux}}],$ we denote
its input to the inverse mapping $\sigma_\lambda$
by $h_\ell(\vec \zeta)$. 
The intended inverse mapping conditions can then be written as
\begin{equation}\label{eq:aux-inverse-mapping}
    \zeta_\ell=\sigma_\lambda(h_\ell(\vec\zeta)),
    \quad \forall \ell\in[d_{\mathrm{aux}}].
\end{equation}
Moreover, each $x_v$
is a linear combination of the auxiliary variables.
Since $s_v(\vx)$ is also affine in $\vx$, we conclude that
every inverse mapping input $h_\ell(\vec\zeta)$ is affine in $\vec\zeta$. In the reduction, we will construct an affine VI whose solutions satisfy \eqref{eq:aux-inverse-mapping} up to a small error.

\subsubsection{The affine VI construction}
We proceed to define the affine operator using the inputs
$h_\ell(\vec\zeta)$ specified above. Notably, the auxiliary variable equation in \eqref{eq:aux-inverse-mapping} alone does not determine
the common multiplier $\lambda$. Recall that \Cref{lem:compiler} holds when $\lambda$ is sufficiently close to 1, i.e., $|\lambda -1| \leq \kappa$.
To ensure this condition, we introduce an additional
anchor coordinate $z_0$ along with the
$d_{\mathrm{aux}}$ auxiliary coordinates. Specifically, consider the vector
\[
    \vec z=(z_0,\vec\zeta)\in\R^{d_{\mathrm{aux}}+1}.
\]
Note that the scalar value $x_v$ of each node is
computed from the coordinates of $\vec z$ as in
\eqref{eq:node_v_realization}, rather than a separate coordinate. Let $R \in \Z_+$ be the radius, we construct an operator on $B_p^{d_{\mathrm{aux}}+1}(R)$ with
\begin{align}
    F_0(\vec z)&=c_R,\label{eq:F-anchor}\\
    F_\ell(\vec z)&=h_\ell(\vec\zeta)-\zeta_\ell,
        \quad \ell\in[d_{\mathrm{aux}}].
        \label{eq:F-neuron}
\end{align}
Here, $F_0$ and $F_\ell$ denote the operator components corresponding to $z_0$ and $\zeta_\ell$
respectively, and $c_R>0$ is a scalar to be specified later.
Since each $h_\ell$ is affine, the entire operator $F$ is affine.
We next explain how its VI conditions enforce the intended
inverse mapping conditions in \eqref{eq:aux-inverse-mapping}.

Since $F_0(\vec z)=c_R>0$,
\Cref{lem:exact-normal} implies that every exact VI solution
$\vec z^*=(z_0^*,\vec\zeta^*)$ lies on the boundary and satisfies
$F(\vec z^*)=\lambda J_p(\vec z^*)$ for a common multiplier $\lambda>0$.
Thus for each $\ell \in [d_{\mathrm{aux}}]$,
\begin{align*}
    h_\ell(\vec\zeta^*)-\zeta_\ell^*
    =\lambda|\zeta_\ell^*|^{p-2}\zeta_\ell^*.
\end{align*}
Combining the above equation with \eqref{eq:h-sigma} and the definition of $\sigma_\lambda,$ we then obtain
\[ \zeta_\ell^*=\sigma_\lambda(h_\ell(\vec\zeta^*)).
\]
Thus the VI conditions enforce the inverse mapping
equations in \eqref{eq:aux-inverse-mapping}.

For the anchor coordinate, the same argument yields $c_R=\lambda(z_0^*)^{p-1}$ with $z_0^*>0$, and hence
\[
    \lambda=\frac{c_R}{{z^*_0}^{p - 1}} = \frac{c_R}{R^{p-1}}
             \left(\frac{R}{z_0^*}\right)^{p-1}.
\]
The above equation shows that keeping $\lambda$ close to $1$
requires both $c_R$ to be close to $R^{p-1}$ and
the anchor coordinate $z_0^*$ to be close to $R$.
We will choose $c_R$ to be a rational approximation
of $R^{p-1}$ and
establish the latter condition by choosing $R$
sufficiently large so that the auxiliary coordinates
contribute only a small fraction of the norm budget $R^p$, the detailed proof is deferred to \Cref{lem:anchor-multiplier}.

\subsubsection{Approximate VI solutions}
We now extend the preceding argument to approximate VI
solutions. Let $\vec z^*\in B_p^{d_{\mathrm{aux}}+1}(R)$ be an $\eps_R$-approximate VI solution, that is, $\vec z^*$ satisfies
\begin{equation}\label{eq:gap-assumption-smooth}
        \Gap_{B_p^{d_{\mathrm{aux}}+1}(R)}(\vec z^*;F)\le\eps_R.
\end{equation}
Let $
        \vec s=S_R(F(\vec z^*))
$
be the support point. \Cref{lem:support_point} suggests that there is a unique $\lambda>0$ such that
\begin{equation}\label{eq:F-lambda-zeta}
        F(\vec z^*)=\lambda J_p(\vec s).
\end{equation}
Let $\vec\zeta^*\in\R^{d_{\mathrm{aux}}}$ denote
the vector of auxiliary coordinates of $\vec z^*$. We proceed with the following lemma.
\begin{lemma}\label{lem:neuron-residual}
For every auxiliary coordinate $\ell\in[d_{\mathrm{aux}}]$, we have
\begin{equation}\label{eq:inverse mapping-residual}
        \abs{\zeta^*_\ell-\sigma_\lambda(h_\ell(\vec\zeta^*))}
        \le2\norm{\vec z^*-\vec s}_p.
\end{equation}
\end{lemma}
\begin{proof}
The $\ell$-th coordinate of \eqref{eq:F-lambda-zeta} is
\[
        h_\ell(\vec\zeta^*)-\zeta_\ell^*=\lambda J_p(s_\ell).
\]
Hence from the definition of $h_\lambda$ in \eqref{eq:h-sigma},
\begin{align}\label{eq:inverse_map_s}
    h_\lambda(s_\ell)
        =s_\ell+\lambda J_p(s_\ell)
        =h_\ell(\vec\zeta^*)+(s_\ell-\zeta_\ell^*).
\end{align}
Using the fact that $\sigma_\lambda$ is $1$-Lipschitz continuous,
\begin{align*}
 \abs{\zeta^*_\ell-\sigma_\lambda(h_\ell(\vec\zeta^*))}&\le\abs{\zeta^*_\ell-s_\ell}
 +\abs{s_\ell
       -\sigma_\lambda(h_\ell(\vec\zeta^*))}\\
 &\le\abs{\zeta_\ell^*-s_\ell}
 +\abs{\sigma_\lambda(h_\ell(\vec\zeta^*)+s_\ell-\zeta^*_\ell)
       -\sigma_\lambda(h_\ell(\vec\zeta^*))}\\
 &\le2\abs{\zeta^*_\ell-s_\ell}
 \le2\norm{\vec z^*-\vec s}_p,
\end{align*}
where the second inequality follows from \eqref{eq:inverse_map_s} and the fact that $\sigma_\lambda(h_\lambda(s_\ell)) = s_\ell.$ The proof is complete.
\end{proof}

In the previous lemma, the VI approximation error $\eps_R$ is captured implicitly through the distance $\norm{\vec z^* - \vec s}_p.$ We next show that this distance also controls the error in the realization of each node's scalar value.
Specifically, let
\begin{align}
    H = \sum_{j\in[m]}\abs{\alpha_{\Nor,j}}+ \sum_{j \in [m]}\abs{\alpha_{\Purify,j}},
    \label{eq:Sstar}
\end{align}
we have the following lemma.
\begin{lemma}
\label{lem:realization-error}
For every node $v$, the corresponding scalar value
$x_v$ in \eqref{eq:node_v_realization} satisfies
\begin{equation}\label{eq:X-network-close}
    \abs{x_v-N^v_\lambda(s_v(\vx))}
    \le 12H\norm{\vec z^*-\vec s}_p.
\end{equation}
\end{lemma}

\begin{proof}
By \Cref{lem:neuron-residual}, equations in \eqref{eq:network-first-layer} and \eqref{eq:network-second-layer} have additive error at most
$2\norm{\vec z^*-\vec s}_p$.
Applying \Cref{lem:network-stability} gives the desired inequality.
\end{proof}

\subsubsection{The common multiplier}
We next show that the anchor coordinate can be used
to ensure $\lambda$ is close to 1.

\begin{lemma}\label{lem:anchor-multiplier}
Let $\kappa>0$ be the corresponding accuracy \Cref{lem:compiler} with $\eta=1/128$. There exist an integer radius $R=O_p(n)$ and a
positive coefficient $c_R$ such that the affine operator $F$ defined in
\eqref{eq:F-anchor} and \eqref{eq:F-neuron} satisfies the following:
If the approximate VI solution $\vec z^*\in B_p^{d_{\mathrm{aux}}+1}(R)$ satisfies
\[
    \norm{\vec z^*-\vec s}_p
    \le\min\left\{\frac12,\frac{\eta}{12H}\right\},
\]
the associated multiplier $\lambda={\norm{F(\vec z^*)}_q}/{R^{p-1}}$
satisfies $$|\lambda-1|\le\kappa.$$
\end{lemma}
\begin{proof}
Combining \Cref{lem:kernel}~(i) and
\eqref{eq:Sstar}, for any $t \in \R$ and $\lambda > 0,$ we have 
\begin{align} \label{eq:realization_upper_bound}
    \abs{N^{\Nor}_\lambda(t)}\le 17H \quad \text{and} \quad \abs{N^{\Purify}_\lambda(t)}\le 17H.
\end{align}  
Since $\norm{\vec z^*-\vec s}_p\le\eta/(12H)$ by hypothesis, \Cref{lem:realization-error}, together with
\eqref{eq:realization_upper_bound}, gives
\begin{equation*}
        \abs{x_v}\le 17H +\eta
        \quad\forall v \in [n].
\end{equation*}
Since each gate input $s_v(\vx)$ is the sum of at most two scalar values, or one
scalar value shifted by $1/4$, we obtain
\begin{equation*}
        \abs{s_v(\vx)}\le 2(17H+\eta)+\frac14 = S.
\end{equation*}
Combining
\Cref{lem:network-stability} with \Cref{lem:realization-error}, it yields
\begin{align}
\abs{\zeta_\ell^*} & \le5+4\norm{\vec z^*-\vec s}_p+
        \max_{\mathrm{gate}\in\{\Nor,\Purify\},j}
        \left(a_{\mathrm{gate},j}S+\abs{\beta_{\mathrm{gate},j}}\right) \notag\\
        & \leq 7 + \max_{\mathrm{gate}\in\{\Nor,\Purify\},j}
        \left(a_{\mathrm{gate},j}S+\abs{\beta_{\mathrm{gate},j}}\right). \label{eq:zeta_coordinate_upper}
\end{align}
% We need one bound for both the auxiliary coordinates of $\vec z$ and the
% corresponding coordinates of its support point $\vec s$. Since
% $\norm{\vec z^*-\vec s}_p\le1/2$, choose the fixed constant
% \[
%         U=8+\max_{\gamma\in\{\Nor,\Purify\},j}
%         \left(a_{\gamma,j}T_*+\abs{\beta_{\gamma,j}}\right).
% \]
% Using also $\abs{s_\ell-\zeta_\ell}\le\norm{\vec z-\vec s}_p$, we obtain
% \begin{equation}\label{eq:coordinate-bound}
%         \abs{\zeta_\ell}\le U-1,
%         \quad
%         \abs{s_\ell}\le U
%         \quad \ell\in[d_{\mathrm{aux}}].
% \end{equation}

Recall that support point $\vec s$ always lies on the boundary of the ball, we have
\begin{align*}
        s_0^p
        =R^p-\sum_{\ell=1}^{d_{\mathrm{aux}}}\abs{s_\ell}^p
        & \ge R^p-\sum_{\ell=1}^{d_{\mathrm{aux}}}\left(|\zeta_\ell^*| + |s_\ell - \zeta_\ell^*|\right)^p \\
        & \ge R^p - d_{\mathrm{aux}}\left(8+ \max_{\mathrm{gate}\in\{\Nor,\Purify\},j}
        \left(a_{\mathrm{gate},j}S+\abs{\beta_{\mathrm{gate},j}}\right)\right)^p,
\end{align*}
where the last step follows from \eqref{eq:zeta_coordinate_upper} and $\abs{s_\ell-\zeta^*_\ell}\le\norm{\vec z^*-\vec s}_p \leq 1.$ Moreover, for the anchor coordinate $z^*_0$, \eqref{eq:F-lambda-zeta} gives
$c_R=\lambda|s_0|^{p-2}s_0$. Since $c_R>0$ and $\lambda>0$, we have $s_0>0$ and $c_R=\lambda s_0^{p-1}.$ Thus
\begin{align}\label{eq:lambda-anchor-bound}
    \lambda = \frac{c_R}{s_0^{p - 1}} = \frac{c_R}{R^{p - 1}}\left(1-\frac{\sum_{\ell=1}^{d_{\mathrm{aux}}}
    \abs{s_\ell}^p}{R^p}\right)^{-(p-1)/p}.
\end{align}

We now choose $R$ sufficiently large to ensure that the
second term in \eqref{eq:lambda-anchor-bound} is close to 1. Write 
$$M_\ell = 8+ \max_{\mathrm{gate}\in\{\Nor,\Purify\},j} \left(a_{\mathrm{gate},j}S+\abs{\beta_{\mathrm{gate},j}}\right).$$
Since $d_{\mathrm{aux}}=6mn$ and $n\ge2$, we have
\[
    \frac{\sum_{\ell=1}^{d_{\mathrm{aux}}}\abs{s_\ell}^p}{R^p}
    \le\frac{d_{\mathrm{aux}}M_\ell^p}{R^p}
    =\frac{6mM_\ell^p}{n^{p-1}}\left(\frac nR\right)^p
    \le\frac{6mM_\ell^p}{2^{p-1}}\left(\frac nR\right)^p.
\]
Therefore, choosing
\begin{align}\label{eq:choice_R}
    R=n\left\lceil
    \max\left\{1,
    \left(\frac{24mM_\ell^p}{2^{p-1}\kappa}\right)^{1/p}
    \right\}
    \right\rceil
\end{align}
ensures that
\[
    \frac{\sum_{\ell=1}^{d_{\mathrm{aux}}}\abs{s_\ell}^p}{R^p}
    \le\frac{\kappa}{4}.
\]
Since $p$, $m$, $M_\ell$, and $\kappa$ are constants
independent of $n$, we have $R=O_p(n)$ and is computable in polynomial time.

For the first term in \eqref{eq:lambda-anchor-bound}, we first choose a rational constant $\tau$ satisfying
$0<\tau\le\kappa/4$.
\Cref{lem:anchor-coefficient} shows that one can compute
a positive rational $c_R$ satisfying ${c_R}/{R^{p-1}}
    \in[1-\tau,1+\tau]$
in polynomial time.

Putting everything together, we conclude from
\eqref{eq:lambda-anchor-bound} and $(p-1)/p<1$ that
\begin{equation}\label{eq:lambda-near-one}
    1-\kappa
    \le 1-\tau
    \le \lambda
    \le (1+\tau)
    \left(1-\frac{\kappa}{4}\right)^{-(p-1)/p}
    \le \frac{1+\kappa/4}{1-\kappa/4}
    \le 1+\kappa.
\end{equation}
The proof is complete.
\end{proof}

Fix $\tau$ and $R$ as in the proof of
\Cref{lem:anchor-multiplier}, and compute $c_R$
using \Cref{lem:anchor-coefficient}.
By \Cref{lem:compiler}, the coefficients of each
affine input $h_\ell(\vec\zeta)$ are rational constants. Moreover, $d_{\mathrm{aux}}=6mn=O_p(n)$,
$R=O_p(n)$, and $c_R$ can be computed in polynomial
time with polynomial bit length.
Thus, the affine operator $F$ defined in
\eqref{eq:F-anchor} and \eqref{eq:F-neuron}
can be constructed in polynomial time, with rational
coefficients of polynomial bit length.

% For now, let $\tau\in(0,1)$ be a rational constant independent of
% $n$, and let $R\ge2$ be an integer radius. We leave their choices
% to the proof of \Cref{lem:anchor-multiplier} below.
% With $c_R$ computed by \Cref{lem:anchor-coefficient},
% \eqref{eq:F-anchor}--\eqref{eq:F-neuron} define a rational affine
% operator. We first record its coefficient bounds and construction
% cost in terms of the radius.
% Fix $\tau$ and $R$ as in the proof of \Cref{lem:anchor-multiplier},
% and compute $c_R$ using \Cref{lem:anchor-coefficient}.
% These choices along with the choice of coefficients in $h_\ell$ complete the construction of the affine VI.
% \begin{proposition}\label{prop:affine-bits}
% The affine operator $F$ in \eqref{eq:F-anchor} and \eqref{eq:F-neuron} is rational and can be
% constructed in polynomial time.
% \end{proposition}
% \begin{proof}
% All coefficients of the weighted approximation functions in \Cref{lem:compiler} are rational constants depending only on $p$. Thus, each $h_\ell(\vec\zeta)-\zeta_\ell$ term contains a constant number of rational terms. By \Cref{lem:anchor-coefficient}, the anchor coefficient $c_R$ can be computed in polynomial time. Since $d_{\mathrm{aux}}=6mn=O_p(n)$ and by $R=O_p(n)$ \Cref{lem:anchor-multiplier}, the entire construction takes time polynomial in the circuit size $n$.
% \end{proof}

\subsection{Completing the proof of \Cref{thm:main}}\label{sec:noneuclidean_final}
We now complete the proof of \Cref{thm:main}~(i)
for $p\in(1,\infty)\backslash\{2\}$.
The endpoint cases $p=1$ and $p=\infty$ are handled
separately in \Cref{sec:endpoint-affine}.

\begin{proof}[Proof of \cref{thm:main}~(i)]
The $\PPAD$-membership is given in \Cref{app:ball-vi-ppad}. For the $\PPAD$-hardness, from \Cref{lem:anchor-multiplier}, it remains to choose $\eps_R$ that makes
$\norm{\vec z^*-\vec s}_p$ sufficiently small. Define
\begin{equation}\label{eq:rho0}
    \rho_0=\min\left\{\frac12,\frac{\eta}{12H}\right\}.
\end{equation}
We need to ensure $\norm{\vec z^*-\vec s}_p\le\rho_0.$
Recall $\mu_p$ and $r_p$ from \eqref{eq:mu-r}, for a given $p$, we choose a
positive constant $c_\eps$ such that
\begin{equation*}
        c_\eps\le\min\{1,\mu_p(1-\tau)\rho_0^{r_p}\},
\end{equation*}
and set
\begin{equation}\label{eq:epsR}
 \eps_R=
 \begin{cases}
 c_\eps/R,&1<p<2,\\
 c_\eps,&p>2.
 \end{cases}
\end{equation}
Since $\norm{F(\vec z^*)}_q\ge c_R\ge(1-\tau)R^{p-1}$,
\Cref{lem:approx-normal} and \eqref{eq:gap-assumption-smooth} give
\begin{align}
    \norm{\vec z^*-\vec s}_p
    &\le
    \left(
    \frac{\eps_R R^{r_p-p}}{\mu_p(1-\tau)}
    \right)^{1/r_p}
    \le
    \left(
    \frac{c_\eps}{\mu_p(1-\tau)}
    \right)^{1/r_p}
    \le\rho_0,
    \label{eq:rho-small}
\end{align}
where the last inequality follows from the choice of $c_\eps$.
By \eqref{eq:rho-small} and
\Cref{lem:realization-error}, we have
\[
    \abs{x_v-N^v_\lambda(s_v(\vx))}
    \le 12H\norm{\vec z^*-\vec s}_p
    \le\eta.
\]
Moreover, \Cref{lem:anchor-multiplier} implies
$\abs{\lambda-1}\le\kappa$, and thus the condition in \Cref{lem:compiler} holds.
Recall that for each node $v \in [n]$, $N^v_\lambda$ denotes the corresponding weighted approximation
function.
Let $\mathrm{gate}_v \in \{\Nor, \Purify\}$ denote the corresponding gate which outputs $v$, we have\begin{align}
    \abs{x_v-\hat\phi_{\mathrm{gate}_v}(s_v(\vx))}
    &\le
    \abs{x_v-N^v_\lambda(s_v(\vx))}
    +\norm{N^v_\lambda-\hat\phi_{\mathrm{gate}_v}}_\infty
    \notag\\
    &\le 2\eta<\eps_{\rm pc}.
    \label{eq:compact-fixed-point}
\end{align}
Since $\hat\phi_{\mathrm{gate}_v}$ is a compactly supported
modification of the original \PureCircuit{} threshold
function, we still need to verify that they agree
at the actual gate inputs. From the range of $\hat\phi_{\mathrm{gate}_v}$, for every node $v \in [n]$, it holds that
$$
        -2\eta\le x_v\le1+2\eta.
$$
For a $\Nor$ gate, its scalar input $s_v(\vx) = x_u + x_w$ therefore lies in $$[-4\eta,2+4\eta]\subseteq[-1,3].$$
For a $\Purify$ gate, its input $s_v(\vx) = x_u \pm 1/4$ lies in
\[
        [-2\eta-1/4,1+2\eta+1/4]\subseteq[-1,2].
\]
From the construction, we have $\hat\phi_{\Nor}=\norfun$ and $\hat\phi_{\Purify}=\purfun$ on these intervals.  Using \eqref{eq:gate-input}, \eqref{eq:compact-fixed-point} becomes
\[
        \norm{\vx-T(\vx)}_\infty\le2\eta<\eps_{\rm pc}.
\]
By \Cref{lem:rounding}, $\vx$ yields a valid \PureCircuit{} assignment.

% \begin{theorem}\label{thm:smooth-radius}
% For every fixed rational $p\in(1,\infty)\backslash\{2\}$, a \PureCircuit{} instance with $n$ nodes can be transformed in polynomial time into a rational affine operator $F$ over $B_p^{d_{\mathrm{aux}}+1}(R)$, where $d_{\mathrm{aux}}=\Theta_p(n)$ is the number of inverse mapping coordinates, $d=d_{\mathrm{aux}}+1$ is the dimension of the ball, and $R=\Theta_p(n)$, such that every $\eps_R$-approximate VI solution, with $\eps_R$ defined in \eqref{eq:epsR}, yields a valid \PureCircuit{} solution in polynomial time.
% \end{theorem}
% \begin{proof}
% The construction and bit complexity are given in \Cref{prop:affine-bits}.  The bounds above show that every $\eps_R$-approximate VI solution yields a $\eps_{\rm pc}$-approximate fixed point of the robust circuit map, which rounds by \Cref{lem:rounding}.
% \end{proof}
% \subsubsection{Rescaling to the unit ball}

The construction so far uses a ball of radius $R$ in \eqref{eq:choice_R}. To obtain the
stated result on the unit ball, we rescale the operator and the variables.
% Let $H_p$ be the bound in \Cref{prop:affine-bits}, and set
% \begin{equation}\label{eq:smooth-normalization-factor}
%  \Lambda_p(R)=\left\lceil\max\{1,H_pR,c_R\}\right\rceil.
% \end{equation}
% After substituting $\vec z=R\widehat{\vec z}$, every nonconstant coefficient is multiplied by $R$, while the constant coefficients are unchanged. Thus $\Lambda_p(R)$ bounds every coefficient of $F(R\widehat{\vec z})$. Moreover,
% \[
%  \Lambda_p(R)=O_p\!\left(R^{\max\{1,p-1\}}\right),
% \]
% and this positive integer can be computed in polynomial time with $O_p(\log n)$ bits.
Set $\tilde{\vec z} = \frac{\vec z}{R} \in B^{d_{\mathrm{aux}}+1}_p$ and $\tilde{F}(\tilde{\vec z}) = F(R \tilde{\vec z}).$
Then
\begin{align}
 \Gap_{B_p^{d_{\mathrm{aux}}+1}}(\tilde{\vec z}; \tilde{F})
 &=\max_{\norm{\vec z'}_p\le1}
   \ip{\vec z'-\tilde{\vec z}}{F(R\tilde {\vec z})} \notag \\
 &=
   \frac{1}{R}\max_{\norm{\vec z'}_p\le R}\ip{\vec z'- \vec z}{F(\vec z)}\notag\\
 &=\frac{1}{R}\Gap_{B_p^{d_{\mathrm{aux}}+1}(R)}(\vec z;F).
 \label{eq:rescale-gap}
\end{align}
Setting $\epsilon' = \epsilon_R / R$ establishes the \PPAD-hardness result over the unit ball $B^{d_{\mathrm{aux}}+1}_p$. Finally, since all coefficients of $\tilde F$
are polynomially bounded in $n$, we can rescale
the entire operator so that it is $1$-Lipschitz.
The resulting approximation error remains
inverse polynomial in $n$, i.e., $\eps = \poly(1/n)$. Since $d_{\mathrm{aux}} = 6mn,$
this proves \Cref{thm:main}~(i) for $p \in (1, \infty) \backslash \{2\}$.
\end{proof}

\subsubsection{The endpoints \texorpdfstring{$p=\infty$ and $p=1$}{p=infinity and p=1}}
\label{sec:endpoint-affine}
\label{sec:cube-affine}
\label{sec:l1-affine}
The endpoint cases $p=\infty$ and $p=1$ follow from known
hardness results for linear variational inequalities over $[0,1]^d$ and for computing symmetric Nash
equilibria. For completeness, we include
a brief reduction for each case.

\paragraph{The case $p=\infty$.}
$\PPAD$-membership is established in \Cref{app:ball-vi-ppad}. We use the constant accuracy hardness of \LinVI{} on $[0,1]^d$. A $\LinVI{}$ instance over the hypercube $[0, 1]^d$ is specified by a rational matrix $\mat{D} \in [-1,1]^{d\times d}$ such that $\norm{\mat{D}}_{1} \leq 1 $ and $\norm{\mat{D}}_{\infty} \leq 1 $, a vector $\vec{b} \in [-1,1]^d$, and an approximation accuracy $\eps \in(0,1]$. The $\LinVI{}$ problem asks for a point $\vec{z} \in [0,1]^d$ such that
\[
(\mat{D} \vec{z} + \vec{b} )_i(z_i'-z_i)\ge -\eps
\quad\forall i\in[d],\ \forall z_i'\in[0,1].
\]

\begin{lemma}[\citealt{Rubinstein15:Inapproximability}, \citealt{bernasconi2024role}]
    \label{lem:LinVI}
    $\LinVI{}$ is \PPAD-hard even when $\eps > 0$ is an absolute constant.
\end{lemma}

% For some absolute constant $\eps_\infty>0$, it is $\PPAD$-hard,
% given a rational matrix $D$ with $\max_i\sum_j|D_{ij}|\le1$
% and a rational vector $c\in[-1,1]^d$, to find $x\in[0,1]^d$ such that
% \[
%     (Dx+c)_i(t-x_i)\ge-\eps_\infty
%     \quad\text{for every }i\in[d]\text{ and }t\in[0,1].
% \]
Set $\vec y=2\vec z-\one$ and define
\[
    F(\vec y)=-\frac12\left(\mat{D}\frac{\vec y+\one}{2}+\vec b\right).
\]
Following similar analysis in \eqref{eq:rescale-gap}, we have
\begin{equation*}
    \Gap_{B_\infty^d}(\vec y;F)
    =\max_{\vec z'\in[0,1]^d}\ip{\vec z-\vec z'}{\mat{D} \vec z+\vec b}.
\end{equation*}
All the coefficients remain bounded in $[-1, 1],$ using \Cref{lem:LinVI} we prove \Cref{thm:main}~(ii).

\paragraph{The case $p=1$.}
Membership in $\PPAD{}$ is proved in \Cref{app:ball-vi-ppad}. By \citet{Chen09:Settling}, computing an
inverse-polynomial approximate Nash equilibrium of a
bimatrix game is $\PPAD$-hard. Applying the standard
symmetrization reduction
\citep[Theorem~2.4]{Nisan07:Algorithmic}
gives \PPAD-hardness for symmetric Nash equilibria with inverse-polynomial accuracy.
Thus, we consider a symmetric game $(\mat{A},\mat{A}^\top)$
with rational $\mat{A}\in[0,1]^{d\times d}$.

Define the affine operator on $B_1^d$ by
\begin{equation*}
    F(\vec z)=\one+\frac12 \mat{A} \vec z.
\end{equation*}
All coefficients lie in $[-1,1]$, and
$1/2\le F_i(\vec z)\le3/2$ for every $\vec z\in B_1^d$.
Moreover,
\begin{equation*}
    \Gap_{B_1^d}(\vec z;F)
    =\max_i F_i(\vec z)-\vec z^\top F(\vec z).
\end{equation*}
Suppose $\Gap_{B_1^d}(\vec z;F) \leq \eps < 1/2$, then $\vec z$ must have at least one positive coordinate. Let $\vec z^+$ denote the positive part
of $\vec z$, that is, $z^+_i = z_i$ if $z_i \geq 0$, otherwise $z^+_i = 0.$ Set $\vec x = \vec z^+/\norm{\vec z^+}_1$.
We have
\begin{align*}
\eps
 \ge \langle \vec x-\vec z,F(\vec z)\rangle
\ge \frac12\sum_{i=1}^d(x_i-z_i)
=\frac12\|\vec z-\vec x\|_1.
\end{align*}
Since $F_i(\vec x) \leq 3/2$ for any $\vec x \in \Delta^d,$ we have
\begin{align*}
|\vec x^\top F(\vec x)-\vec z^\top F(\vec z)|
&\le |(\vec x-\vec z)^\top F(\vx)|
   +|\vec z^\top(F(\vx)-F(\vec z))|\\
&\le \frac32\|\vec x-\vec z\|_1
   +\frac12\|\vec x-\vec z\|_1 \leq 4\eps.
\end{align*}
Thus
\begin{align*}
\max_i(\mat A\vx)_i-\vx^\top\mat A\vx
&=2\left(\max_iF_i(\vx)-\vx^\top F(\vx)\right)\\
&\le
2\left(
\max_iF_i(\vec z)-\vec z^\top F(\vec z)
+\norm{F(\vx)-F(\vec z)}_\infty
+4\eps
\right)\\
&\le 10\eps+\norm{\vx-\vec z}_1 \le 12\eps.
\end{align*}

Thus, by setting $\epsilon$ sufficiently small, we can solve an inverse-polynomial approximate symmetric Nash
equilibrium. The claimed $\PPAD$-hardness follows by rescaling the operator to be $1$-Lipschitz.
\section{Affine VIs over the Euclidean ball}\label{sec:affine-euclidean}
In this section we provide a polynomial-time algorithm for affine VIs on the Euclidean ball. Throughout this section we assume
\[
    F(\vec x)=\mat A\vec x+\vec b,
    \quad \mat A\in\Q^{d\times d},\quad \vec b\in\Q^d,
\]
and the domain is the unit ball $B_2^d=\{\vec x\in\R^d:\norm{\vec x}\le1\}$. We write $\norm{\cdot}$ for the Euclidean norm. Let $\epsilon\in(0,1)$ be a rational accuracy parameter, and let $L$ denote the total binary encoding length of the input $(\mat A,\vec b,\epsilon)$. The main result of the section is stated below:

\begin{theorem}\label{thm:affine}
Given rational $\mat A\in\Q^{d\times d}$, $\vec b\in\Q^d$, and $\epsilon\in(0,1)$, one can compute, in time $\poly(d,L,\log(1/\epsilon))$, a rational vector $\vec x\in B_2^d$ with
\[
    \Gap_{B_2^d}(\vec x;\mat A\vec x+\vec b)\le\epsilon.
\]
\end{theorem}

\subsection{Optimality conditions and the zero case}\label{sec:affine-zero}
By Corollary~\ref{lem:normal-form} in Section~\ref{sec:ball-geometry}, an exact solution either satisfies
\[
    \mat A\vec x+\vec b=\vec 0,\quad \norm{\vec x}\le1,
\]
or satisfies
\begin{equation}
    \mat A\vec x+\vec b=\lambda\vec x,
    \quad \lambda>0,\quad \norm{\vec x}=1.
    \label{eq:affine-boundary}
\end{equation}
Indeed, in the interior of the ball the normal cone contains only the zero vector, while at a boundary point every nonzero outward normal is a positive multiple of that point. This is why the Euclidean case reduces to linear equations once $\lambda$ is fixed.

We first consider the affine equation $\mat A\vec x=-\vec b$. If it is feasible, its minimum-norm solution $\vec x_0$ is the minimizer of the convex quadratic program
\[
    \min_{\vec x\in\R^d}\norm{\vec x}^2
    \quad\text{subject to}\quad \mat A\vec x=-\vec b.
\]
Convex quadratic programming is solvable in polynomial time~\citep{KozlovTarasovKhachiyan1980}. For this program, the minimizer is rational and can be computed exactly since the constraints and the first-order optimality conditions form a linear system with rational coefficients. If $\norm{\vec x_0}\le1$, the algorithm returns $\vec x_0$, and the gap is zero.

Henceforth we assume that $\mat A\vec x=-\vec b$ has no solution in the ball. Every exact VI solution must therefore satisfy~\eqref{eq:affine-boundary}.

\subsection{Bisection on the multiplier}\label{sec:affine-bisection}
For a given rational $\lambda\ge0$, we test whether the linear system
\begin{equation}
    (\lambda\mat I-\mat A)\vec x=\vec b
    \quad\text{has a solution with}\quad \norm{\vec x}\le1.
    \label{eq:affine-ball-test}
\end{equation}
As in the zero case, we minimize $\norm{\vec x}^2$ subject to these equations. The test succeeds precisely when the program is feasible and its minimum is at most one. It therefore takes polynomial time, including when $\lambda\mat I-\mat A$ is singular.

From the discussion of the previous section, we may assume that the test fails at $\lambda=0$. To obtain a value where it succeeds, we take
\[
    \lambda_{\max}=1+\sum_{i,j}|A_{ij}|+\sum_i|b_i|.
\]
Since $\lambda_{\max}>\norm{\mat A}_2$ where $\norm{\mat A}_2$ is the Euclidean operator norm, the matrix $\lambda_{\max}\mat I-\mat A$ is invertible. The solution $\vec x_{\lambda_{\max}}$ of the linear system
$(\lambda_{\max}\mat I-\mat A)\vec x=\vec b$ satisfies
\[
    \lambda_{\max}\norm{\vec x_{\lambda_{\max}}}
    =\norm{\mat A\vec x_{\lambda_{\max}}+\vec b}
    \le\norm{\mat A}_2\norm{\vec x_{\lambda_{\max}}}+\norm{\vec b}.
\]
Therefore
\[
    \norm{\vec x_{\lambda_{\max}}}
    \le\frac{\norm{\vec b}}{\lambda_{\max}-\norm{\mat A}_2}
    \le\frac{\sum_i|b_i|}{1+\sum_i|b_i|}<1.
\]

Bisection starts with $\lambda_\ell=0$ and $\lambda_u=\lambda_{\max}$. At each iteration, we test the midpoint $\lambda_m=(\lambda_\ell+\lambda_u)/2$. If~\eqref{eq:affine-ball-test} succeeds at $\lambda_m$, we replace $\lambda_u$ by $\lambda_m$ and retain a corresponding solution $\vec x_u$ in the ball. Otherwise, we replace $\lambda_\ell$ by $\lambda_m$. Thus, throughout the algorithm, the test fails at $\lambda_\ell$ and succeeds at $\lambda_u$. We stop when
\[
    \lambda_u-\lambda_\ell\le\frac{\epsilon}{4}.
\]
This takes at most $\lceil\log_2(4\lambda_{\max}/\epsilon)\rceil$ iterations. The test need not be monotone in $\lambda$. The only property we maintain is that the two endpoints have different outcomes. The next step explains why two nearby endpoints with this property are sufficient.

\subsection{From nearby multipliers to an approximate solution}
At the end of bisection, let $\vec x=\vec x_u$ be the retained point in the ball, and write $\mat M=\lambda_\ell\mat I-\mat A$. Since $(\lambda_u\mat I-\mat A)\vec x=\vec b$, we have
\[
    \mat M\vec x-\vec b=-(\lambda_u-\lambda_\ell)\vec x,
    \textrm{ and } \norm{\mat M\vec x-\vec b}\le\frac{\epsilon}{4}.
\]
Thus $\vec x$ approximately satisfies the linear equations at $\lambda_\ell$, but its norm may be smaller than one. We next find a point $\vec y$ outside the ball such that the same error bound holds along the entire segment from $\vec x$ to $\vec y$.

If $\mat M$ is invertible, we take $\vec y=\mat M^{-1}\vec b$. The test failed at $\lambda_\ell$, so $\norm{\vec y}>1$. For every $s\in[0,1]$,
\[
    \mat M\bigl((1-s)\vec x+s\vec y\bigr)-\vec b
    =(1-s)(\mat M\vec x-\vec b).
\]
If $\mat M$ is singular, Gaussian elimination gives a nonzero rational vector $\vec w$ with $\mat M\vec w=\vec 0$. We scale it so that $\max_i|w_i|=1$ and take $\vec y=\vec x+3\vec w$. Since $\norm{\vec w}\ge1$, we have
\[
    \norm{\vec y}\ge3\norm{\vec w}-\norm{\vec x}\ge2.
\]
In this case, every point $\vec z$ on the segment satisfies $\mat M\vec z=\mat M\vec x$. This argument applies whether or not the linear system at $\lambda_\ell$ is feasible. In both cases, we have found a rational point $\vec y$ with $\norm{\vec y}>1$ such that
\begin{equation}
    \norm{\mat M\vec z-\vec b}\le\frac{\epsilon}{4}
    \quad\text{for every $\vec z$ on the segment $[\vec x,\vec y]$.}
    \label{eq:affine-small-error}
\end{equation}

We now bisect this segment to obtain a rational point $\vec z$ satisfying
\begin{equation}
    1-\frac{\epsilon}{2\lambda_{\max}}\le\norm{\vec z}\le1.
    \label{eq:affine-near-sphere}
\end{equation}
Initially, $\vec x$ is the inside endpoint and $\vec y$ is the outside endpoint. If the inside endpoint already satisfies~\eqref{eq:affine-near-sphere}, we return it. Otherwise, we take the midpoint. If its norm is at most one, it becomes the inside endpoint; if its norm is greater than one, it becomes the outside endpoint. All comparisons use squared norms, so this procedure uses only rational arithmetic.

To verify the gap bound, we write $F(\vec z)=\lambda_\ell\vec z+\vec e$. By~\eqref{eq:affine-small-error}, $\norm{\vec e}\le\epsilon/4$. Using~\eqref{eq:ball-gap} and the fact that $\lambda_\ell\ge0$, we obtain
\[
\begin{aligned}
    \Gap_{B_2^d}(\vec z;F)
    &=\norm{\lambda_\ell\vec z+\vec e}-\inner{\vec z}{\lambda_\ell\vec z+\vec e}\\
    &\le\lambda_\ell\norm{\vec z}(1-\norm{\vec z})
        +(1+\norm{\vec z})\norm{\vec e}\\
    &\le \lambda_{\max}\frac{\epsilon}{2\lambda_{\max}}+2\frac{\epsilon}{4}=\epsilon.
\end{aligned}
\]
The first error term comes from being slightly inside the sphere, and the second comes from approximately satisfying the linear equations.

\subsection{Algorithm and proof of Theorem~\ref{thm:affine}}
\begin{algorithm}[t]
\caption{Affine VI on the Euclidean ball}\label{alg:affine-ball-vi}
\begin{enumerate}[leftmargin=2em,itemsep=3pt,topsep=3pt]
    \item Apply the zero test of Section~\ref{sec:affine-zero}; return its point if the test succeeds.
    \item Bisect $[0,\lambda_{\max}]$ using~\eqref{eq:affine-ball-test}, keeping failure at $\lambda_\ell$ and success at $\lambda_u$, until $\lambda_u-\lambda_\ell\le\epsilon/4$. Let $\vec x$ be the solution retained at $\lambda_u$.
    \item For $\mat M=\lambda_\ell\mat I-\mat A$, take $\vec y=\mat M^{-1}\vec b$ if $\mat M$ is invertible; otherwise take $\vec y=\vec x+3\vec w$, where $\mat M\vec w=\vec 0$ and $\max_i|w_i|=1$.
    \item Bisect $[\vec x,\vec y]$ and return a point $\vec z$ with $1-\epsilon/(2\lambda_{\max})\le\norm{\vec z}\le1$.
\end{enumerate}
\end{algorithm}

\begin{proof}[Proof of Theorem~\ref{thm:affine}]
The zero case returns a feasible rational point with gap zero. Otherwise, the first bisection produces nearby multipliers $\lambda_\ell,\lambda_u$ and a rational solution at $\lambda_u$ in the ball. The construction above gives an outside point and bounds the error throughout the connecting segment. The second bisection returns a point satisfying~\eqref{eq:affine-near-sphere}, whose gap is at most $\epsilon$ by the preceding calculation.

It remains to bound the running time. The rational number $\lambda_{\max}$ has polynomial encoding length, and the first bisection uses at most $\lceil\log_2(4\lambda_{\max}/\epsilon)\rceil$ iterations. Each midpoint therefore has polynomial encoding length. The convex quadratic programs and the linear systems used to compute $\vec x$, $\vec y$, and $\vec w$ have rational data of polynomial encoding length and can be solved in polynomial time. Their rational outputs also have polynomial encoding length.

For the second bisection, the initial distance between the endpoints is at most $\sum_i|y_i-x_i|$, and this distance is halved at every iteration. Since the outside endpoint has norm greater than one, the inside endpoint has norm at least $1-\epsilon/(2\lambda_{\max})$ once their distance is at most $\epsilon/(2\lambda_{\max})$. Thus at most
\[
    \left\lceil\log_2\left(\frac{2\lambda_{\max}(1+\sum_i|y_i-x_i|)}{\epsilon}\right)\right\rceil
\]
iterations suffice. The coordinates of $\vec x$ and $\vec y$ have polynomial encoding length, so this bound is polynomial in $d$, $L$, and $\log(1/\epsilon)$. All norm comparisons are comparisons of rational squared norms, and all returned points are rational. This completes the proof.
\end{proof}
\section{Quadratic VIs over the Euclidean ball}\label{sec:euclidean_degree_2}

In this section, we complement the polynomial-time algorithm for affine VIs over
the Euclidean ball established in \Cref{sec:affine-euclidean} by
showing that this tractability result is sharp with respect
to the degree of the operator. Specifically, we establish
$\PPAD$-hardness of computing an inverse-polynomial approximate
VI solution when the operator is quadratic.

Our reduction starts from \PureCircuit{} and builds on its
continuous realization introduced in \Cref{sec:purecircuit}.
As in the non-Euclidean reduction, we approximate the threshold functions through discretization. While the previous construction exploits the boundary geometry of the $\ell_p$ ball to obtain a weight function with a sufficiently strong tail bound, here we introduce selector variables to implement the desired weights directly through the quadratic operator.

\subsection{The construction} \label{sec:degree_2_construction}

Let $n=|V|$ be the number of nodes in a \PureCircuit{}
instance, we set $ \delta=1/n$, $m=8n+1,$ and define the grid
\[
    c_r=-4+(r-1)\delta
    \quad\text{for }r=1,\ldots,m.
\]
Then $c_1=-4$, $c_m=4$, and $c_r$ forms a grid over $[-4, 4]$ with spacing $\delta$.
As in \Cref{sec:non-euclidean}, we use $s_v$ and $\phi_v$ from \eqref{eq:gate-input}, so $T_v(\vx)=\phi_v(s_v(\vx))$.
For each pair $(v,r)$, where $v\in [n]$ is a node and $r\in{1,\ldots,m}$ is a grid index, we introduce two additional \emph{selector variables}
\[
    (\alpha_{v,r},\beta_{v,r})= \vec y_{v,r}\in\R^2.
\]
We let $\vec y$ be the vector of all selector variables and further define $t_{v,r}$ to measure the displacement between the target point $s_v(\vx)$ and the sample point $c_r$, normalized by the grid spacing $\delta$. Specifically, we have
\begin{align} \label{eq:selector_dist}
    t_{v,r}(\vec x)= \frac{s_v(\vx) - c_r}{\delta} = n(s_v(\vec x)-c_r).
\end{align}
As discussed in the technical overview, directly imposing $\alpha_{v,r}(1+t_{v,r}(\vx)^2)=1$ would introduce degree-3 terms in the operator. We instead introduce the following selector matrix, which implicitly realizes the desired weight while keeping the operator quadratic. Let
\[
    \mat A_{v,r}(\vec x)=
    \begin{pmatrix}
        1&-t_{v,r}(\vec x)\\
        t_{v,r}(\vec x)&1
    \end{pmatrix}.
\]
Our reduction first uses a ball of radius $R=4n$, which will later be rescaled to the unit ball in \Cref{sec:rescaling}. Let $\vec z=(\vec x,\vec y)\in B_2^d(R)$
denote the concatenation of all variables, where $d=n+2nm$ is the dimension of the VI. For the selector block, we define
\begin{align}\label{eq:selector_operator}
    F_{\alpha_{v,r}}(\vec z)= C_{\rm select}\bigl(1-\alpha_{v,r}+t_{v,r}(\vec x)\beta_{v,r}\bigr), \quad F_{\beta_{v,r}}(\vec z)=C_{\rm select} \bigl(-t_{v,r}(\vec x)\alpha_{v,r}-\beta_{v,r}\bigr),
\end{align}
where $C_{\rm select}>0$ is a scaling parameter whose value will be
specified later.
Moreover, for every node $v$ in~\PureCircuit{}, the quadratic operator corresponding to $x_v$ is
\begin{equation}
    F_{x_v}(\vec z)=\sum_{r=1}^m \left(\alpha_{v,r}(\phi_v(c_r)-x_v)\right).
    \label{eq:Fx}
\end{equation}

As in \Cref{sec:non-euclidean}, we approximate
$T_v(\vx)=\phi_v(s_v(\vx))$ by a weighted average
of the sampled values $\phi_v(c_r)$, assigning
larger weights to grid points $c_r$ closer to
the target input $s_v(\vx)$. In particular,
given a vector $\vx \in \R^n$, consider a vector $\hat{\vec y}(\vx) \in \R^{2mn}$ satisfying
\begin{align}\label{eq:comparison_point_y}
    \begin{pmatrix}1\\0\end{pmatrix} - \mat A_{v,r}(\vec x) \hat{\vec y}_{v,r}(\vx)= 0.
\end{align}
This system is exactly the condition $F_{\vec y}(\vx,\hat{\vec y}(\vx))=0$ with the given $\vx$, and $\hat{\vec y}(\vx)$ is the
corresponding selector block. Solving the system, for every $v \in [n]$ and $r \in [m],$ we get
\begin{equation}
    \hat{\alpha}_{v,r}(\vec x)= \frac{1}{1+t_{v,r}(\vec x)^2},
    \quad
    \hat{\beta}_{v,r}(\vec x)= -\frac{t_{v,r}(\vec x)}{1+t_{v,r}(\vec x)^2}.
    \label{eq:cauchy-selector}
\end{equation}

For a fixed vector $\vx\in\R^n$, define the weighted average
\begin{align}\label{eq:comparison_point_x}
 \bar{x}_v(\vx)=\frac{\sum_{r=1}^m\hat\alpha_{v,r}(\vx)\phi_v(c_r)}{\sum_{r=1}^m\hat\alpha_{v,r}(\vx)}.
\end{align}
All weights $\hat{\alpha}_{v, r}(\vx)$ are positive and all sampled values $\phi_v(c_r) \in [0,1]$, so $\bar\vx(\vx)\in[0,1]^n$ for any $\vx \in \R^n$.
The next lemma shows that at an approximate VI solution $(\vx^*, \vec y^*),$ the selector $\vec y^*$ is close to $\hat{\vec y}(\vx^*).$

\begin{lemma}\label{lem:y_closeness}
Let
$\vec z^*=(\vx^*,\vec y^*)$ be an $\epsilon$-approximate VI solution with the operator defined in \eqref{eq:selector_operator} and \eqref{eq:Fx}. Then
\[
 \norm{\vec y^*-\hat{\vec y}(\vx^*)}_2\le\sqrt{\frac{2\epsilon}{C_{\rm select}}},
\]
where $\hat{\vec y}(\vx^*) = (\hat{\vec{\alpha}}(\vx^*), \hat{\vec{\beta}}(\vx^*))$ is defined in \eqref{eq:cauchy-selector}.
\end{lemma}
\begin{proof}
We use the comparison point $\hat{\vec z}=(\bar\vx(\vx^*),\hat{\vec y}(\vx^*))$. First, \eqref{eq:cauchy-selector} gives
\begin{equation*}
 \norm{\hat{\vec y}_{v,r}(\vx^*)}_2^2
 =\hat\alpha_{v,r}(\vx^*)^2+\hat\beta_{v,r}(\vx^*)^2
 =\frac{1}{1+t_{v,r}(\vx^*)^2}\le1.
\end{equation*}
Since $\bar\vx(\vx^*)\in[0,1]^n$, it follows that
\[
 \norm{\hat{\vec z}}_2^2\le n+nm=8n^2+2n\le16n^2=R^2.
\]
Thus $\hat{\vec z}$ is a feasible comparison point in $B^d_2(R)$.

Moreover, we have 
    \begin{align} \label{eq:vi_decomposition}
        \left\langle \hat{\vec z} - \vec z^*, F(\vec z^*)\right\rangle = \left\langle \bar{\vec x} - \vec x^*, F_{\vx}(\vec z^*)\right\rangle + \left\langle \hat{\vec y}(\vx^*) - \vec y^*, F_{\vec y}(\vec z^*)\right\rangle.
    \end{align}
For the first term in \eqref{eq:vi_decomposition}, \eqref{eq:Fx} and \eqref{eq:comparison_point_x} imply
\begin{align*}
F_{x_v}(\vx^*,\hat{\vec y}(\vx^*))
&=\sum_{r=1}^m\hat\alpha_{v,r}(\vx^*)
\bigl(\phi_v(c_r)-x_v^*\bigr)\\
&=\sum_{r=1}^m\hat\alpha_{v,r}(\vx^*)\phi_v(c_r)
-x_v^*\sum_{r=1}^m\hat\alpha_{v,r}(\vx^*)\\
&=\left(\sum_{r=1}^m\hat\alpha_{v,r}(\vx^*)\right)
\left(
\frac{\sum_{r=1}^m\hat\alpha_{v,r}(\vx^*)\phi_v(c_r)}
{\sum_{r=1}^m\hat\alpha_{v,r}(\vx^*)}
-x_v^*
\right)\\
&=\left(\sum_{r=1}^m\hat\alpha_{v,r}(\vx^*)\right)
\bigl(\bar x_v(\vx^*)-x_v^*\bigr).
\end{align*}
Hence
\begin{align}\label{eq:x_norm_first_term_bound}
 \inner{\bar\vx(\vx^*)-\vx^*}{F_{\vx}(\vx^*,\hat{\vec y}(\vx^*))}=\sum_{v=1}^n\left(\sum_{r=1}^m\hat\alpha_{v,r}(\vx^*)\right)
             \bigl(\bar x_v(\vx^*)-x_v^*\bigr)^2.
\end{align}
Because $|x_v^*|\le R$ and $c_r\in[-4,4]$, \eqref{eq:selector_dist} yields $|t_{v,r}(\vx^*)|\le n(2R+4)$. Let
\[
 \sigma=1+n^2(2R+4)^2.
\]
We have $\hat{\alpha}_{v, r} \geq 1/\sigma$. Thus, \eqref{eq:x_norm_first_term_bound} gives
\begin{equation}\label{eq:x_norm_first_term_bound_2}
 \inner{\bar\vx(\vx^*)-\vx^*}{F_{\vx}(\vx^*,\hat{\vec y}(\vx^*))}
 \ge\frac{1}{\sigma}\norm{\bar\vx(\vx^*)-\vx^*}_2^2.
\end{equation}
It remains to bound the difference between $F(\vec z^*)$ and $F(\vx^*, \hat{\vec y}(\vx^*))$. For every $v$,
\begin{align*}
 \left|F_{x_v}(\vx^*,\vec y^*)-F_{x_v}(\vx^*,\hat{\vec y}(\vx^*))\right|&=\left|\sum_{r=1}^m\bigl(\alpha^*_{v,r}-\hat\alpha_{v,r}(\vx^*)\bigr)
                    \bigl(\phi_v(c_r)-x_v^*\bigr)\right|\\
 &\le(1+R)\sum_{r=1}^m\left|\alpha^*_{v,r}-\hat\alpha_{v,r}(\vx^*)\right|.
\end{align*}
The Cauchy--Schwarz inequality therefore gives
\[
 \norm{F_{\vx}(\vx^*,\vec y^*)-F_{\vx}(\vx^*,\hat{\vec y}(\vx^*))}_2
 \le\sqrt m(1+R)\norm{\vec y^*-\hat{\vec y}(\vx^*)}_2.
\]

For the second term in \eqref{eq:vi_decomposition}, observe that
\begin{align}\label{eq:VI_y_norm_positive}
 \inner{\hat{\vec y}_{v,r}(\vx^*)-\vec y^*_{v,r}}{F_{\vec y_{v,r}}(\vec z^*)}\notag
 &=C_{\rm select}\inner{\hat{\vec y}_{v,r}(\vx^*)-\vec y^*_{v,r}}{\mat A_{v,r}(\vx^*)\bigl(\hat{\vec y}_{v,r}(\vx^*)-\vec y^*_{v,r}\bigr)}\notag\\
 &=C_{\rm select}\norm{\hat{\vec y}_{v,r}(\vx^*)-\vec y^*_{v,r}}_2^2,
\end{align}
where the last step follows from the fact that the off-diagonal part of $\mat A_{v, r}$ is skew-symmetric.

Set $a=\norm{\bar\vx(\vx^*)-\vx^*}_2$ and $b=\norm{\hat{\vec y}(\vx^*)-\vec y^*}_2$. Combining \eqref{eq:x_norm_first_term_bound_2} and \eqref{eq:VI_y_norm_positive}, the VI condition of $\vec z^*$ yields
\begin{align}
 \epsilon& \ge\inner{\hat{\vec z}-\vec z^*}{F(\vec z^*)} \notag \\
 & \ge\frac{\norm{\bar{\vx}(\vx^*) - \vx^*}_2^2}{\sigma}+C_{\rm select}\norm{\hat{\vec y}(\vx^*) - \vec y^*}_2^2-\sqrt m(1+R)\norm{\bar{\vx}(\vx^*) - \vx^*}_2\norm{\hat{\vec y}(\vx^*) - \vec y^*}_2.\label{eq:VI_comparison_decomposition}
\end{align}
Young's inequality bounds the last term by
\[
 \sqrt m(1+R)\norm{\bar{\vx}(\vx^*) - \vx^*}_2\norm{\hat{\vec y}(\vx^*) - \vec y^*}_2\le\frac{\norm{\bar{\vx}(\vx^*) - \vx^*}_2^2}{2\sigma} +\frac{\sigma m(1+R)^2}{2}\norm{\hat{\vec y}(\vx^*) - \vec y^*}_2^2.
\]
Thus, it holds that
\begin{align}\label{eq:joint-comparison-bound}
 \epsilon\ge\frac{\norm{\bar{\vx}(\vx^*) - \vx^*}_2^2}{2\sigma}
 +\left(C_{\rm select}-\frac{\sigma m(1+R)^2}{2}\right)\norm{\hat{\vec y}(\vx^*) - \vec y^*}_2^2.
\end{align}
Choosing
$C_{\rm select}\ge\sigma m(1+R)^2$ and taking the square root completes the proof.
\end{proof}
The above proof also gives the following lemma.
\begin{lemma}\label{lem:x_comparison_clossness}
Let $\vec z^*=(\vx^*,\vec y^*)$ be an $\epsilon$-approximate VI solution. We have
\[
 \norm{\vx^*-\bar\vx(\vx^*)}_2
 \le2\sqrt{\bigl(1+n^2(2R+4)^2\bigr)\epsilon}.
\]
\end{lemma}
\begin{proof}
Dropping the second term in \eqref{eq:joint-comparison-bound} and 
substituting $\sigma=1+n^2(2R+4)^2$ proves the claim.
\end{proof}

\subsection{Approximating the threshold gates}
The next lemma shows that the weighted average $\bar x_v(\vx^*)$ approximates $\phi_v(s_v(\vx^*))$.
\begin{lemma}\label{lem:selector_approx_operator}
Let $(\vx^*,\vec y^*)$ be an $\epsilon$-approximate solution to the VI constructed in \eqref{eq:selector_operator} and \eqref{eq:Fx}. Suppose $\epsilon\le n^{-5}$ and $n$ is sufficiently large. Then, for every $v\in[n]$,
\[
 \left|\bar x_v(\vx^*)-\phi_v(s_v(\vx^*))\right|
 \le\frac{24+48\log(m)}{n}.
\]
\end{lemma}
\begin{proof}
Fix $v\in[n]$. For $\epsilon\le n^{-5}$ and sufficiently large $n$, \Cref{lem:x_comparison_clossness} gives $\norm{\vx^*-\bar\vx(\vx^*)}_\infty\le1$. Since $\bar\vx(\vx^*)\in[0,1]^n$, every coordinate of $\vx^*$ lies in $[-1,2]$, and hence $s_v(\vx^*)\in[-4,4]$.

Let $c_k$ be a nearest grid point to $s_v(\vx^*)$. The spacing is $\delta=1/n$, so $|c_k-s_v(\vx^*)|\le\delta/2$. Therefore
\[
 \hat\alpha_{v,k}(\vx^*)=\frac{1}{1+n^2(s_v(\vx^*)-c_k)^2}\ge\frac12.
\]
Using that $\phi_v$ is $6$-Lipschitz, we get
\begin{align}\label{eq:bound_on_approx_operator}
 \left|\bar x_v(\vx^*)-\phi_v(s_v(\vx^*))\right|
 &\le\frac{6\sum_{r=1}^m\hat\alpha_{v,r}(\vx^*)|c_r-s_v(\vx^*)|}
              {\sum_{r=1}^m\hat\alpha_{v,r}(\vx^*)}\notag\\
 &\le12\sum_{r=1}^m \left(\hat\alpha_{v,r}(\vx^*)|c_r-s_v(\vx^*)|\right).
\end{align}
We group the grid points by nonnegative integer $h$ such that $h \delta \leq |c_r - s_v(\vx^*)| < (h + 1)\delta.$ There are at most two grid points in each group. Moreover, for each grid point $c_r$ in group $h$, it holds that
    \begin{align*}
        \hat{\alpha}_{v, r}(\vx^*) = \frac{1}{1 + n^2(s_v(\vx^*) - c_r)^2} \leq \frac{1}{1 + h^2}.
    \end{align*}
Thus
    \begin{align*}
        \sum_{r = 1}^m \left( \hat{\alpha}_{v, r} |c_r - s_v(\vx^*)|\right) \leq 2\delta \sum_{h = 0}^m \frac{h+1}{1 + h^2} \leq 2\delta (1 + 2 \log(m)).
    \end{align*}
    Combining with \eqref{eq:bound_on_approx_operator} we get the desired bound.
\end{proof}

Combining~\Cref{lem:x_comparison_clossness} and~\Cref{lem:selector_approx_operator}, we are ready to prove the main theorem of this section.
\begin{theorem}\label{them:degree_two_hardness}
Let $\epsilon\leq n^{-5}$, $d = 16n^2 + 3n,$ and $R=4n$. It is \PPAD-hard to compute an $\epsilon$-approximate solution $(\vx^*, \vec y^*) \in B_2^d(R)$ to the VI induced by a quadratic operator.
\end{theorem}
\begin{proof}
    From \cref{lem:x_comparison_clossness} and \cref{lem:selector_approx_operator}, for any coordinate $v \in [n],$ we have
    \begin{align*}
        \left|\phi_{v}(s_v(\vx^*)) - x_v^*\right| & \leq \left|x_v^* - \bar{x}_v(\vx^*)\right| + \left|\bar{x}_v(\vx^*) - \phi_v(s_v(\vx^*))\right| \\
        & \leq 2\sqrt{(1 + n^2(2R+4)^2)\epsilon} + \frac{24+ 48\log(m)}{n}.
    \end{align*}
    By setting $\epsilon \leq n^{-5}$ and with sufficiently large $n$, we have
    \begin{align*}
        2\sqrt{(1 + n^2(2R+4)^2)\epsilon} \leq \frac{1}{64} \quad \text{and} \quad \frac{24+ 48\log(m)}{n} \leq \frac{1}{64}.
    \end{align*}
    Thus
    \begin{align*}
        \norm{T(\vx^*) - \vx^*}_\infty \leq \frac{1}{32}.
    \end{align*}
    The \PPAD-hardness then follows from \cref{lem:rounding} and the fact that finding a satisfying assignment for \PureCircuit{} is \PPAD-hard. The proof is complete.
\end{proof}
\subsection{Rescaling to the unit ball}\label{sec:rescaling}
Finally, we rescale the results in \Cref{them:degree_two_hardness} to the unit ball $B^d_2$ and normalize the coefficients.

\begin{theorem}\label{thm:degree2-unit}
There exists a constant $c>0$ such that
\problem{Ball$_2$-VI} is $\PPAD$-complete
for $\epsilon=d^{-c}$, even when restricted
to quadratic operators.
\end{theorem}
\begin{proof}
The $\PPAD$-membership is established in \Cref{app:ball-vi-ppad}. For the hardness, consider the VI constructed in
\Cref{them:degree_two_hardness}. Following the same
rescaling argument as in \Cref{sec:noneuclidean_final},
we rescale the domain to the unit Euclidean ball
and normalize the operator to be 1-Lipschitz. The resulting approximation accuracy remains inverse-polynomial in the dimension $d$. 
\end{proof}
\section{Conclusions}
In this paper, we characterize the computational complexity of VIs over $\ell_p$ balls with affine and quadratic operators. We show that computing an approximate solution to  affine VIs is $\PPAD$-complete for every fixed $p\in[1,\infty)\backslash\{2\}$ as well as for $p=\infty$. In contrast, the Euclidean ball exhibits a different complexity landscape: affine VIs admit a polynomial-time algorithm, whereas quadratic VIs are $\PPAD$-complete. For $p\in[1,\infty)$, our hardness results hold at inverse-polynomial accuracy. One immediate open question is whether our results can be extended to constant accuracy.

More generally, our techniques illustrate how the geometry of the feasible set and the structure of the operator can be exploited to establish both hardness and tractability. We hope that these techniques will find broader use in characterizing the complexity of other equilibrium and optimization problems.

\paragraph{Acknowledgments.} Ioannis Panageas and Jingming Yan were supported by National Science Foundation grant CCF-2454115.

\paragraph{AI methodology.} We worked on these problems over several months and used OpenAI’s GPT-5.6 and GPT-6.0 during that time. We developed the polynomial-time algorithm for affine variational inequalities on the Euclidean ball without using GPT.

For the hardness proof for quadratic operators on the Euclidean ball, we used GPT to explore several proof strategies. These included a formulation as a linear complementarity problem and an argument that treated interior and boundary solutions separately. The proof in this paper draws on the selector construction from an earlier work on team zero-sum games~\citep{Anagnostides26:Team}. For affine operators on non-Euclidean \(\ell_p\) balls, GPT helped us to extend the discretization from the Euclidean case. We also used both models to check individual proof steps and refine the exposition of the corresponding subsections. We checked the arguments ourselves, rewrote the relevant passages, and take full responsibility for the correctness and originality of the entire paper, including its references.

\bibliographystyle{plainnat}
\bibliography{references,refs}

\newpage
\appendix
\section{Convexity and continuity of the normal map}
\label{appendix:convexity}

For completeness, we establish the convexity and
duality-map continuity bounds used in
\Cref{lem:approx-normal}.

\begin{lemma}
\label{lem:uniform-convexity-appendix}
Let $1<p<\infty$ and $\vec u,\vec v\in\R^d$.
The following inequalities hold
\begin{enumerate}[label=(\roman*)]
    \item For $p\ge2$, let
    $\Phi(\vec u)=\norm{\vec u}_p^p/p$. Then
    \begin{equation}
    \label{eq:p-uniform-convexity-appendix}
        \Phi(\vec u)
        \ge \Phi(\vec v)
        +\ip{J_p(\vec v)}{\vec u-\vec v}
        +\frac{2^{2-p}}{p}\norm{\vec u-\vec v}_p^p.
    \end{equation}

    \item For $1<p\le2$, let
    $\Psi(\vec u)=\norm{\vec u}_p^2/2$. Then
    \begin{equation}
    \label{eq:p-strong-convexity-appendix}
        \Psi(\vec u)
        \ge \Psi(\vec v)
        +\ip{\norm{\vec v}_p^{2-p}J_p(\vec v)}
             {\vec u-\vec v}
        +\frac{p-1}{2}\norm{\vec u-\vec v}_p^2.
    \end{equation}
\end{enumerate}
\end{lemma}

\begin{proof}
At $p=2$, both inequalities coincide and follow
directly from $J_2(\vec u)=\vec u$ and the strong convexity of the Euclidean norm. We consider the remaining cases separately.
\begin{itemize}
    \item Case $p > 2.$ We first prove the scalar inequality
    \begin{equation}
    \label{eq:scalar-monotone}
        (\abs{a}^{p-2}a-\abs{b}^{p-2}b)(a-b) \ge 2^{2-p}\abs{a-b}^p.
    \end{equation}
    We consider the
    normalized inequality. When $a-b\ne0$, dividing
    by $\abs{a-b}^p$ and setting $t = b/(a-b)$ reduces
    the problem to
    \[
    \abs{t+1}^{p-2}(t+1)-\abs{t}^{p-2}t \geq 2^{2 - p}.
    \]
    Differentiating the left side shows that the expression attains
    its global minimum at $t=-1/2$. Substituting this value yields $2^{2-p}$, proving \eqref{eq:scalar-monotone}.

    Summing \eqref{eq:scalar-monotone} coordinatewise gives
    \begin{equation}
    \label{eq:Jp-monotone-appendix}
        \ip{J_p(\vec u)-J_p(\vec v)}{\vec u-\vec v}
        \ge 2^{2-p}\norm{\vec u-\vec v}_p^p.
    \end{equation}
    Recall that $\Phi(\vec u)=\norm{\vec u}_p^p/p$.
    Let $\vec h=\vec u-\vec v$, for
    $t\in[0,1]$, define
    \[
    G(t):=
    \Phi(\vec v+t\vec h)-\Phi(\vec v)
    -t\ip{J_p(\vec v)}{\vec h}
    -\frac{2^{2-p}}{p}t^p\norm{\vec h}_p^p.
    \]
    For $t>0$, applying
    \eqref{eq:Jp-monotone-appendix} to
    $\vec v+t\vec h$ and $\vec v$ gives
    \[
        \ip{J_p(\vec v+t\vec h)-J_p(\vec v)}{\vec h}
        \ge 2^{2-p}t^{p-1}\norm{\vec h}_p^p.
    \]
    Differentiating $G$, we get
    \[
    G'(t)
    =\ip{J_p(\vec v+t\vec h)-J_p(\vec v)}{\vec h}
     -2^{2-p}t^{p-1}\norm{\vec h}_p^p
    \ge0.
    \]
    Thus, $G$ is nondecreasing on $[0,1]$. Since $G(0)=0$, we have $G(1)\ge0$, which yields
    \eqref{eq:p-uniform-convexity-appendix}.
    \item Case $1 < p <2.$ We consider $\Psi(\vec u)=\norm{\vec u}_p^2/2$.
    \citet[Proposition 3]{BallCarlenLieb1994} showed that, for any
    $\vec x,\vec y\in\R^d$ and $p \in [1, 2]$, we have
    \[
        \frac{\norm{\vec x+\vec y}_p^2
        +\norm{\vec x-\vec y}_p^2}{2}
        \ge
        \norm{\vec x}_p^2
        +(p-1)\norm{\vec y}_p^2.
    \]
    Setting
    $\vec x=(\vec u+\vec v)/2$ and
    $\vec y=(\vec u-\vec v)/2$ yields
    \[
    \Psi(\vec u)+\Psi(\vec v)
    \ge
    2\Psi\left(\frac{\vec u+\vec v}{2}\right)
    +\frac{p-1}{4}\|\vec u-\vec v\|_p^2.
    \]
    Thus
    \[
        \Psi\left(\frac{\vec u+\vec v}{2}\right)
        \le
        \frac{\Psi(\vec u)+\Psi(\vec v)}{2}
        -\frac{p-1}{8}\norm{\vec u-\vec v}_p^2.
    \]
    Since the midpoint inequality holds for arbitrary
    vectors, repeated application and the continuity of
    $\Psi$ imply that $\Psi$ is $(p-1)$-strongly convex. Therefore
    \[
        \Psi(\vec u)
        \ge
        \Psi(\vec v)
        +\ip{\nabla\Psi(\vec v)}{\vec u-\vec v}
        +\frac{p-1}{2}\norm{\vec u-\vec v}_p^2.
    \]
    Using
    $\nabla\Psi(\vec v)
    =\norm{\vec v}_p^{2-p}J_p(\vec v)$
    yields \eqref{eq:p-strong-convexity-appendix}.
\end{itemize}
\end{proof}

We next establish the continuity properties of the
duality map $J_p$ needed in \Cref{lem:approx-normal}.

\begin{lemma}
\label{lem:duality-continuity-appendix}
Let $1<p<\infty$ and $q=p/(p-1)$ be the dual
exponent. The following inequalities hold.
\begin{enumerate}[label=(\roman*)]
    \item For $1<p\le2$ and any
    $\vec u,\vec v\in\R^d$,
    \begin{equation}
    \label{eq:Jp-holder-appendix}
        \norm{J_p(\vec u)-J_p(\vec v)}_q
        \le 2^{2-p}\norm{\vec u-\vec v}_p^{p-1}.
    \end{equation}

    \item For $p>2$ and any
    $\vec u,\vec v\in B_p^d(R)$,
    \begin{equation}
    \label{eq:Jp-lipschitz-appendix}
        \norm{J_p(\vec u)-J_p(\vec v)}_q
        \le (p-1)R^{p-2}\norm{\vec u-\vec v}_p.
    \end{equation}
\end{enumerate}
\end{lemma}

\begin{proof}
We again consider the two cases separately.
\begin{itemize}
    \item Case $1<p\le2$. Put $\alpha=p-1\in(0,1]$.
    If $a,b$ have the same sign, concavity of
    $t^\alpha$ on $[0,\infty)$ gives
    \[
        \bigl|\abs{a}^{p-2}a-\abs{b}^{p-2}b\bigr|
        \le \abs{a-b}^{p-1}.
    \]
    If they have opposite signs, the same concavity gives
    \[
        \abs{a}^\alpha+\abs{b}^\alpha
        \le2^{1-\alpha}(\abs{a}+\abs{b})^\alpha
        =2^{2-p}\abs{a-b}^{p-1}.
    \]
    Applying these scalar bounds coordinatewise and
    using $(p-1)q=p$, we obtain
    \eqref{eq:Jp-holder-appendix}.
    \item Case $p>2$. Let $\vec u,\vec v\in B_p^d(R)$, put
    $\vec h=\vec u-\vec v$, and define
    $\vec z(t)=\vec v+t\vec h$.
    For any $\vec w$ with $\norm{\vec w}_p\le1$,
    differentiation and H\"older's inequality give
    \begin{align*}
    \left|\frac{d}{dt}
        \ip{\vec w}{J_p(\vec z(t))}\right|
    &=(p-1)\left|
        \sum_i w_i\abs{z_i(t)}^{p-2}h_i
        \right|\\
    &\le(p-1)\norm{\vec w}_p
        \norm{\vec z(t)}_p^{p-2}\norm{\vec h}_p\\
    &\le(p-1)R^{p-2}\norm{\vec u-\vec v}_p.
    \end{align*}
    The mean value theorem and duality of the $p$
    and $q$ norms therefore imply
    \eqref{eq:Jp-lipschitz-appendix}.
\end{itemize}
\end{proof}

\section{Membership in \PPAD{} for $p\geq 1$ or $p=\infty$}
\label{app:ball-vi-ppad}

In this section, we prove membership in $\PPAD{}$ for the circuit
formulation of \problem{Ball$_p$-VI}.

\begin{theorem}
\label{thm:ball-vi-ppad}
For every fixed rational $p\geq 1$, and also for $p=\infty$,
\problem{Ball$_p$-VI} belongs to $\PPAD{}$.
This holds for a $1$-Lipschitz operator $F$ represented by a
well-behaved arithmetic circuit and every rational $\epsilon\in(0,1)$
encoded in binary.
\end{theorem}

\begin{proof}
Let $q$ denote the dual exponent of $p$, as in \Cref{sec:preliminaries}.
We construct a continuous map $H\colon[-1,1]^d\to[-1,1]^d$ whose
values can be approximated in polynomial time. We choose a rational
$\delta>0$ such that, given any rational $\vec{z}\in[-1,1]^d$ with
$\norm{H(\vec{z})-\vec{z}}_\infty\leq\delta$, we can compute in polynomial
time a rational point in $B_p^d$ with VI gap at most $\epsilon$.
Finding such a $\vec{z}$ is in $\PPAD{}$ by the approximate fixed point
theorem of Etessami and Yannakakis~\citep{EtessamiYannakakis2010}, in
the form given by Chen, Paparas, and Yannakakis~\cite[Theorem~6.8]{ChenPaparasYannakakis2017},
and the claim will follow. Below we give the construction of $H$.

\paragraph{The projected map.}
We denote by $\Pi=\Pi_{B_p^d}$ the Euclidean projection onto $B_p^d$.
For $\vx\in B_p^d$, we define
\[
    T(\vx)=\Pi\bigl(\vx+F(\vx)\bigr).
\]
By the characterization of Euclidean projection, $\vx=T(\vx)$ if and
only if $F(\vx)\in N_{B_p^d}(\vx)$. As observed after
\Cref{def:vi-gap}, this is equivalent to~\eqref{eq:VI}.

The projection inequality also bounds the VI gap when $\vx$ is close
to $T(\vx)$. Indeed, writing $\vec{y}=T(\vx)$, we have
\[
    \inner{\vec{u}-\vec{y}}{\vx+F(\vx)-\vec{y}}\leq 0
    \qquad\text{for every }\vec{u}\in B_p^d.
\]
It follows that
\begin{align*}
    \inner{\vec{u}-\vx}{F(\vx)}
    &\leq \inner{\vec{u}-\vec{y}}{\vec{y}-\vx}
          +\inner{\vec{y}-\vx}{F(\vx)} \\
    &\leq \bigl(2\sqrt d+\norm{F(\vx)}_2\bigr)\norm{\vec{y}-\vx}_2,
\end{align*}
where the last inequality uses $\norm{\vec{u}-\vec{y}}_2\leq 2\sqrt d$.
Since $\vx\in B_p^d\subseteq[-1,1]^d$, we have $\norm{\vx}_2\leq\sqrt d$.
The triangle inequality and the $1$-Lipschitz continuity of $F$ give
\[
    \norm{F(\vx)}_2
    \leq \norm{F(\vx)-F(\vec{0})}_2+\norm{F(\vec{0})}_2
    \leq \norm{\vx}_2+\norm{F(\vec{0})}_2
    \leq \sqrt d+\norm{F(\vec{0})}_1.
\]
Thus the coefficient $2\sqrt d+\norm{F(\vx)}_2$ in the preceding bound
is at most $3\sqrt d+\norm{F(\vec{0})}_1\leq 3d+\norm{F(\vec{0})}_1$.
Setting $M=3d+\norm{F(\vec{0})}_1$ and recalling that $\vec{y}=T(\vx)$,
we obtain
\begin{equation}
\label{eq:membership-projected-gap}
    \Gap_{B_p^d}(\vx;F)
    =\max_{\vec{u}\in B_p^d}\inner{\vec{u}-\vx}{F(\vx)}
    \leq M\norm{T(\vx)-\vx}_2.
\end{equation}

\paragraph{The map on the cube.}
We extend $T$ to the cube by defining
\[
    H(\vec{z})=T(\Pi(\vec{z}))
    \textrm{ with } \vec{z}\in[-1,1]^d.
\]
Since $\Pi$ is nonexpansive and $F$ is $1$-Lipschitz, both $T$ and
$H$ are $2$-Lipschitz in the Euclidean norm.

We choose $\delta=\epsilon/(4d^2M)$. Given a rational $\vec{z}$ with
$\norm{H(\vec{z})-\vec{z}}_\infty\leq\delta$, we return the rational point
\[
    \vx=\frac{\vec{z}}{1+d\delta}.
\]
To see that $\vx$ is feasible, we use $H(\vec{z})\in B_p^d$ to obtain
\[
    \norm{\vec{z}}_p
    \leq \norm{H(\vec{z})}_p+\norm{\vec{z}-H(\vec{z})}_p
    \leq 1+d\delta.
\]
Thus $\vx\in B_p^d$ and $H(\vx)=T(\vx)$. Moreover,
$\norm{\vx-\vec{z}}_2=d\delta\norm{\vx}_2\leq d\sqrt d\,\delta$.
Using the Lipschitz bound on $H$, we obtain
\begin{align*}
    \norm{T(\vx)-\vx}_2
    &\leq \norm{H(\vx)-H(\vec{z})}_2
          +\norm{H(\vec{z})-\vec{z}}_2+\norm{\vec{z}-\vx}_2 \\
    &\leq 3\norm{\vx-\vec{z}}_2+\sqrt d\,\delta
     \leq 4d^2\delta.
\end{align*}
By~\eqref{eq:membership-projected-gap}, the VI gap at $\vx$ is
at most $4d^2M\delta=\epsilon$.

\paragraph{Computing the projection.}
It remains to show that $H$ can be approximated in polynomial time.
For a rational $\vec{b}\in\R^d$, its projection
$\vec{y}=\Pi(\vec{b})$ is the unique minimizer of the convex function $g:\R^d\to \R$ 
\[
    g(\vec{u})=\tfrac12\norm{\vec{u}}_2^2+\norm{\vec{b}-\vec{u}}_q.
\]
Indeed, $\vec{b}-\vec{y}\in N_{B_p^d}(\vec{y})$, so $\vec{y}$ attains
the support value in direction $\vec{b}-\vec{y}$. Hence
$\inner{\vec{b}-\vec{y}}{\vec{y}}=\norm{\vec{b}-\vec{y}}_q$
by~\eqref{eq:p-gap} for $1<p<\infty$ and by
$\ell_1$--$\ell_\infty$ duality at the endpoints.
Since $\vec{y}\in B_p^d$, H\"older's inequality also gives
$\norm{\vec{b}-\vec{u}}_q\geq\inner{\vec{b}-\vec{u}}{\vec{y}}$
for every $\vec{u}\in\R^d$.
Combining these two relations, we obtain
\[
    g(\vec{u})-g(\vec{y})
    \geq \tfrac12\bigl(\norm{\vec{u}}_2^2-\norm{\vec{y}}_2^2\bigr)
          +\inner{\vec{y}-\vec{u}}{\vec{y}}
    =\tfrac12\norm{\vec{u}-\vec{y}}_2^2.
\]

For fixed rational $1<q<\infty$, the norm $\norm{\cdot}_q$ can be
approximated in polynomial time by bisection for fixed rational powers.
For $q=1$ and $q=\infty$, it can be evaluated exactly on rational
inputs. We can therefore apply the weak convex minimization theorem
of Gr\"otschel, Lov\'asz, and Schrijver~\cite[Theorem~4.3.13]{Grotschel93:Geometric}
to $g$ on $[-2,2]^d$. Since $\vec{y}\in[-1,1]^d$ is at distance at
least one from the boundary of this cube, for any rational $0<\rho<1$
the theorem gives a rational $\vec{w}$ with
\[
    g(\vec{w})\leq g(\vec{y})+\rho^2/2
\]
in time polynomial in the input length and $\log(1/\rho)$. The preceding
bound on $g(\vec{w})-g(\vec{y})$ implies $\norm{\vec{w}-\vec{y}}_2\leq\rho$.

Finally, $\norm{\vec{w}}_p\leq 1+d\rho$, so $\vec{w}/(1+d\rho)$ is a
rational point in $B_p^d$ within Euclidean distance
$(1+d\sqrt d)\rho$ of $\vec{y}$. We can therefore approximate the
inner projection in $H$ by a feasible rational point, evaluate the
circuit for $F$ at that point, and approximate the outer projection.
The Lipschitz bounds on $F$ and $\Pi$ ensure that polynomially many
bits suffice throughout. The approximate fixed point theorem now
applies to $H$, proving the claim.
\end{proof}
\section{Affine ball VIs need not satisfy the Minty condition}\label{appendix:minty}
The affine Euclidean algorithm does not assume that a Minty solution exists. With the maximization convention in \eqref{eq:VI}, the Minty condition asks for a point $\vx^*\in B_2^d$ such that
\[
 \inner{\vx^*-\vxp}{F(\vxp)}\ge0\quad\text{for every }\vxp\in B_2^d.
\]
Consider first the one-dimensional ball $B_2^1=[-1,1]$ and the affine operator $F(x)=x$. The condition at $x'=1$ requires $x^*\ge1$, whereas the condition at $x'=-1$ requires $x^*\le-1$. No point satisfies both requirements.

The same obstruction holds in higher dimensions. On $B_2^d\subset\R^d$, take $F(\vx)=\vx$. For any candidate $\vx^*$, choose a unit vector $\vec u$ with $\inner{\vec u}{\vx^*}<1$ and set $\vxp=\vec u$. Then
\[
 \inner{\vx^*-\vxp}{F(\vxp)}
 =\inner{\vx^*}{\vec u}-1<0.
\]
Thus affine VIs over the Euclidean ball need not satisfy the Minty condition.

\section{Omitted technical lemmas}
\begin{lemma}\label{lem:polynomial-tail}
For every $\beta>1$ and positive integer $i$, we have
\[
    \sum_{r=i+1}^{\infty}r^{-\beta}
    \le \frac{i^{1-\beta}}{\beta-1}.
\]
\end{lemma}
\begin{proof}
Since $x^{-\beta}$ is decreasing for $\beta>1$, it holds that
\[
    \sum_{r=i+1}^{\infty}r^{-\beta}
    \le \int_i^\infty x^{-\beta}\,dx
    =\frac{i^{1-\beta}}{\beta-1}.
\]
The proof is complete.
\end{proof}

\begin{lemma}\label{lem:anchor-coefficient}
For every fixed rational $p>1$ and fixed rational
$\tau\in(0,1)$, there is a polynomial-time algorithm that, given
a positive integer $R \in \Z_+$, computes a positive rational $c_R$ such that
\begin{equation}\label{eq:cR-relative}
    1-\tau\le\frac{c_R}{R^{p-1}}\le1+\tau.
\end{equation}
\end{lemma}
\begin{proof}
Write $p=a/b$, where $a>b\ge1$ are fixed integers.
Choose an integer $k\ge1$ such that $2^{-k}\le\tau$,
and define
\[
    M=\left\lfloor
    \left(2^{kb}R^{a-b}\right)^{1/b}
    \right\rfloor,
    \quad c_R=\frac{M}{2^k}.
\]
By construction,
\[
    0\le R^{p-1}-c_R<2^{-k}\le\tau.
\]
Since $R\ge1$, we have $R^{p-1}\ge1$, and hence
\[
    1-\tau\le\frac{c_R}{R^{p-1}}\le1.
\]
Finally, since $a$, $b$, and $k$ are constants,
the integer $M$ can be computed in polynomial time.
The resulting rational $c_R$ also has polynomial bit length.
\end{proof}

\end{document}